\documentclass[12pt,preprint]{aastex}

\newcommand{\Ha}{H$\alpha$}

\newcommand{\Hb}{H$\beta$}

\newcommand{\htwo}{H$_{2}$}

\newcommand{\kms}{km~s{$^{-1}$}}

\newcommand{\Msun}{M$_{\hbox{$\odot$}}$}

\newcommand{\tc}{{$\theta^1$~Ori~C}}

\newcommand{\tE}{{$\theta^1$~Ori~E}}

\newcommand{\oiii}{[O~III]}

\newcommand{\oi} {[O~I]}
\newcommand{\nii}{[N~II]}
\newcommand{\sii}{[S~II]}
\newcommand{\siii}{[S~III]}

\newcommand{\vt}{V$\rm _{T}$}
\newcommand{\vrad}{V$\rm _{r}$}

\newcommand{\vsun}{V$_{\hbox{$\odot$}}$}
\newcommand{\Vomc}{V$_{\hbox{OMC}}$}

\newcommand{\Te}{T$\rm_{e}$}

\newcommand{\hr}{Huygens Region}

\slugcomment{Submitted to the  AJ}

\shorttitle{Outflows in the Central ONC}
\shortauthors{O'Dell}

\begin{document}
\title{The Nature and Frequency of Outflows from Stars in the Central Orion Nebula Cluster\
\footnote{
Based on observations with the NASA/ESA Hubble Space Telescope,
obtained at the Space Telescope Science Institute, which is operated by
the Association of Universities for Research in Astronomy, Inc., under
NASA Contract No. NAS 5-26555.}
\footnote{Based on observations at the San Pedro Martir Observatory operated by the  Universidad Nacional Aut\'onoma de M\'exico.}}

\author{C. R. O'Dell}
\affil{Department of Physics and Astronomy, Vanderbilt University, Box 1807-B, Nashville, TN 37235}

\author{G. J. Ferland}
\affil{Department of Physics and Astronomy, University of Kentucky, Lexington, KY 40506}

\author{W. J. Henney}
\affil{Instituto de Radioastronom\'{\i}a y Astrof\'{\i}sica, Universidad Nacional Aut\'onoma de M\'exico, Apartado Postal 3-72,
58090 Morelia, Michoac\'an, M\'exico}

\author{M. Peimbert}
\affil{Instituto de Astronomia, Universidad Nacional Aut\'onoma de M\'exico, Apdo, Postal 70-264, 04510 M\'exico D. F., M\'exico}

\author{Ma.T. Garc\'ia-D\'iaz}
\affil{Instituto de Astronomia, Universidad Nacional Aut\'onoma de M\'exico,  Km 103 Carretera Tijuana-Ensenada, 22860 Ensenada, B.C., M\'exico}

\and

\author{Robert H. Rubin\
\footnote{Deceased}}
\affil{NASA/Ames Research Center, Moffett Field, CA 94035-0001}

\email{cr.odell@vanderbilt.edu}

\begin{abstract}

Recent Hubble Space Telescope images have allowed the determination with unprecedented accuracy of motions and changes of shocks within the inner Orion Nebula. These originate from collimated outflows from very young stars, some within the ionized portion of the nebula and others within the host molecular cloud. We have doubled the number of Herbig-Haro objects known within the inner Orion Nebula. We find that the best-known Herbig-Haro shocks originate from a relatively few stars, with the optically visible X-ray source COUP 666 driving many of them. 

While some isolated shocks are driven by single collimated outflows, many groups of shocks are the result of a single stellar source having jets oriented in multiple directions at similar times. This explains the feature that shocks aligned in opposite directions in the plane of the sky are usually blue shifted because the redshifted outflows pass into the optically thick Photon Dominated Region behind the nebula. There are two regions from which optical outflows originate for which there are no candidate sources in the SIMBAD data base.

\end{abstract}
\keywords{HII regions--ISM:individual(Orion Nebula, NGC 1976, M42, Orion-South--protoplanetary disks)
}

\section{BACKGROUND AND INTRODUCTION}
\label{sec:intro}

The proximity of the Orion Nebula and its associated Orion Nebula Cluster of young stars makes this the best
test object for studying regions of massive star formation, the structure of the surrounding remaining gas and dust, and testing the assumptions made in the study of more distant HII regions. In the investigation reported upon here, we give new and important information on the nebula and on the outflows from the low-mass stars belonging to the cluster.

The 3-D structure of the Orion Nebula has been the subject or by-product of
numerous studies, the most recent and thorough being those of \citet{ode08b,ode10}. Most of the emission comes from a relatively thin blister of ionized gas on the observer's side
of the Orion Molecular Cloud. Since the radial velocity of the ionized gas becomes progressively more blue shifted with increasing ionization state \citep{zuc73,bal74}, the ionizing source (\tc) must lie between the observer and the molecular cloud. This ionized gas decreases in density as it flows away from the Main Ionization Front (MIF) and towards the observer. In the opposite direction, beyond the MIF, is a thin but high density Photon Dominated Region (PDR). In the near foreground  are layers of mostly neutral material known as the 
Veil (c.f. \citet{pvdw} and references therein), which reveals itself through HI absorption lines at 21-cm, its extinction of visible light (the Dark Bay to the east of the Trapezium asterism is the best exemplar), and spectroscopic 
absorption lines occurring from ultraviolet through radio wavelengths. The 
optically brightest part of the nebula is the nearly edge-on  \citep{md11} ionization front enveloping a dense molecular cloud (Orion-S) that lies between the MIF and the foreground
Veil. The dominant ionizing star \tc\ lies at the middle of the Orion Nebula Cluster (ONC) of stars.
 Although Orion-S is likely to be a "free-floating" condensed structure \citep{ode10}, it is presently
impossible to rule out that it is a dense feature with an ionization shadow projecting away from \tc. This would be similar to the "Pillars of Creation" images in NGC~6611 except that in the Orion Nebula we view the pillar at a large angle with respect to the plane of the sky.

The existence of a new set of images of a major portion of the inner Orion Nebula at the highest resolution yet realized and the ability to astrometrically compare those images with some of the earliest images made with the Hubble Space Telescope (HST) presents the opportunity to determine motions and changes with unprecedented accuracy. Those data can be compared with the results of recent high resolution radial velocity mapping of the same region. The access to these two sets of complementary data justifies a new study of the inner Orion Nebula.

\subsection{Our Basic Approach}
\label{sec:approach}

The images used in this program were made with the WFPC2 camera of the Hubble Space Telescope and its successor the WFC3 camera. The 
narrow-band emission line filters F487N, F502N, F656N, F658N, and F673N
are very similar in both cameras and isolate well emission from
\Hb\ 486.1 nm, \oiii\ 500.7 nm, \Ha\ 656.3 nm, \nii\ 658.4 nm and the 
\sii\ doublet at 671.6+673.1 nm. The intermediate width filter F547M,
which is free of strong emission lines and is used less frequently. For convenience, we commonly indicate the signal in a filter by the name
of the filter, for example the \Ha\ image is simply called the F656N
image and the ratio of the \Ha\ to \Hb\ images (usually normalized so that each is near unity) as F656N/F487N.  We call ratio images those that are the ratio
of the signal from two filters. F656N/F487N is a good measure of the line of sight 
reddening. Since the Balmer lines originate from throughout any ionized zone  F502N/F487N (or F502N/F656N when the reddening does not vary 
significantly over the FOV) is a good measure of the high ionization gas, F658N/F656N is a good measure of the low ionization gas, and F673N/F656N is a good measure of very low ionization gas lying very near an ionization front. F547M/F487N is a good measure of the relative strength of the nebular continuum to
hydrogen line emission. Although we did not use ACS camera images in this study, they can be compared with those used here. It should be noted that the ACS filter designated as F658N passes both the \Ha\ 656.3 nm and \nii\ 658.4 nm lines and the ACS  F660N filter isolates the \nii\ 658.4 nm line.
 
\begin{figure}
\epsscale{1.0}
\plotone{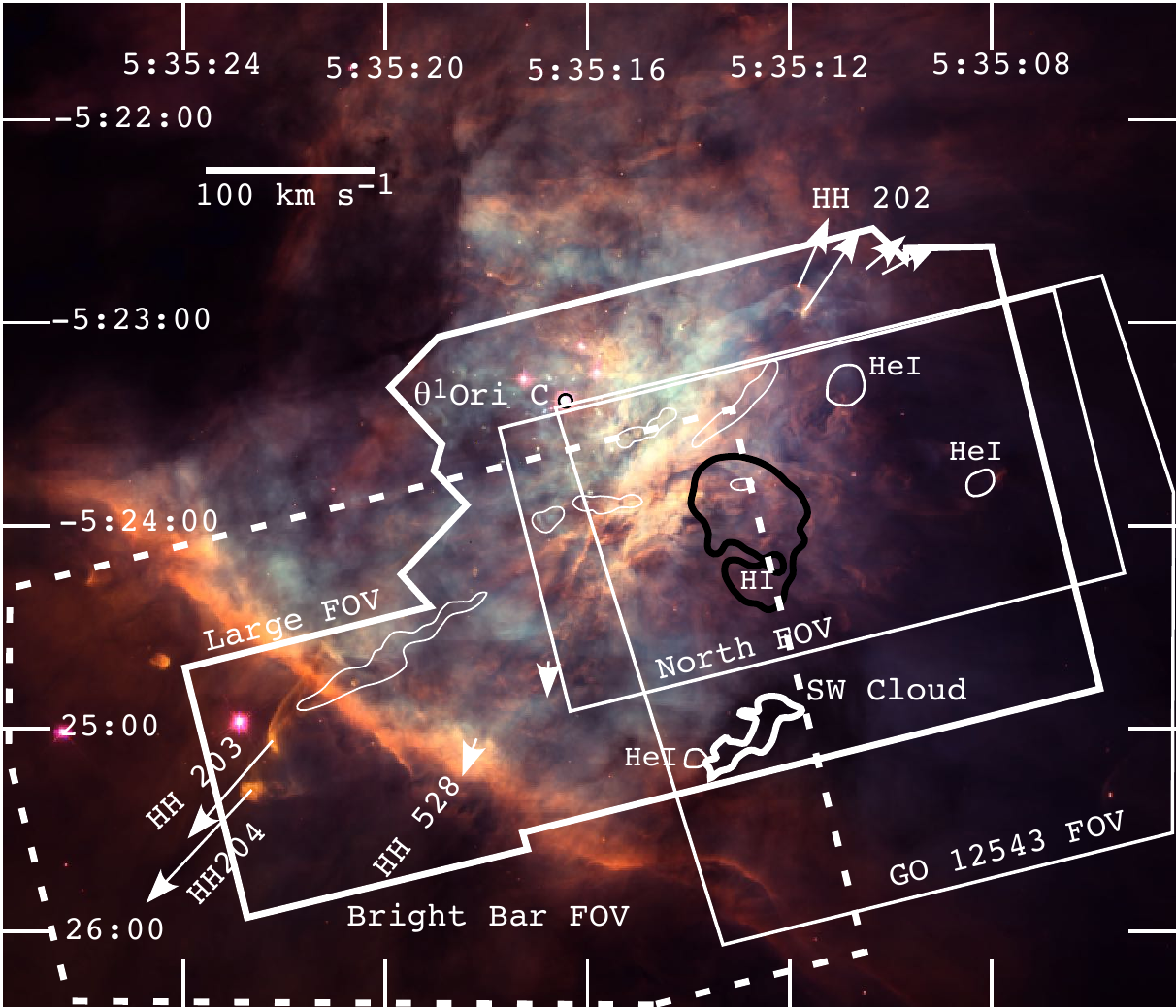}
\caption{This 348\arcsec $\times$ 299\arcsec\ image  with north up is extracted from \citet{ode96} and shows the fields, sometimes cropped, examined in this study together with related nearby features. The 2000 coordinates are shown.  The full field of the GO~12543 images is shown to the southwest from the dominant ionizing star \tc. The exact location of the smaller fields overlapping with the North FOV are shown in better detail in Figure~\ref{fig:fig2}. The North FOV extends east of the GO12543 and includes only WFPC2 images, as explained in Section~\ref{sec:NewObs}.
The Bright Bar FOV is used for Figure~\ref{fig:fig24}. The Large FOV is used for Figures~\ref{fig:fig21}, \ref{fig:fig22}, and \ref{fig:fig23}.
Motions of objects lying outside of the present tangential velocity study are shown for HH~203 and HH~204, HH~528, and HH~202  
\citep{ode08a}.  The light lines denote high radial velocity features probably feeding these shocks \citep{doi04} except for the three features 
labeled He~I for reasons addressed in Section~\ref{sec:intro}.The irregular structure shown with heavy black line and labeled HI is 
described in Section\ref{sec:EarlierObs}. The heavy white line  designated as SW Cloud indicates the region of high and isolated extinction in 
the study of \citet{ode00b}. }
\label{fig:fig1}
\end{figure}

\placefigure{fig:fig1}

The region studied in this investigation is shown in Figure~\ref{fig:fig1}.
The primary source of new information is from the GO 12543 observations. 
The northern portion of this FOV was most useful and was extended to the east 
by using data from GO~5469 and GO~11038.  We have added to the figure the positions of several well-known and well-studied 
Herbig-Haro
objects. We have also shown the high
velocity jets identified by \citet{doi04}i and strong features found in the
the low velocity resolution study of the 1083 nm He~I line by \citet{tak02}. They are included here because \citet{hen07} established that these features are visible because they are Doppler shifted off the absorption core of this optically thick line .

The irregular structure shown with heavy black line and labeled HI is the H~I 21 cm line absorption feature \citep{pvdw} associated with Orion-S. 

A primary aim of this study is to determine the motion and changes of nebular
features, jets, and shocks in emission lines. The methodology we use as a guide 
for identifying motions and changes is to divide an aligned first epoch image by
the aligned second epoch image. We call this a motion image and it has the
characteristic that a single bright moving object will be dark in the direction of motion
and light in the trailing direction. The opposite pattern will be seen in the case of a moving or changed dark object. A motion image will be of constant value of 
unity (since each image is normalized to unity) and only the moving and changed
features will appear.  

\subsection{Nomenclature and Reference Numbers}
\label{sec:nomen}

Throughout this paper a number of designations, acronyms, and abbreviations will be used. Some are in common usage and some are not.
In this section we identify them. 

Herbig-Haro objects will be called HH objects. The molecular cloud to the southwest of the Trapezium stars that is seen in H$\rm _{2}$CO and H~I absorption (\citet{pvdw} and references therein) will be called 
Orion-S. The brightest part of the Orion Nebula, NGC~1976, will be designated as the \hr\ (the region shown in the first published drawing made by Christian Huygens in 1659), while the fainter larger elliptical region to the southwest will be the Extended Orion Nebula \citep{gud08} or EON. The Position angle will be PA and the Field of View will be FOV.

Where positions are given they are in epoch 2000 and of the form 5:35:16.46 -5:23:22.85 (the coordinates of the dominant ionization source \tc). Radial velocities (\vrad) are in the Heliocentric system (\vsun) 
and for easy comparison with the Local Standard of Rest, one subtracts 18.1  \kms\ from \vsun.
The radial velocity of the host molecular cloud is 25.8 \kms~\citep{ode08b}. This velocity is used in derivation of the spatial motion (\Vomc) of objects with respect to the host cloud. The angle of this velocity vector with respect to the plane of the sky will be $\theta$ and positive values indicate motion towards the observer. This notation for positive values of $\theta$ is different from that in other studies but is appropriate here because the optically thick OMC forms an opaque background.

\citet{ode94} introduced a position-based designation system that
has come into common use. It drew on the fact that most of the objects
within the \hr\ fall near \tc\ and the designations are simply the abbreviated coordinates. For
example \tc\ would be designated as 165-323, that is the Right Ascension values were rounded to 0.1$\rm ^{s}$  and 5:35 is deleted, and Declination values
to 1\arcsec, with -5:2 deleted. In many cases we now have positions to better accuracy and when this is the case we introduce here a system rounding off to 0.01$\rm ^{s}$ and 0.1\arcsec, so that \tc\ becomes 164.6-322.8 instead of 165-323.
Usually an object designated as XXX-XXX is a shock or other extended feature.

We have relied on the SIMBAD data base for positions of stellar sources. Publications that have been particularly important sources of positions are used with the designations HC \citep{hil00}, MAX \citep{rob05}, and COUP \citep{get05} where these are more convenient than the position-based designations. Almost always a source has been detected in multiple studies. If it has a COUP designation, we will usually use this, but when the object was already well known by a different name, we use that.
Designations of the compact objects discussed in the text are summarized in Table~\ref{tab:SIMBAD}. The epoch of the coordinates is 2000.0. Errors in this catalog are noted in Appendix~\ref{AppendixA}. The spectral types and sometime spectral type ranges are from \citet{hil13}. 

Since the position of large features and regions are often expressed relative
to other objects, when the precision of a PA value is not necessary, we use the 
abbreviations of directions, for example SSW for south southwest. All of our 
images are displayed with  14\arcdeg\ along the positive Y axis unless
otherwise noted. This orientation reflects that of our new images with the WFC3
in program GO~12543. We will use CW and CCW to indicate clockwise and counterclockwise changes of PA.
The abbreviation PA is usually omitted when expressing angles in degrees of arc.

\newpage
\placetable{tab:SIMBAD}
\begin{deluxetable}{llcl}
\tabletypesize{\tiny}
\tablecaption{Compact Sources in the Optical Outflow Source Region Mentioned in this Study\label{tab:SIMBAD}}
\tablewidth{0pt}
\tablehead{
\colhead{Designation in Text} &
\colhead{RA \& DEC Designation} & 
\colhead{Spectral Type} &
\colhead{Other Designations*}}
\startdata
COUP 9   & ** &  K3III-IVe  & JW 46, 2MASS, H 46 \\
V2202 Ori     & 105.4-416.5 & K8-M0 & COUP 385, JW 349, MAX 9, MLLA 220, HC 146, HHH 349, LML 52\\
LQ Ori     & 107.3-344.6 & K2-M1 & COUP 394, DRS 21, HC 224, JW 352, MAX 12, MLLA 316, ZRK 4\\ 
d109.4-326.7 & 109.4-326.7 & --- & HC 286, MLLA 379 \\
COUP 419    & 113.1-426.5 & --- &  HC 127, LML 77, MLLA 194\\
COUP 423     & 114.9-351.9 & --- &  MAX 19, MLLA 289, HC 203, LML 80 \\
COUP 443    & 117.0-351.3 &M0-M3 & BOM d117-352, COUP 443, JW 368, MLLA 290, HC 205, HHH 368, LML 91 \\    
V1228 Ori & 122.8-348.0 &K1-M0& MAX 27, SB 8, HC 215, HHH378a \\
COUP 478  & 123.4-352.4 & --- & FBG 291, HC 206, LML 108, LR 3, MLLA 292\\
COUP 480   & 124.6-404.1 &---&  MLLA 245, HC 169, LML 109\\
----               & 129.9-401.6 & ---&  2MASS \\
----                & 133.6-359.6 & ---&2MASS  \\
V1398 Ori    & 134.4-340.2 &$<$M0& COUP 545, JWLR. 409, MAX 41, MLLA 327, HC 240, HHH409 \\
COUP 554   & 135.6-355.3 & ---&MAX 43, MLLA 276, SB 4, HC 192, LML 139 \\
----                 & 135.6-402.6 & ---&2MASS\\
EC 13                 & 135.7-408.2 & ---& ZRK 137-408\\
COUP 555    & 136.0-359.0 & ---&  MAX 42, MLLA 263, HC 178, LML 138, SB 5, ZRK 136-359\\                 
COUP 564    & 136.8-345.3 &---&  EC-MM8, HC 222, MLLA 313, LML 141, ZRK 137-347\\
EC 9& 137.2-350.6 & ---& ---\\
MAX 46    & 137.8-340.0 &---&  MLLA 328, SB 1, LM 78, HC 242, LML 144, LR 32, PMF 35\\
COUP 582        & 138.6-407.1  &---& ---\\
COUP 593  & 139.2-320.3 & ---&HC 314, LML 149, MAX 52, MLLA 413, OW 139-320\\
EC 14                  & 139.3-409.4 & ---& MM14, ZRK 139-409\\
COUP 602      & 140.4-338.3 & M0-M3&  JW 431, MLLA 336, TCC 1, HC 247, HHH 431, LML 152\\
--- & 140.9-351.2 &---& F 053246.64-052544.72\\
COUP 607  & 141.7-357.0 & --- & --- \\
V1328  Ori   & 142.8-424.6 & ---&COUP 616, JW 437, MAX 57, MLLA 203, HC 135, LML 156\\
DR 1186  & 142.9-353.1 &---&  ZRK 15\\
EC16 & 143.6-354.6 & ---&---\\
COUP 632 & 144.0-350.9 & ---&MAX 61, MLLA 293, SB 2, LM 1, LML 162\\
H 20051 & 144.6-353.8 & ---& 2MASS \\
HC193 &  145.3-355.1 & --- & MLLA 273B\\
H 20045 & 145.6-349.4 &---& ---\\
HC 209 & 145.7-350.8 & ---& FBG 421,  LML 170, MLLA 295, ZRK 141-351\\
H 20044 &147.7-352.0 & ---& ---\\
COUP 666      & 148.0-346.0 &K6-M4& HBJ 20030, JW 453, HHH 453, MLLA 306,\\  
      ---          &    --- &---&TCC 14, DR 623, HC 220, LML 181, LR 72, SI 47\\
H 20041/2  & 148.3-351.3 &---& ---\\
H 20036 & 148.6-350.1 & --- & ---\\
COUP 679  & 150.1-354.0 & --- &  HC 195, LML 193, LR 83, MLLA 280, MC 20 \\
H 20018 &150.7-341.0 &---& ---\\
ZRK 24 & 150.8-353.0 & --- & --- \\
H 20027 & 151.3-349.7 & --- & --- \\
ZRK 25 & 151.5-353.6 & --- & --- \\
EC 17 & 151.6-340.6 & ---& --- \\
--- & 152.3-340.4 & ---&H 20014 \\
COUP 691& 152.5-349.9 &--- &  MLLA 297, HC 211, LML 204, LR 92\\
DR 769 & 153.0-355.8 & --- & --- \\
MAX 77 & 154.9-352.3 & --- & --- \\
COUP 717   &155.1-337.2 & K & BOM d155-338, HC 251, MAX 79, LML 212, ZRK 34\\
COUP 725 & 156.8-339.0 &---&  MLLA 334, TCC 37, HC 246, LM 49, LML 221, LR 114\\
COUP 728 & 156.8-533.2 & K8e & 2MASS,  DR 241, JW 482, MLLA 56, V2274\\
HC 236 & 157.0-341.9  & ---& LML 222, MLLA 38, LR 115\\
COUP 734 & 157.3-337.9 & --- & HC 248, LML 224, TCC 41\\
----  & 157.6-338.4 &---& Discovered in this study. \\
$\theta^1$~Ori~E & 157.7-310.0 & mid-G-Giants & COUP 732, HC 344, MLLA 48, TCC 40\\
LV 6   & 157.9-326.7 & G4-K5& HC 287, HHH 489, JW 489, MAX 86, MLLA 381, TCC 42, ZRK 40\\
COUP 747     & 158.5-325.5 &---& MLLA 385, TCC 47, HC 291, LML 232, ZRK 43\\
COUP 757 & 158.7-337.6 & --- & DR 1007, HC 250, LML 237, LR 128, MLLA 339, TCC 50, ZRK 44\\
V2279 Ori***     &  159.3-349.9 &G4-M2& COUP 758, HC 455, JW 499, HHH 499, MAX 92, MLLA 296\\
AC Ori           &   159.8-352.7 & F2-K7&COUP 768, JW 503, HC 202, HHH 503, MAX 94, MLLA 288, ZRK 47\\ 
LV 4 & 160.5-324.4 & --- & HC 296, LML 245, LR 134, MAX 98, MLLA 389, TCC 54, ZRK 48 \\
COUP 769   &   160.7-353.3 &---&  LML 248\\
LV 3  & 162.8-316.5  &---&  COUP 787, JW 512, MAX 105, MLLA 422, TCC 63, HC 322, OW 163-317, ZRK 52\\
HC 292 & 163.2-325.3 & --- & LML 269, LR 154, MLL 386, TCC 66\\
HC 341 & 164.0-311.3 & --- &DOH~5, MLLA 445, TCC 67 \\
COUP 820  & 166.1-316.1 & ---& MLLA 426, TCC 70, HC 325, OW 166-316, ZRK 58\\
LV 2  & 167.2-316.6 &G5-K5& $\theta^1$~Ori~G, COUP 826, JW 524, MAX 116, MLLA 424, HC 323, HHH 524\\
COUP 827 & 167.6-328.0  &---&  MLLA 375, HC 284, OW 168-328, TCC 74, ZRK 60 \\
COUP 900 & 175.7-324.7 & $>$G6 & HC 295, LML 325, LR 207, MAX 131, MLLA 387, TCC 95\\
HC~271 & 180.4-330.9 & --- & LML 347, LR 234, MLLA 363, TCC 106, ZRK 74\\
COUP 943 & 180.5-401.0 & --- & DR 458, JW 575, HC 177, LM 350, LR 235\\
\enddata
\tablecomments{~*Other Designations: 2MASS, \citep{2mass}; BOM, \citet{bal00}; COUP, \citet{get05}; DOH, \citet{doi04}; EC, \citet{eis06}; DR, \citet{dar09}; FBG, \citet{fei02}; F, \citet{fel93}; H, \citet{rod09}; HBJ,\citet{her02}; HC, \citet{hil00}; HHH,\citet{hil13}; JW, \citet{jw88}; LM, \citet{lad00}; LML, \citet{lad04}; LR, \citet{luh00}; 
LV, \citet{lv79}; MAX, \citet{rob05}; MC, \citet{mcc94}; MLLA, \citet{mlla}; NW, \citep{nut07}; OW, \citet{ode94}; SB, \citet{smi04}; SI, \citet{sim99}; PMF, \citet{pri08}; TCC,\citet{mcc94}; ZRK, \citet{zap04a}; **Position 5:34:39.89 -5:26:42.1; ***fainter companion 0.51\arcsec at  47\arcdeg}
\end{deluxetable}

\newpage
\subsection{A Short Primer on Photoionization Physics}
\label{sec:physics}

The basic physics of an astronomical photo-ionized gas is a mature subject
and well described in the popular text of \citet{ost06}. In a blister type nebula
one finds, when proceeding from the molecular cloud towards the ionizing star, first a mostly
molecular and dust rich PDR. This is most visible in its radio and infrared emission lines and the scattered star light that produces a continuum that
is much stronger than what is expected of an atomic gas.  At the very thin H$\rm ^{o}$- H$\rm ^{+}$ MIF one begins to get photoionization of hydrogen atoms and this continues throughout
the ionized zone. \sii\ emission arises from very near the MIF because sulfur
rapidly becomes doubly ionized when moving away from the MIF towards the star. Immediately
beyond the MIF is a low-ionization zone where helium is neutral and nitrogen is singly ionized and easily visible in \nii. Further out helium becomes singly ionized and the easiest way to trace this zone is the coexisting doubly ionized oxygen's \oiii\ 
emission.  The hottest star in the Orion Nebula  is unable to doubly ionize helium and hence higher ionization states of oxygen are not found. This means that even though the observer of the Orion Nebula is looking at a column that passes through several levels of ionization, we know where the emission originates 
along the line of sight.  When the ionization front is tilted more nearly along the line of sight, as found in the Orion Bright Bar, the ionization layers become even more obvious, with \oiii\ being closest to the 
ionizing star, followed by \nii\ and finally \sii.

The \hr\ hosts many stellar jets and outflows, and these may simply be dominated by photoionization. However, in the case of the highest velocity features
collisional ionization can occur, a useful reference number is that the kinetic energy in eV
of a hydrogen atom is 0.00518$\times$V$\rm _{gas}^{2}$, where 
V$\rm _{gas}$ is the gas velocity in \kms. Anticipating the results of this study, one can say that collisional ionization is important in many HH objects. \htwo\ emission can arise from the PDR, where it can become quite visible when a jet is passing through and the gas becomes warmer. 

There is a nebular continuum that arises from atomic processes such as recombination, two-photon, and free-free emission. 
In the case of the Orion Nebula, the observed continuum is much stronger than the expected atomic continuum because of starlight scattered by the dust within the PDR. The light in the extended area to the south of the brightest part of the nebula is primarily scattered light coming from the \hr\ \citep{ode10}.

The new observations in this study do not include measurements of the weak auroral transitions of \nii\ or \oiii, which are needed to determine the electron temperature. However, in the absence of variable extinction the \Ha /\Hb\ ratio would decrease with increasing electron temperature and in the absence of strong scattered stellar continuum, one expects that the continuum would grow stronger relative to the \Hb\ line with increasing electron temperature (meaning that the signal ratio F547M/F487N should increase.

\subsection{Outline of This Paper}
\label{sec:outline}

The order of presentation follows.  The observational material and the data processing are presented in Section~\ref{sec:data}. East-West flows are described in Section~\ref{sec:eastwest}; Large-scale features in the \hr\ in Section~\ref{sec:connections}, smaller Herbig-Haro systems in Section~\ref{sec:smallHH}, and ionization shadows in Section~\ref{sec:shadow}.

In Section~\ref{sec:discussion} the results are discussed in the following order: the nature and origin of the major flow systems (Section \ref{sec:fans}, 
the relation of our results with respect to 21-cm absorption line studies (Section~\ref{sec:21cm}), 
the large-scale outflows (Section~\ref{sec:Biggies}),
outflows coming from blank areas (Section~\ref{sec:blanks}),
shocks that are not the results of collimated outflows (Section~\ref{sec:NonHH}), 
individual sources (Section~\ref{sec:sources}).

\section{OBSERVATIONAL DATA}
\label{sec:data}

We are able to draw on both new and existing observational data for both imaging and spectroscopy. Since the ionization range of the Orion Nebula is quite low, we use the high signal to noise (S/N) ratio F658N and F502N  images in our analysis (although occasionally good quality F673N images are available and are used). There was already an excellent set of high resolution (10 \kms) spectroscopy mapping the \hr\ (Garc\'ia-D\'iaz et al. 2008 and references therein). We have been able to supplement these spectra with superior data for the \oiii\ 500.7 nm line in the highest ionization part of the nebula.

\subsection{New Imaging Observations}
\label{sec:NewObs}

Our new imaging observations were made with HST's WFC3 as part of program GO~12543. The field of view covered is shown in Figure~\ref{fig:fig1}. Observations were made (2012 January 7) with the narrow-band emission line filters F487N (\Hb\ 409 s), F502N (\oiii\ 348 s), F656N (\Ha\ 349 s), F658N (\nii 602 s), and F673N (\sii\ 700 s) in addition to observations with the continuum sampling intermediate width filter F547M (348 s). 
The characteristics of these filters and their calibration have been described by \citet{ode13a}.  Although the images are of high quality, the gap between the two detectors in the WFC3 have poorer cosmic ray event canceling and these have generally been left in the images, except where hand-editing was necessary.  Since the present study does not try to do spectrophotometry, we usually normalized the images to the same signal level.

These are the highest angular resolution (0.04 \arcsec/pixel sampling ) optical images of a portion of the \hr. When used alone, we employed the original images in combination with one another.  When used for comparison with earlier (under-sampled) WFPC2 images (0.0996\arcsec/pixel) we processed them with IRAF \footnote{IRAF is distributed by the National Optical Astronomy Observatories, which is operated by the Association of Universities for Research in Astronomy, Inc.\ under cooperative agreement with the National Science foundation.} task ``gauss'' to match their broader image cores.

\begin{figure}
\epsscale{0.8}
\plotone{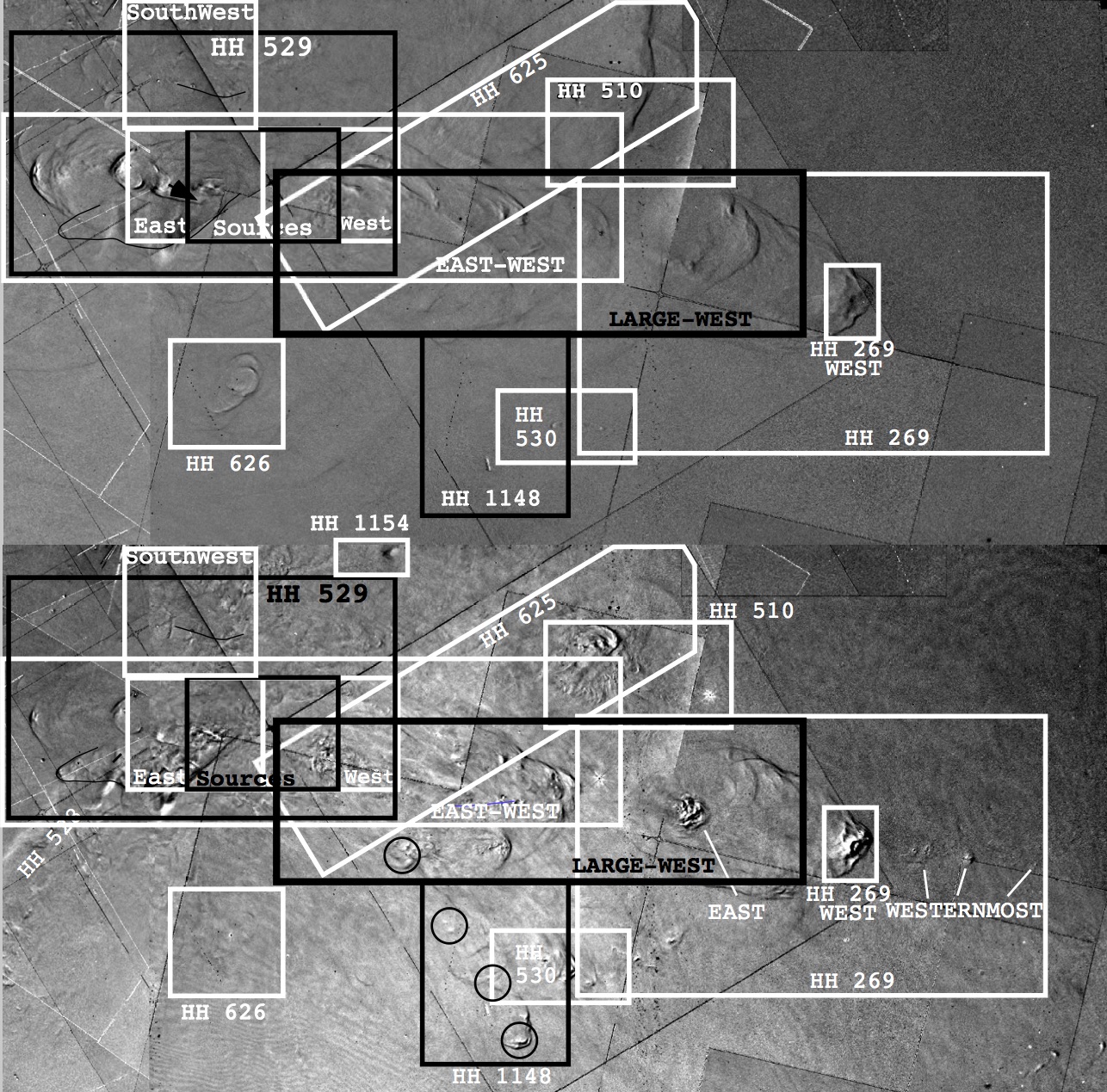}
\caption{This pair of 196.4\arcsec $\times$91.8\arcsec\ images isolates the northern region of the images examined in detail. It has PA = 14\arcdeg\ up, which is the case for all other images in this paper unless otherwise noted. The region is designated as ``North  FOV'' in Figure~\ref{fig:fig1}. Each image is centered at 5:35:11.0 -5:23:53. It shows motion images, with F502N used for the upper panel and F658N for the lower. This will be the pattern in the remaining images in this study. The black and white outlines show the FOV's used for individual flows: HH~269 (Section~\ref{sec:HH269}); HH~269-West (Section~\ref{sec:HH269West}); HH~510 (Section~\ref{sec:HH510}); HH~625 (Section~\ref{sec:HH625}), East-West \ref{sec:eastwest}, West (Section~\ref{sec:OOSwest}), East (Section~\ref{sec:OOSeast}), Sources~(\ref{sec:sources}), HH~529 (Section~\ref{sec:HH529}), HH~530 (Section~\ref{sec:HH530redefined}), HH~626 (Section~\ref{sec:HH626}), 
Large-West (Section~\ref{sec:LargeWest}), SouthWest(\ref{sec:nearSW}), and HH~1148 (Section~\ref{sec:HH1148}). In the lower panel the circles designate the shocks argued in Section~\ref{sec:HH1148} to compose the large-scale flow HH~1148.}

\label{fig:fig2}
\end{figure}

\subsection{Earlier Imaging Observations}
\label{sec:EarlierObs}

We have been able to use earlier HST observations made with the WFPC2. In order to compare these images with the new WFC3 images, we aligned AstroDrizzle processed versions of them with the GO 12543 AstroDrizzle processed images using stars common to both. Table~\ref{tab:images} summarizes the data sets we used.  For the FOV of the WFC3 we matched the WFPC2 images by magnifying them by 2.5247 as determined from measuring a large set of common stars to an error of $\pm$0.0003, to the same pixel scale as the WFC3.  In a small region to the east of the GO 12543 FOV we could only use WFPC2 images. The east extension of our primary area of study is shown in Figure~\ref{fig:fig1}. Data from the GO 12543 FOV was used whenever possible, with the lower-resolution data from the WFPC2 being used only in the small east extension. The Guest Observer program numbers of the data used are given in Table~\ref{tab:images}. 

\subsection{Spectroscopic Observations}
\label{sec:spectra}

The most useful data set of spectra is the compilation of north-south orientation long-slit spectra by \citet{gar08}. In addition to their original observations \citep{gar07} in low ionization lines, they recalibrate the high ionization spectra of  \citet{doi04} and present combined results for emission lines from a wide variety of ionization states(\oi\ 630.0 nm, \sii\ 671.6 nm+673.1 nm, \nii\ 658.4 nm, \siii\ 631.2 nm, \Ha\ 656.3 nm, \oiii\ 500.7 nm) calibrated to 2 \kms\ accuracy and a resolution of about 10 \kms. We will refer to this as the Spectroscopic Atlas, or simply the Atlas.

New observations were made at the San Pedro Martir observatory in February 2013 in the \oiii\ 500.7 nm line in essentially the same manner as the earlier low ionization line observations, except that the slit was oriented east-west. The slit center was at 5:35:15.9. Fifteen spectra were obtained in steps of 1.4\arcsec\ starting at 23\arcsec\ south of \tc\ and proceeding south. Their total exposure time in each slit of 300 seconds or 600 seconds, depending on the source brightness, with the 2.1-m telescope's MEZCAL spectrograph gave higher S/N ratio images than the earlier observations made at the Kitt Peak National Observatory with the 4-m telescope's echelle spectrograph by \citet{doi04}.  

\subsection{Determination of Changes of Position and Structure}
\label{sec:motions}

In order to detect changes of position (the tangential motion, which we designate by the tangential velocity calculated with the assumption of a distance to the Orion Nebula of 440 pc \citep{ode08a}) and changes of the structure, the first and second images had to be aligned. This was done using stars common to both FOVs and the IRAF tasks ``geomap'' and ``geotran'', using the GO 102543 images as the reference system. The scatter of the matched images was $\pm$0.5 pixels (2.5 \kms) for the WFC3-WFPC2 combinations and $\pm$0.2 pixels (3 \kms) for the WFPC2-WFPC2 combinations used in the east extension of the GO12543 FOV.

We have chosen to use the ratio of the first epoch image divided by the second epoch image to identify the moving and changing features. 
Figure~\ref{fig:fig2} shows the power of this approach. This figure shows the very different changes in the images of high and low ionization gas and the eye is able to identify large-scale patterns quite well.
A bright linear feature moving orthogonally will appear as a dark/light combination of lines, with the dark line being in the direction of motion. The pattern would be the reverse for a linear dark feature. In the event of a mix of bright and dark features the pattern of light-dark fails to be a guide
to changes and only the measurement of specific features is useful.

The WFPC2 images often have ''scars'' where CCD boundaries occur. These propagate through to the motion images as straight lines of various forms. Likewise, the motions and ratio images formed from a combination of exposures often show changes of the nebula's ''background'' signal because of imperfect flat-field corrections. These too propagate through into the motions and ratio images. Rather than make cosmetic adjustment for either the boundaries or the flat-field correction step (because this involves modifying the scientific information), we have left the images in this scarred, but rigorous form.

We measured individual features by two methods. The first method was a least-squares image-shifting procedure developed by \citet{har01} from an approach originated by \citet{cur96}. In many cases where the features being measured were very complex, we identified individual small structures and determined their motions by direct comparison of the images. This is the same methodology employed in our earlier studies of the \hr\ (O'Dell \&\ Henney 2008a and references therein), the brightest LL Ori objects in the Orion Nebula \citep{hen13}, and the Ring Nebula \citep{ode13c}. The results for individual features are given in Appendix B.

\begin{deluxetable}{lcccc}
\tabletypesize{\scriptsize}
\tablecaption{Data Sets Used in Determination of Tangential Velocities\label{tab:datasets}}
\label{tab:images}
\tablewidth{0pt}
\tablehead{
\colhead{Program} &
\colhead{Camera} &
\colhead{Modified Julian Date} &
\colhead{Program Pair} &
\colhead{Velocity Scale (\kms /pixel)*} }
\startdata
GO 5085(FOV5)  & WFPC2 & 49737 & -----   & ----- \\
GO 5469              & WFPC2 & 49797 & -----   &----- \\
GO 11038(FOV1) & WFPC2 & 54406 & -----  & ----- \\
GO 12543            & WFC3    & 55935 & ----- & ----- \\
-----                       &   ------    & -----     & 5085+12543 & 4.92 \\
-----                       &   ------    & -----     & 5469+12543 & 4.96 \\
-----                       &   ------    & -----     & 5469+11038 &  16.46 \\
\enddata
\tablecomments{~*Velocity Scale was determined using a distance of 440 pc \citep{ode08a} and pixel scales of 0.04\arcsec/pixel (WFC3) and 0.0996\arcsec/pixel (WFPC2).}
\end{deluxetable}

\subsection{Determination of Radial Velocities}
\label{sec:RadVel}

The \nii~658.3 nm and \oiii~ 500.7 nm archived spectra had an instrumental  full width at half maximum intensity (FWHM) of about 10 \kms~and the new \oiii~ spectra's FWHM are slightly smaller. The observed FWHM is the quadratic addition of the instrumental FWHM and the thermal broadening FWHM. The thermal broadening is about  5.7 \kms~for \nii~and 5.4 \kms~ for \oiii~if the electron temperature is 10000 K. 
The quadratic addition of the instrumental and thermal FWHM values gives expected observed FWHM of 11.5 \kms~for \nii~ and 11.4 \kms~for \oiii. In the case of the nebula's background emission, there may also be an additional small broadening due to acceleration of gas away from the nebula's MIF.  

Since the radial velocity features corresponding to the high tangential velocity features are usually faint as compared with the background, how well a radial velocity component can be measured will be determined by its contrast. Those lines of high doppler shift from the nebula's systemic velocity (18 \kms~ for \nii~and 15 \kms~ for \oiii) are the most easily detected, with accuracy of measurement of about 2 \kms. Things get worse as the doppler shift decreases, with a line of about equal intensity to the nebula's emission resolvable down to about a velocity difference of 3 \kms~ and with a comparable uncertainty. These observational biases constrain the results of our radial velocity database analogous to the way that form and change of structure constrain the tangential velocity database.

\begin{figure}
\epsscale{0.8}
\plotone{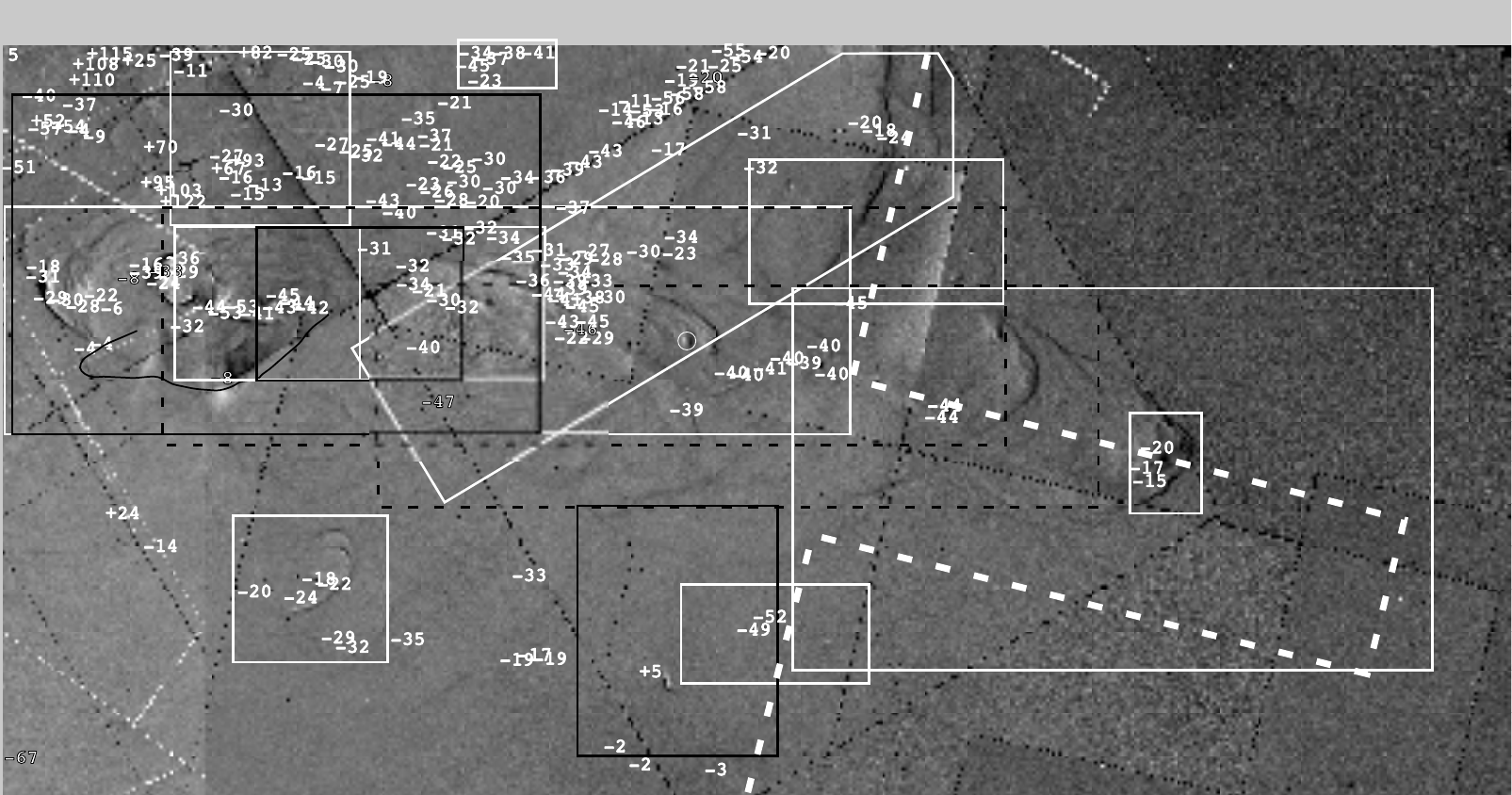}
\caption{Like Figure~\ref{fig:fig2} except showing only the upper (F502N) panel. Radial velocities in \oiii~ are shown. The labels of the samples have been removed for clarity. The irregular heavy dashed line shows the western boundary of the area covered by the combined archived and new slit spectra.}
\label{fig:vradOIII}
\end{figure}

\begin{figure}
\epsscale{0.8}
\plotone{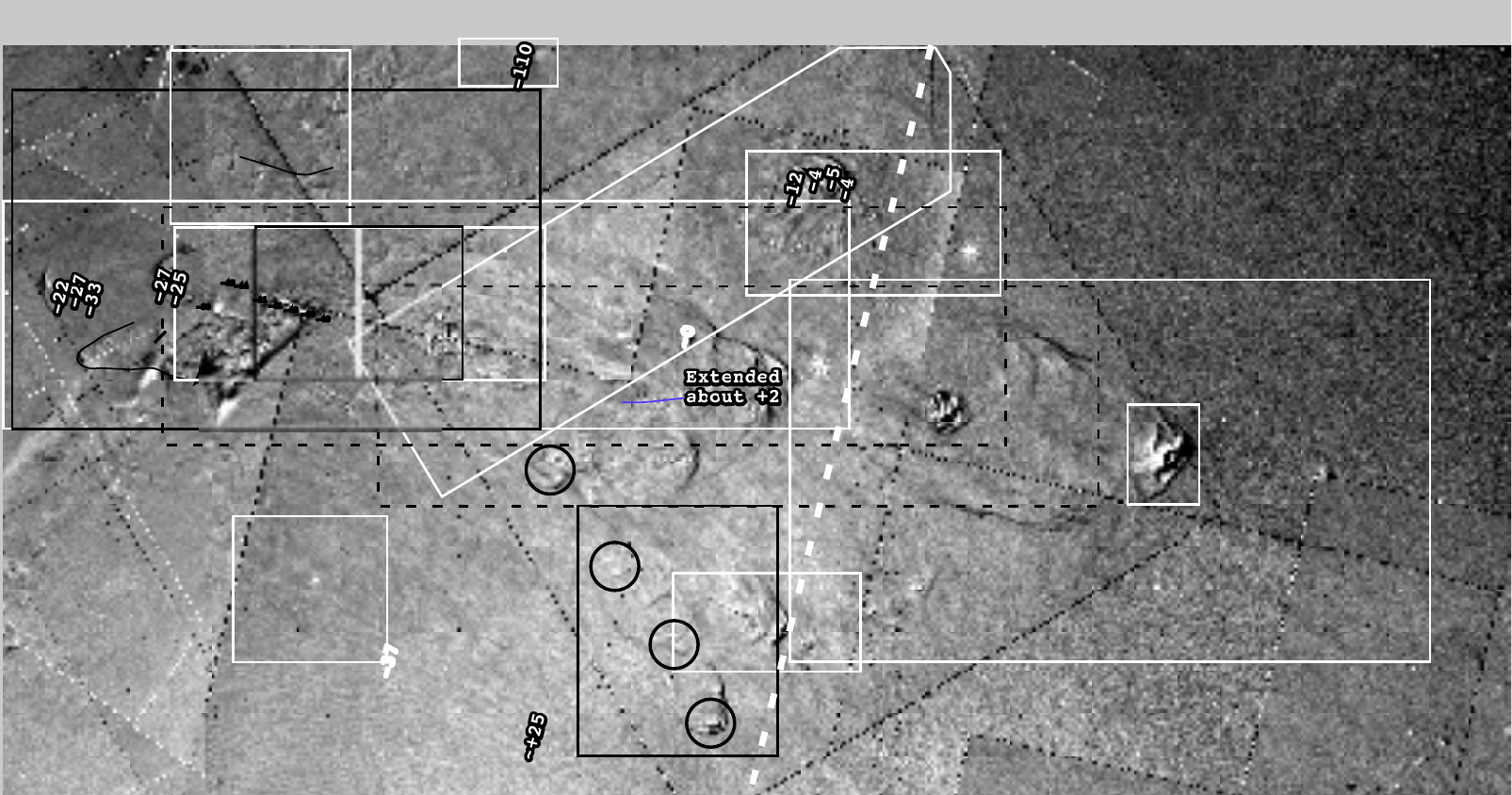}
\caption{Like Figure~\ref{fig:fig2} and Figure~\ref{fig:vradOIII} except showing only F658N results. Radial velocities in \nii~ are shown. The heavy dashed line shows the western boundary of the area covered by the archived slit spectra.The open black circles indicate the position of features associated with HH~1148 that are discussed in Section~\ref{sec:HH1148}. The region identified as ``Extended about +2 is a feature seen throughout much of the Huygens region starting with the study of HH~269 by \citet{wal95}. }
\label{fig:fig4}
\end{figure}

Our slit spectra were examined visually and those indicating the presence of a high velocity feature were analyzed using IRAF task ``splot''. This task can deconvolve the observed spectra into multiple components.  Although the spectral data archive also includes spectra for \oi, \sii, \siii, and \Ha, we did not measure those spectra because we have high quality tangential motion measures for only \nii~and\oiii. Those spectra should be valuable for understanding individual objects.  The derived radial velocities are presented in Figure~\ref{fig:vradOIII} and Figure~\ref{fig:fig4}.
There, as in all places in this article radial velocities (\vrad) are in the Heliocentric rest frame. Correction to the Local Standard of Rest rest frame can be obtained by subtracting 18.1 \kms.  The radial velocity of the host molecular cloud is 25.8 \kms\ \citep{ode08b}. The velocity of the background molecular cloud can vary in different regions up to a few kilometers per second, which does not have an important impact on this study. The astronomical seeing during the spectroscopic observations and the adjustments made in alignment mean that each velocity sample represents a diameter of about 2\arcsec\ and a positional uncertainty of slightly less than 2\arcsec. These are given in Appendix~\ref{AppendixC}.

\placefigure{fig:vradOIII}

\placefigure{fig:fig4}

\newpage
\section{EAST-WEST FLOWS}
\label{sec:eastwest}

Many of the most striking flows in the Huygens region are oriented east-west, with west moving objects on the west and east moving objects on the east.
We first define a sample that covers the middle of the Huygens region, explain the orientation of previously known HH objects with respect to this FOV, then examine the HH objects within and adjacent to this FOV

The East-West sample is shown in Figure~\ref{fig:fig2} and comparison with Figure~\ref{fig:fig1} shows that the 21-cm absorption line tracer \citep{pvdw}  of the Orion-S cloud lies in the middle of the sample. 
This sample was chosen to illustrate the east-west  flow from the region, with the west moving HH~269 being an apparent extension to the west and the east moving HH~529 to the east.  The HH~269 FOV lies to the immediate west and the HH~625 FOV overlaps the East-West sample and extends to the NW. 

\begin{figure}
\epsscale{1.0}
\plotone{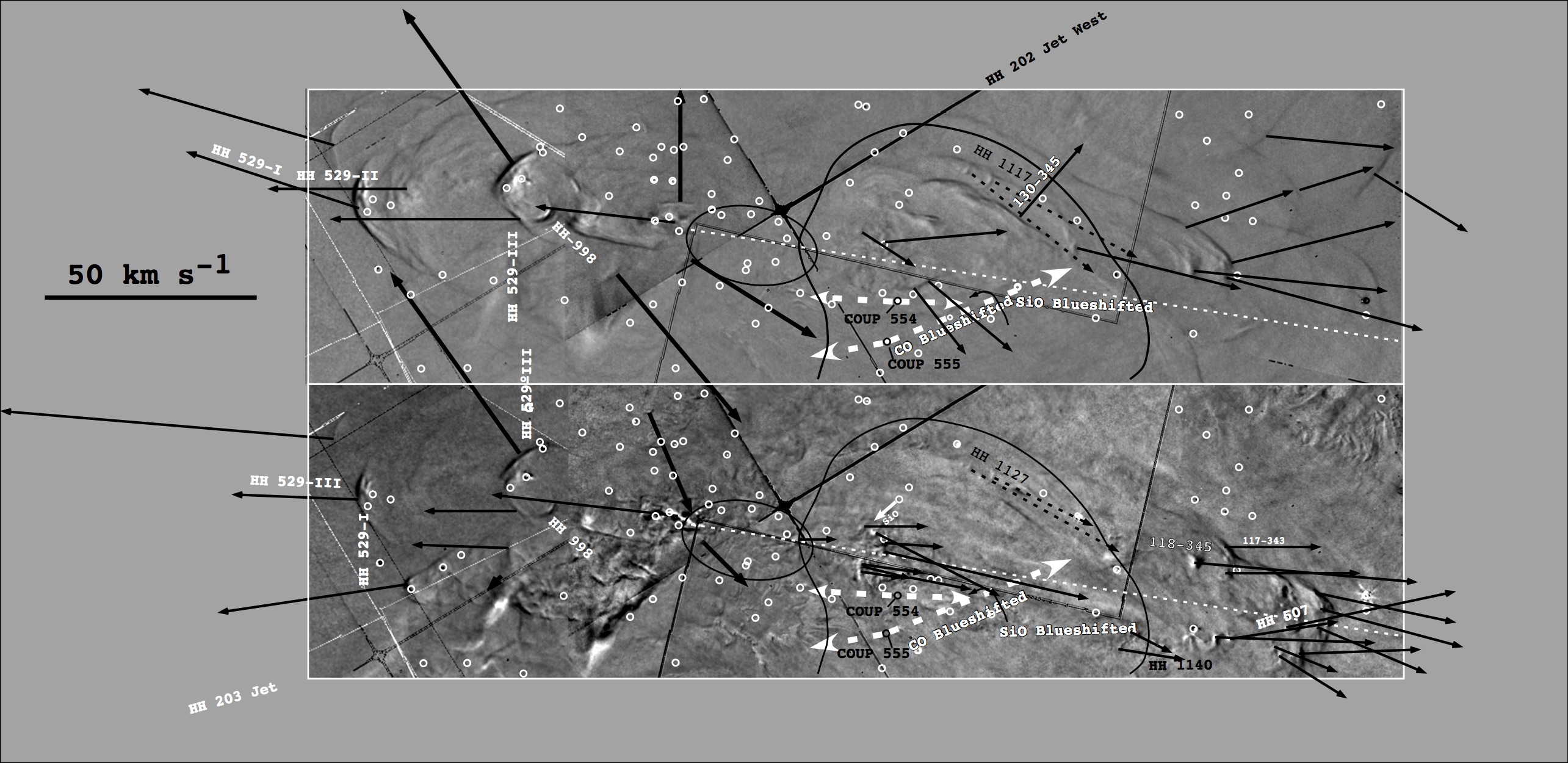}
\caption{
These 103.4\arcsec $\times$27.8\arcsec\ panels depict the motions images for the East-West sample in Figure~\ref{fig:fig2}. They show the east and west motions as one crosses Orion-S.
The ellipse indicates the OOS 
region \citep{ode03a,ode03b} and the irregular line the outer 21-cm line absorption boundary of the Orion-S cloud \citep{pvdw}. Arrows 
indicate the tangential velocity vectors of some of the optical features. Circles indicate the position of known point sources taken from the SIMBAD database. The heavy white dashed arrows indicate molecular outflows, with their length indicating the distance over which they 
 have been measured \citep{zap06}.  The light white dashed line indicates the direction of the HH~269 features.}
\label{fig:fig5}
\end{figure}

The  intersection of the line connecting west moving HH~269 and east moving HH~529 and the well-defined jets feeding HH~202 to the NW and HH~203 to the SE was pointed out by \citet{ode97a} and \citet{ros02}
 from radial velocity mapping. \citet{ode03a}  used information from all these flows, based on astrometric data , to argue for a common
region of origin within an ellipse of 7\arcsec $\times$12\arcsec\ centered at 5:35:14.56 -5:23:54 and designating it as the OOS (the Optical Outflow Source, Figure~\ref{fig:fig5}). The OOS was defined in the north-south direction by the approximate axes of the HH~269 and HH~529 flows and in the east-west direction by the region where there is a reversal of tangential motions, which agrees with the intersection of flows leading to HH~202 to the NW and HH~203+HH~204 to the SE. It lies on the east boundary of the core of Orion-S and Figure~\ref{fig:fig5} shows that it is near the peak surface density of known stars . In the current article we examine the question of the origins of these HH flows. As we will see, it is likely that there are multiple sources of the outflow, all lying nearby but not within the OOS. 

\placefigure{fig5}

\newpage
\subsection{HH~269}
\label{sec:HH269}

The first of the objects now included in the designation HH~269 was noted by \citet{fei76} and this region was noted to have concentrations of  density higher than the local nebula by \citet{wal94}. A big step in characterizing its components was the study of \citet{wal95}, which included imaging with the HST's WF/PC and high and low resolution spectroscopy. At that point it was considered a single elliptical structure with two enclosed bright knots. These knots were first designated as HH~269-East (hereafter in this section East) and HH~269-West (hereafter in this section West) by \citet{bal00}, who also derived tangential velocities from images over a short time-base. 

 \citet{sta02} mapped the entire Orion Nebula region in \htwo. They found several small aligned features. By comparison with our new HST images we find that the \htwo\ feature in their sample 2-6 corresponds to HH~269-East.  In their images we see that there are fainter \htwo\ features, one associated with HH~269-West, and a hint of others to the west.  The study of He~I \citep{tak02} shows a high velocity component at HH~269-West. There is an isolated \htwo\ knot  further west with 153\arcsec\ at  275\arcdeg (this direction falling along a line from HH~269-East to HH~269-West).  
 
 \citet{ode03a} and \citet{ode08a} have derived improved tangential velocities from HST images of higher resolution and longer time-base, establishing that the two bright features were part of a series of low ionization shocks sharing the same direction and probably also having the same source. The best  \vt\ values using the material in the present study are presented in the next section as are new determinations of \vrad.

\begin{figure}
\epsscale{1.0}
\plotone{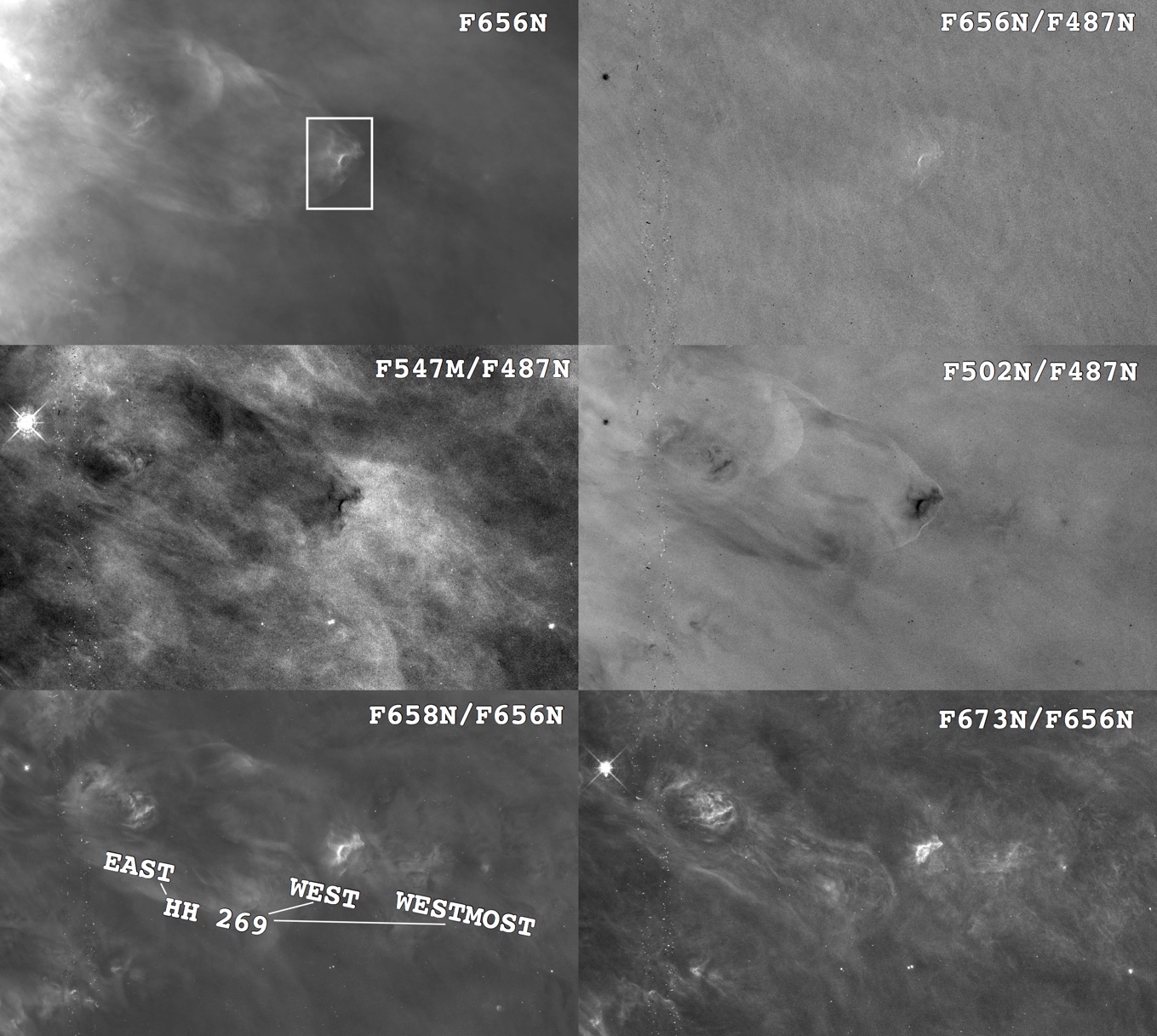}
\caption{This mosaic of 78.3\arcsec$\times$46.7\arcsec\  FOV images is centered on HH~269.
The top left panel is an F656N image and the top right panel shows the extinction sensitive F656N/F487N image. The others are selected to show extinction insensitive but ionization structure sensitive image ratios. The smaller field isolating HH~269-West in Figure~\ref{fig:fig8} is shown in the F656N image.}
\label{fig:fig6}
\end{figure}

We have assembled our images of HH~269 in Figure~\ref{fig:fig6}. The F656N panel shows that the object is not conspicuous. However, the structure becomes more obvious in the line ratio images. The F658N/F656N image shows well that the bright East and West components are of low ionization (as shown in the low resolution spectroscopy of \cite{wal95}) and that there is an open parabolic shock oriented to the west and with HH~269-West at its tip.  This shock contains both high (F502N) and low (F658N) emission.  There also appears to be an unrelated high ionization shock with a larger PA and a tip 12.4\arcsec\ at  310\arcdeg\ from HH~269-East (Figure~\ref{fig:fig7}).  

The F656N/F487N image shows that interstellar reddening does not vary across HH~269, since the F656N/F487N ratio is nearly constant except at the linear feature in middle of HH~269-West (Section~\ref{sec:HH269West}).

\subsubsection{HH~269 Motions}
\label{sec:HH269motions}

We now have up-to-date \vrad\ and \vt\ values for the HH~269 system. We first summarize here the \vrad\  and then the \vt\ results.

The radial velocities obtained by \citet{wal95} are  \vrad\ = -13 \kms\  and -22.5 \kms\ in \nii\ for the East and West components respectively. They did not isolate a component in \oiii. Our new spectra cover portions of both the East and West features as shown in Figure~\ref{fig:vradOIII}. In the West feature our \oiii~velocity (average \vrad=-18 \kms) agrees well with the \citet{wal95} \vrad\ value for \nii~ of --22.5$\pm$2.4 \kms.

However, the \citet{wal95} value of \vrad~=~-13$\pm$1.4 \kms\ derived from \nii\ in East is quite different from our value for a very weak component  in \oiii~of \vrad~=~-44 at two points. In \oiii~this 
feature looks quite different than in \nii~and it is likely that we are seeing two different and superimposed objects. 
 Figure~\ref{fig:vradOIII} shows that the moving 
objects to its east all have \vrad\ about -40 \kms, strengthening the argument that the East \oiii~ \vrad\ measurements are of a different feature than the \nii~feature.

\citet{ode03a} and \citet{ode08a} have derived improved tangential velocities from HST images of higher resolution and longer time-base, establishing that the two bright features were part of a series of low ionization shocks sharing the same direction and probably also having the same source.

\begin{figure}
\epsscale{1.0}
\plotone{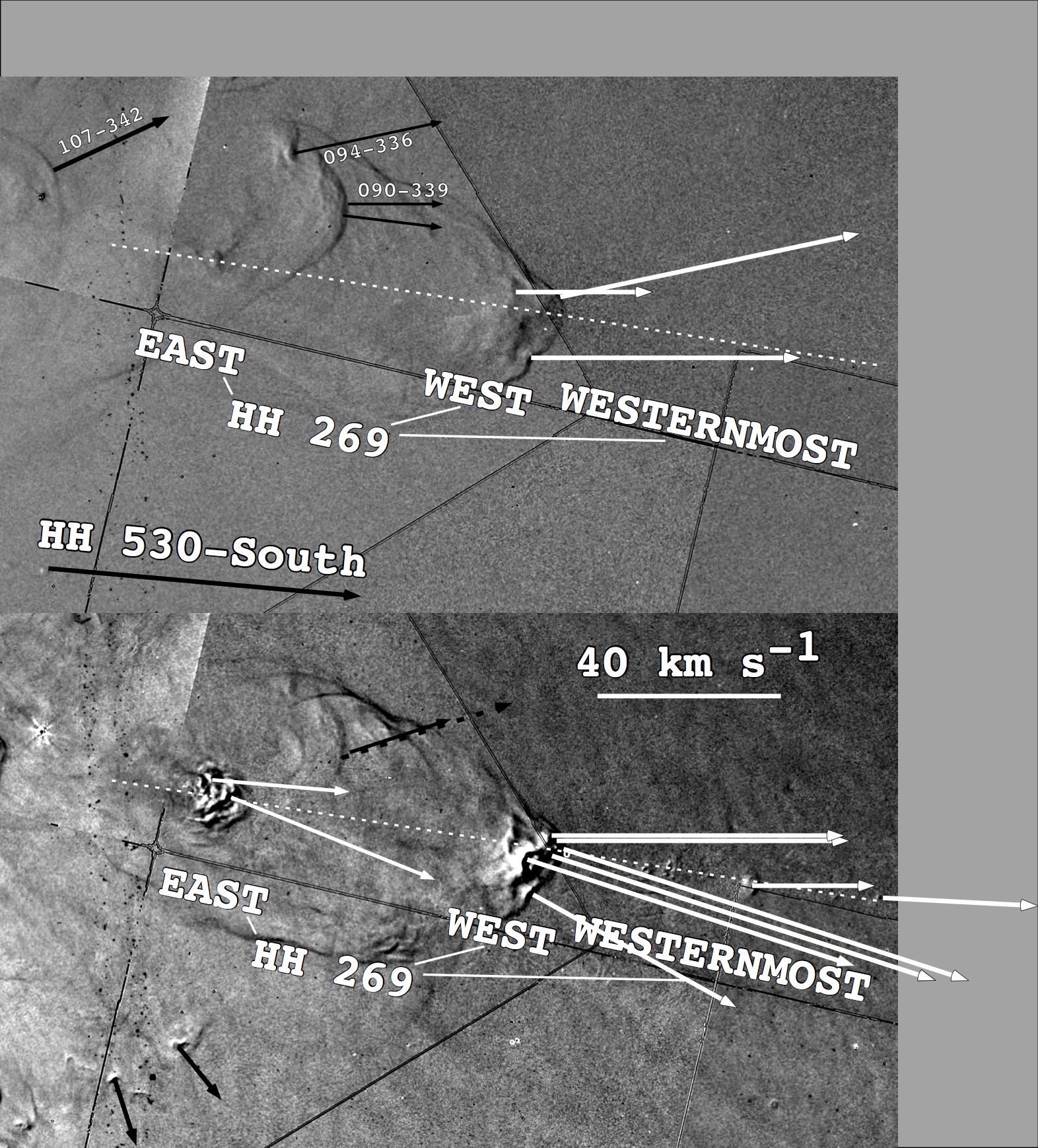}
\caption{The same FOV as Figure~\ref{fig:fig6} is shown but now depicting motion images (F502N in the top panel, F658N in the lower). In this image the white arrows depict the motions associated with the primary flow and the dark arrows are motions not associated with it. The dashed white line indicates the line common to the HH269 East, West, and Westernmost \nii\ features, pointing towards  275\arcdeg.}
\label{fig:fig7}
\end{figure}

In the pair of motion images shown in Figure~\ref{fig:fig7} we see that tangential velocities are well defined in F658N for both the East and West components. We also see in F658N that there are a series of slower moving small knots to the west. These lie exactly along the line projecting through both East, West, and the \htwo\ observed by \citet{sta02}. We call these collectively the HH~269-Westernmost features.  In F502N there are no features that may be associated with HH~269-East, while the features observed nearby share orientation and placement that identify them with the high ionization shock NNW of East. In the lower-left part of the  F502N panel we note the western portion of the HH 1157 bipolar flow discussed in Section~\ref{sec:HH1157}. The dark vectors in the F502N panel are discussed in Section~\ref{sec:LargeWest} where they are assigned to a different flow and not from HH~269. Feature HH~269-East is highly structured and resembles the well studied HH~204 shocks \citep{ode97b}.

Both the East and West components are moving towards the observer with respect to the plane of the sky. Adopting the \citet{wal95} \nii\ radial velocity and our average \nii\ \vt~=~13 \kms\ for the East shocks gives \Vomc~=~57 \kms\ at 43\arcdeg. Adopting the \citet{wal95} \nii\  radial velocity and our average of both the \nii\ and \oiii\ components of \vt~=~66 \kms\ yields \Vomc~=~82 \kms\ at 36\arcdeg\ for the West shocks. If the difference of the angle of approach of these two figures is accurate, then their origin is probably not as far east as is argued in Section~\ref{sec:OOSwest}; however, the derived angles are quite sensitive to the observational data and the 7\arcdeg\ difference is probably within the error of determination. The difference in velocity is more likely to be real..

\subsubsection{HH~269-West}
\label{sec:HH269West}
\begin{figure}
\epsscale{0.8}
\plotone{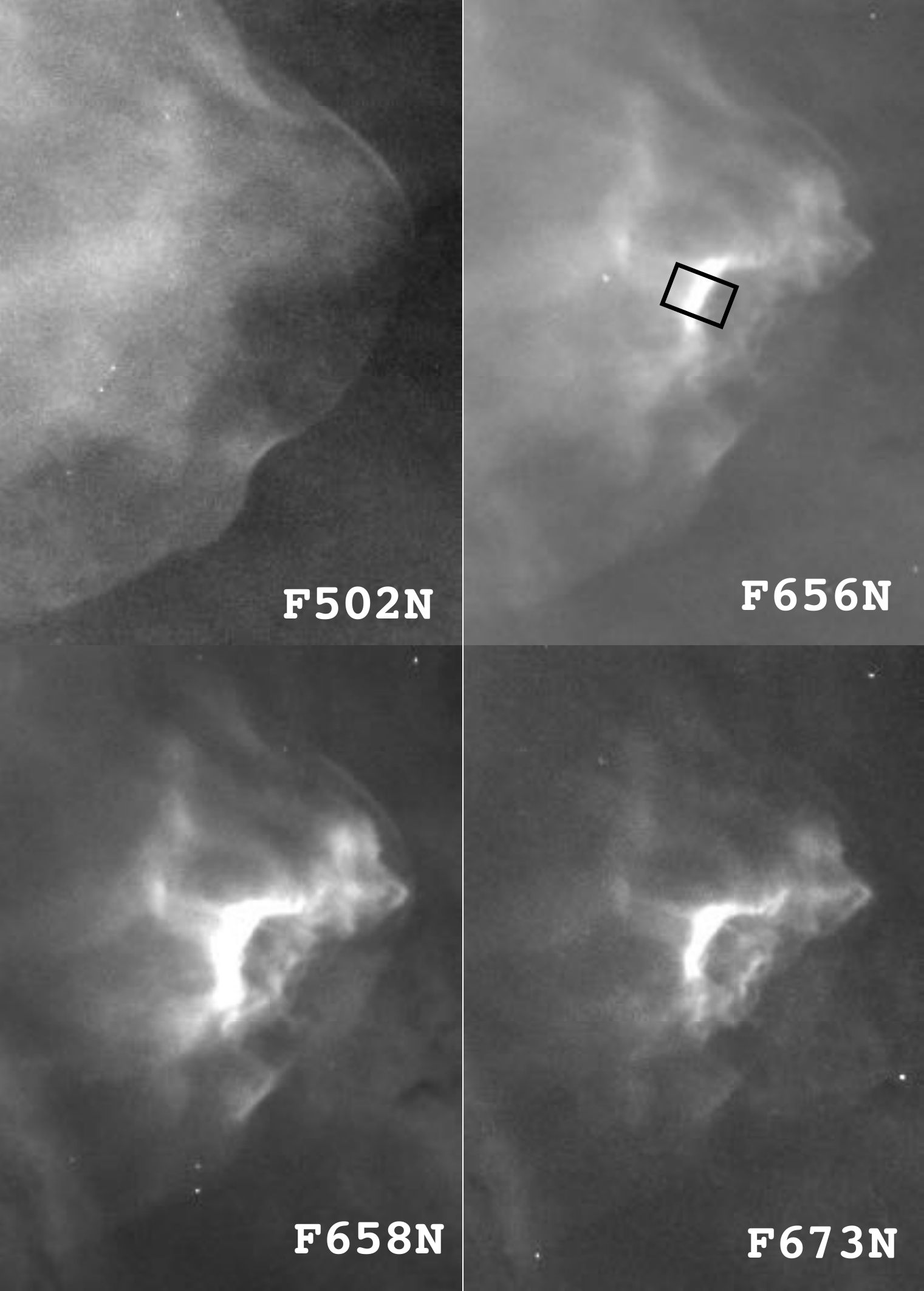}
\caption{This mosaic of 8.76$\times$12.24\arcsec\ single filter images isolating
 the HH~269-West feature again have  14\arcdeg\ up. The small box within the F656N         
 image shows the area sampled in the profiles shown in Figure~\ref{fig:fig9}.}
\label{fig:fig8}
\end{figure}

We show a fine-scale image of HH~269-West in the mosaic of images in Figure~\ref{fig:fig8}. The F502N image shows that West contains a bow shock feature oriented to the west. Within the boundaries of the bow shock the lower ionization emission is highly structured, with the dominant feature being the bright north-south oriented nearly linear feature that appears to be quite different from the nearby gas.  This feature must be related to the shock caused by a collimated jet, but, we have not addressed particularities of this complex structure. Previous studies of Herbig-Haro objects have argued that similar structures are Mach disks \citep{mor92,mor93}.

\begin{figure}
\epsscale{0.5}
\plotone{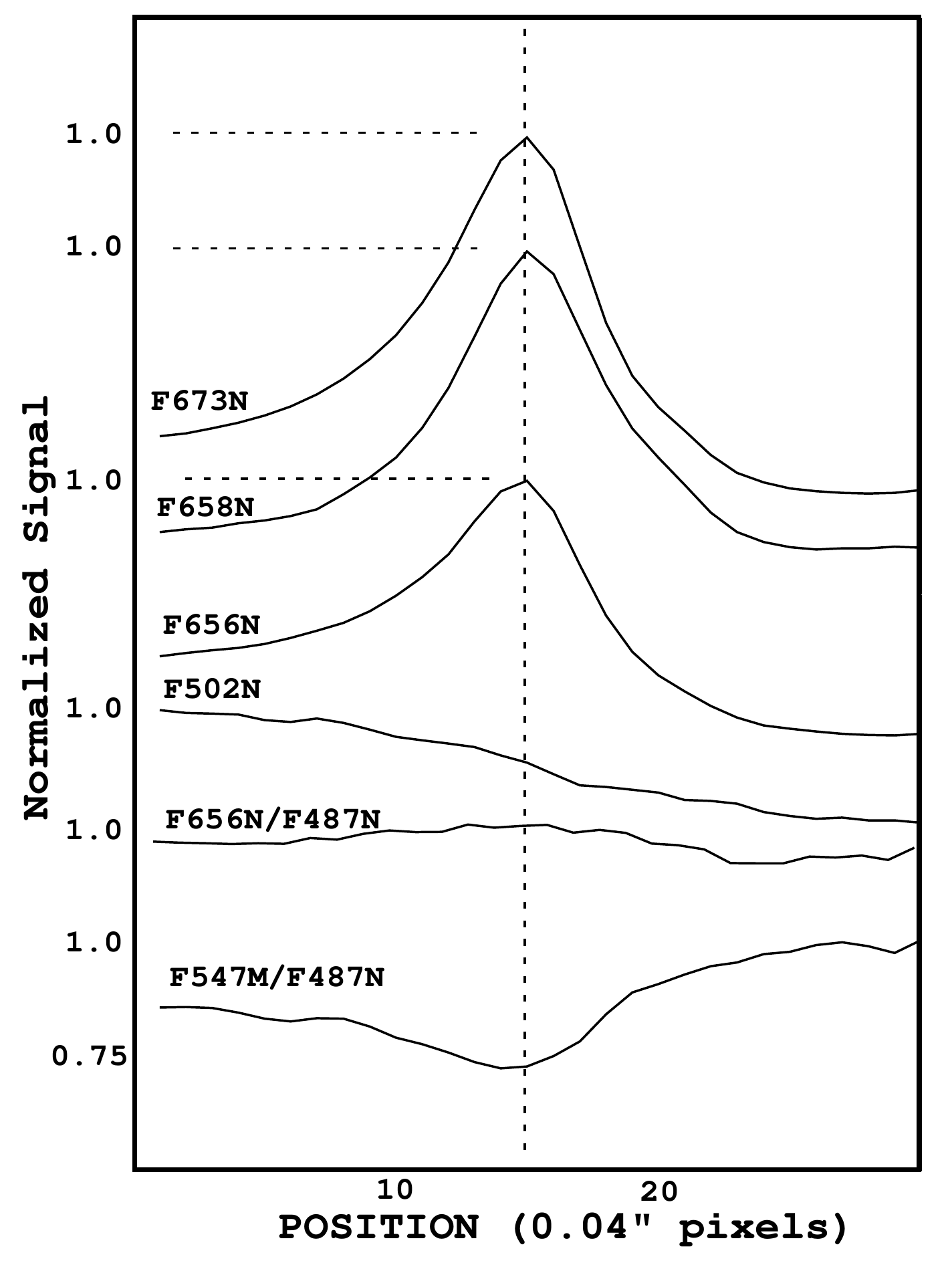}
\caption{This series of profiles is from the left to right in the small box within Figure~\ref{fig:fig8} and is an average along the short dimension of the box. In each case the maximum value has been normalized to 1.0. The scale is the same for each, but they have been shifted for clarity. The vertical dashed line is drawn at a fixed pixel in the sample.}
\label{fig:fig9}
\end{figure}

As part of an effort to understand the bright nearly linear feature, we show In Figure~\ref{fig:fig9} the results of averaging the signal in the 1.1\arcsec\ wide box shown in Figure~\ref{fig:fig8}. It is remarkable that the signal from F656N, F658N, and F673N all peak at exactly (within a fraction of an 0.04\arcsec\ pixel), although the widths are different. F502N does not seem to be produced in the feature. The F656N/F487N ratio increases at the peak of emission and the F547M/F487N ratio decreases there. The ground-based spectra of \citet{wal95} indicate that the HH~269-West feature has about twice the local density. Since those spectra had low spatial resolution, the density enhancement in the locally bright linear feature is probably even greater. This is almost certainly true because the peak surface brightness in F656N is much higher than the local region, even though its width is very small.
The feature is certainly a thin density feature. The drop in the F547M/F487N ratio indicates that either the scattering of stellar continuum is less important (in contrast with the expectation that the dust would have higher density where the gas density is higher) or that the electron temperature is lower. A lower electron temperature is also consistent with the F656N/F487N ratio being higher. 

\newpage
\subsection{Motions Near the OOS Region.}

In this section we
will discuss outflows in the vicinity of the OOS, beginning with a region west (Section~\ref{sec:OOSwest}), then a region east of (Section~\ref{sec:OOSeast}).

\subsubsection{Motions in the Region West of the OOS region}
\label{sec:OOSwest}

\begin{figure}
\epsscale{0.75}
\plotone{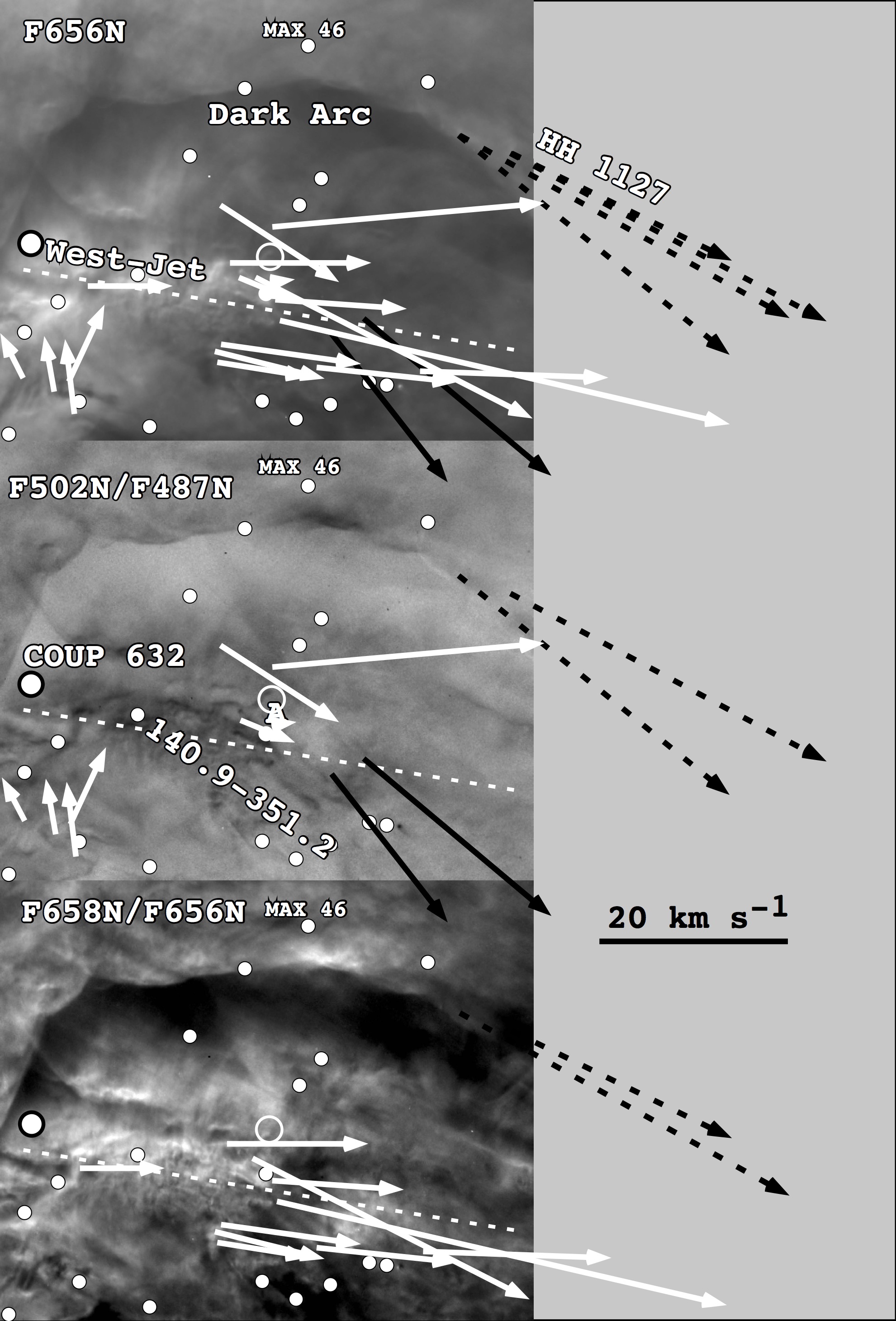}
\caption{These three line and ratio-images show the 22.6\arcsec $\times$18.7\arcsec\ FOV designated as West in Figure-\ref{fig:fig2}. The F656N image has had the velocity vectors for both F502N and F658N added from Figure~\ref{fig:fig11}. The F656N image also shows the location of the possible West-Jet. The other panels show extinction insensitive ratio-images that highlight ionization differences and the appropriate motion vectors have been added.
 SIMBAD stars are shown with filled circles in this and following figures.
Flows from near MAX 46 have been labeled as HH~1127. This object is discussed in Section~\ref{sec:1127}. The dashed white line indicates the adjusted projection of the axis of HH~269.}
\label{fig:fig10}
\end{figure}

The region we designate here as West (Figure~\ref{fig:fig2}) is a complex region already known to be of considerable interest. The northern portion contains a predominant feature designated in Figure~\ref{fig:fig10} and in previous studies as the Dark Arc. The 3-D structure of this feature remains undetermined in spite of attempts to explain it \citep{ode00b}. It is almost certainly part of the Orion-S cloud and must lie on the observer's side of Orion-S (otherwise it would be invisible due to the high extinction in Orion-S). 

The conundrum is that the sharp north edge of the Dark Arc has a narrow rim of bright F658N and F673N emission, indicating that we are seeing an ionization boundary. However, as we see in Figure~\ref{fig:fig10} the high ionization (south) side of this front is in the opposite direction of \tc, which lies to the NE. A clue to the structure may lie in the fact that the series of shocks forming  HH~1127 (c.~f.~Section~\ref{sec:1127}) become visible starting at the rim after passing from the obscured source (either MAX 46 or COUP 602) across the north rim of the Dark Arc. This indicates that the south side of the Dark Arc's sharp northern boundary is an open space.

\begin{figure}
\epsscale{0.8}
\plotone{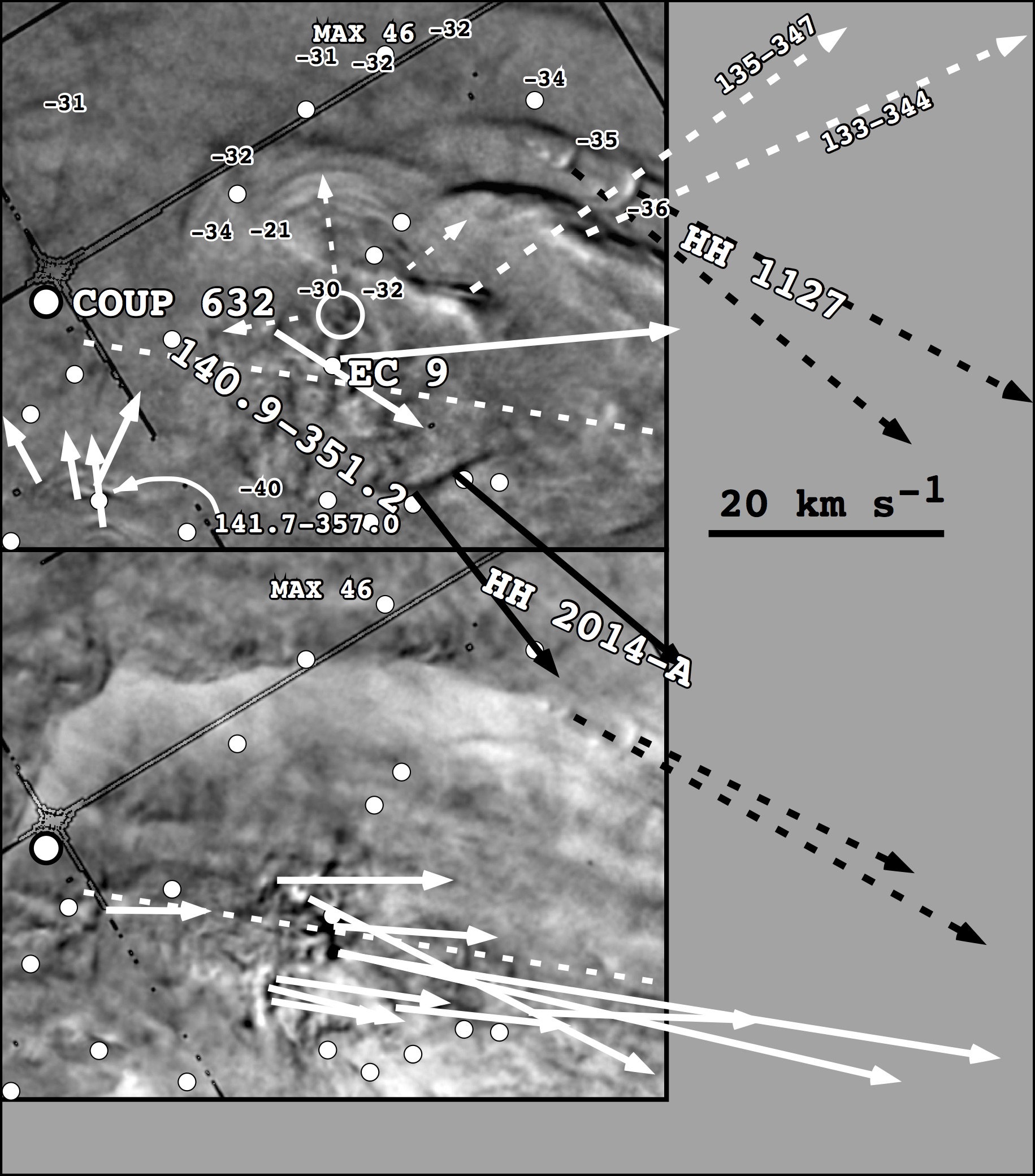}
\caption{The same FOV as Figure~\ref{fig:fig10} is shown except now the motion images are shown. The dark dashed arrows indicate motions in two portions of HH~ 1127, the heavy white arrows motions that are likely to be part of the system driving HH~269. The light white arrows indicate the symmetry axis of faint concentric expanding features seen in F502N. The light dashed white lines with arrows show the direction of motion of concentric rings of motion  visible in F502N. The dashed white line without arrows indicates the adjusted projection of the axis of HH~269.The open white circle indicates the point of convergence of them as discussed in Section~\ref{sec:blanksDarkArc}. \oiii~radial velocities are shown as small negative numbers with black centers and white borders. The four \oiii~vectors in the lower left of the lower panel are not shown for clarity in Figure~\ref{fig:fig10} and are discussed in Section~\ref{sec:GroupShocks}.}
\label{fig:fig11}
\end{figure}

The thin and irregular low-ionization feature designated as the West-Jet has been the subject of multiple studies and has been
 argued (O'Dell et al. 2008a and references therein) to be the easternmost component of the collimated material driving HH~269. It has an orientation of  276\arcdeg. It lies almost on the symmetry axis of the HH~269 features, and there are ten low ionization features nearly along the symmetry axis. 
 The axis of the West-Jet is essentially parallel to the symmetry axis of the HH~269 features. If the HH~269 axis was adjusted to  276\arcdeg, rather than the best value from the HH~269 components of  275\arcdeg, then the east end of the HH~269 axis would be raised north and be in exact alignment with the West-Jet, sharing both the alignment angle and position. Such a small change to the value of the HH~269 axis is allowable because determination of the HH~269 axis is imperfect and we know that jet flows can curve. Figure~\ref{fig:fig10} and Figure~\ref{fig:fig11} show the adjusted HH~269 axis.
 
 However, this putative jet lacks a continuous structure and shows large tangential motions only along its west end. 
 The middle panel of Figure~\ref{fig:fig10} shows that the putative jet is low ionization. There are no identified \nii\ \vrad\ sources associated with this feature. The only two nearby tangential \oiii~motions are displaced from this feature, although they do move approximately along the HH~269 axis.  In \nii~there is a host of motion features whose northern
 boundary is along the west end of the West-Jet, but many lie 4\arcsec~ south and only two lie exactly on the West-Jet.  An additional isolated moving \nii~feature lies to the east right on the West-Jet. 
 The three \nii\ \vt\ features lying on and moving along the axis of HH~269 argue that the West-Jet is an actual collimated outflow. The presence of similar features above and below the axis indicates that either not all of the features are related to the West-Jet or that we are seeing the superposition of two flows of different origin by similar location and movement on the plane of the sky.
 In any event, the easternmost moving feature must belong to a flow that originates to the east of star  140.9-351.2.

The shocks and \oiii~radial velocity features in the upper right of Figure~\ref{fig:fig11} are discussed in Section~\ref{sec:1127}. 

The concentric rings shown in the F502N motions image (the upper panel of Figure~\ref{fig:fig11}), and their central point are discussed in Section~\ref{sec:blanksDarkArc} and Section~\ref{sec:blanksDarkArc}. The \oiii~radial velocity features towards PA~=~67\arcdeg~
start at this spot, after consideration of the size of the spectroscopic samples and it is possible that they represent an otherwise undetected collimated flow from the invisible source. 

\placefigure{fig:fig10}

\placefigure{fig:00SwestMotions}

\newpage

\begin{figure}
\epsscale{0.4}
\plotone{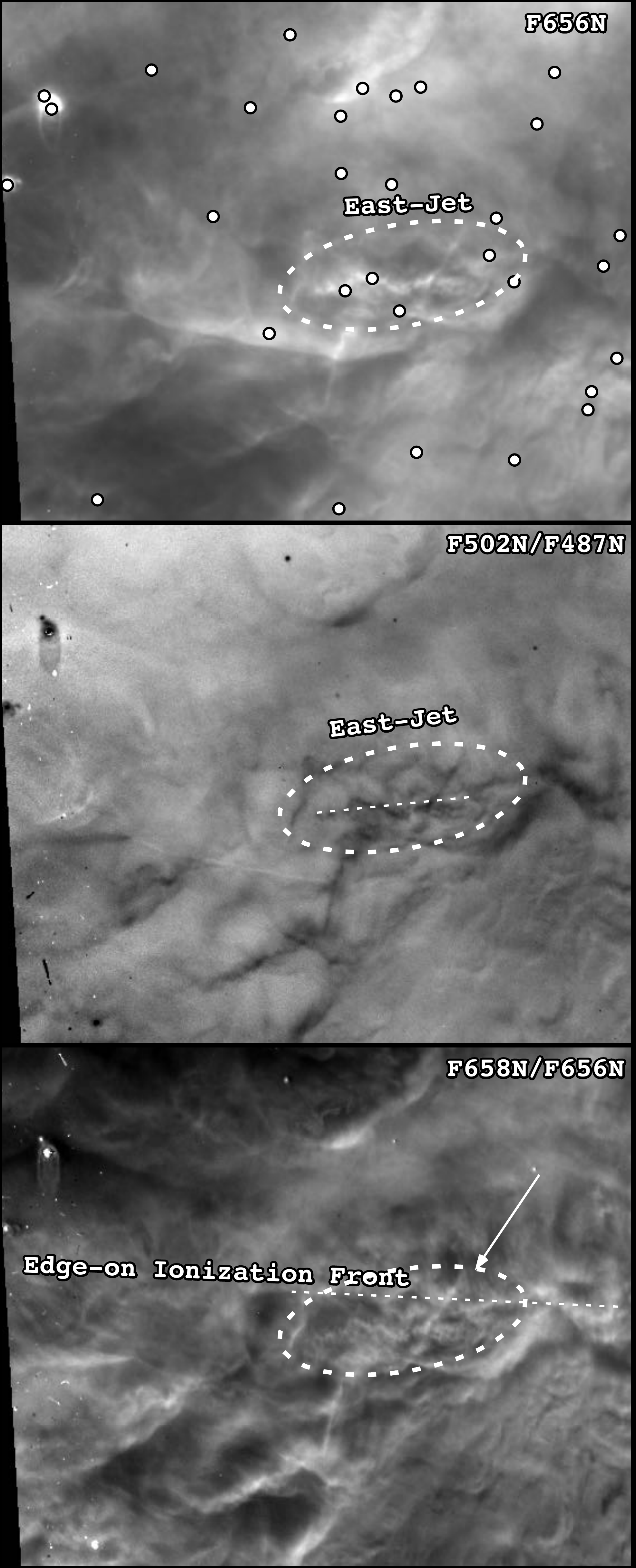}
\caption{This figure shows the line and line-ratio images for a 22.6\arcsec $\times$18.7\arcsec\  FOV  immediately east of the West FOV and designated as East. A well defined low ionization filamentary structure is easily seen and is designated as the East-Jet. The axis of the Jet is shown with a light dashed white line in the F502N/F656N ratio-image while the axis of a nearby edge-on ionization front is shown in the F658N/F656N ratio-image. 
The region including the East-Jet is shown by the white dashed-line ellipse. The light vector points to the linear feature arising from COUP~666.}
\label{fig:fig12}
\end{figure}

\placefigure{fig:fig12}

\placefigure{fig:00SeastOldNew}
\begin{figure}
\epsscale{0.5}
\plotone{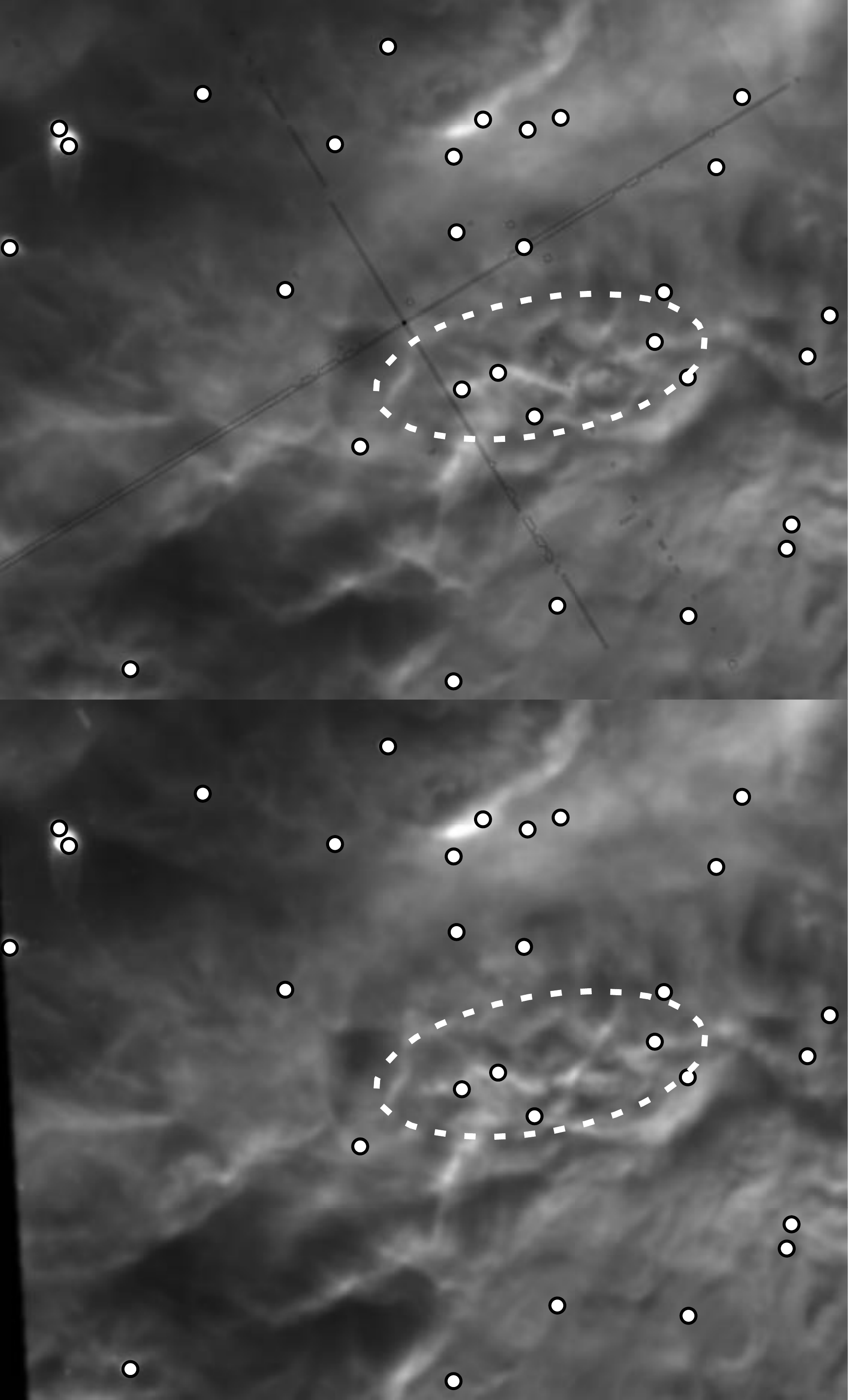}
\caption{First (top, GO5469) and second (bottom, GO12543) F658N images of the East FOV  are shown. The region including the East-Jet is again shown by the white dashed-line ellipse. Examination of the Jet feature demonstrates both its rapid motion to the east and also changes in its structure.}
\label{fig:fig13}
\end{figure}

\begin{figure}
\epsscale{0.5}
\plotone{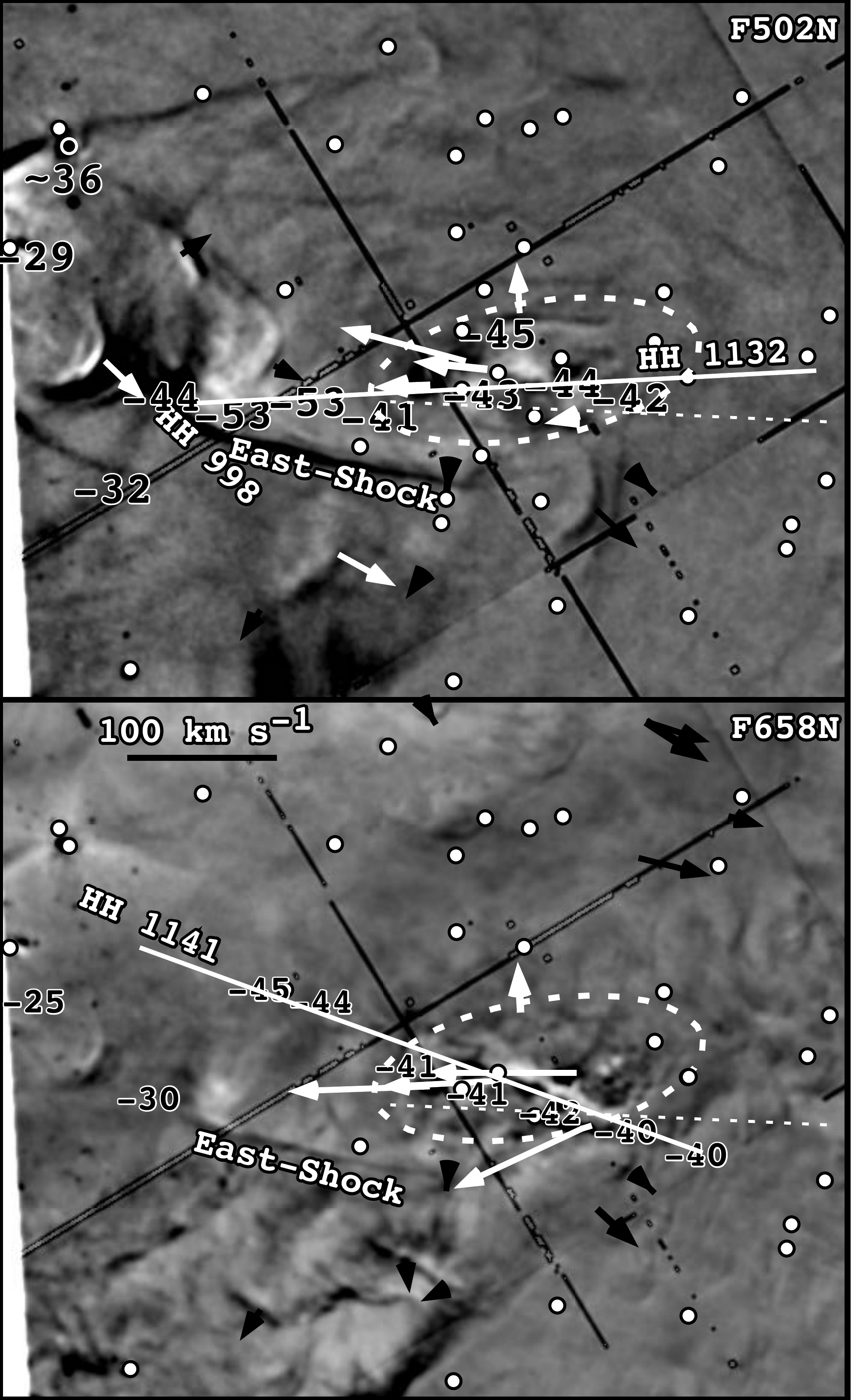}
\caption{These East FOV motion images have the motions of the Jet and features possibly associated with it as solid white arrows, except for the pair of F502N shocks associated with HH~998 (c.f. Section ~\ref{sec:HH998}). Features not associated with either system are shown with black arrows. The region including the East-Jet is again shown by the white dashed-line ellipse. The dashed light lines indicate the position of the approximate  projection of the  axis of the HH~269 features. The white line in the upper panel follows the symmetry axis of the East-Jet along its full extent. The white line in the lower panel follows the symmetry axis of the East-Second-Jet in two segments. The black numbers indicate radial velocities.}
\label{fig:fig14}
\end{figure}

\placefigure{fig:fig14}

\subsubsection{Motions in the Region East of the OOS region}
\label{sec:OOSeast}

Many earlier studies have established that the east-west flow centered on Orion-S reverses in or near the OOS. Figure~\ref{fig:fig12} demonstrates that there is a well defined, irregular, mixed ionization jet there. 

The changes of position and of structure of the features constituting  the East-Jet are illustrated in the F658N images in Figure~\ref{fig:fig13}. The complexity of the region works to make determination of changes of position more difficult; but, fortunately and for the most part, the large size of the motions balances this limitation.

Figure~\ref{fig:fig14} demonstrates the velocity vectors in this region. The center of the East-Jet is at 5:35:15.04 -5:23:53.1. 
 The orientation of the East-Jet is towards  108\arcdeg, while a similar appearance feature to its north is oriented at 101\arcdeg. The fact that most of the north feature is not moving rapidly makes it likely that is an ionization front viewed edge-on and is labeled ``Edge-on Ionization Front'' in Figure~\ref{fig:fig12}.

The F658N motion-image in Figure~\ref{fig:fig14} shows that the motion of the East-Jet is along its axis of symmetry and that an east-west feature about 1.2\arcsec\ south has a similar motion in F658N.  In the F502N panel we see that there is motion or change on this south-lying feature but on the north side of the East-Jet there is clear northerly motion of F502N emitting material.  One moving F502N feature near the center of the ellipse has been measured and has a similar velocity but different direction of the F658N features.  There are multiple other moving F502N features, but they have not been measured because there have been significant changes in their structure. Examination of the moving features indicate that the bulk motion of the East-Jet may be a few degrees smaller in PA than that suggested by the symmetry of the irregular filament. Like the West-Jet there are both F502N and F658N features, although the F502N features are relatively stronger in the East-Jet.

The radial velocity results shown in Figure~\ref{fig:fig14} define two patterns of high radial velocity features.
The \oiii\ features align towards PA~=~107\arcdeg, which agrees with the symmetry axis of the East-Jet (PA~=~108\arcdeg.
These high radial velocity \oiii\ features extend well beyond the ellipse containing the West-Jet.
 The seven \nii\ high tangential velocity features have PA's of 83\arcdeg\ on the east end and  87\arcdeg\ on the better-defined west end. The pattern of positions of the \nii\ high radial velocity features has PA~=~94\arcdeg\ and intersects the \oiii\ pattern on the east end of the outline of the East-Jet features. 
 
Although the velocities are similar in the two ions, it appears that we are observing two different 
series, the East-Jet (best defined in \oiii\ radial velocities and \nii\  tangential velocities) and an East-Second-Jet (defined
by a single \oiii\ tangential velocity feature and seven \nii\ radial velocity features).
Using the East-Jet \oiii\ \vrad\ values and the \nii\ \vt\ values gives a spatial velocity \Vomc~=~116 \kms\ with $\theta$=32\arcdeg. 
Using the East-Secondary-Jet \oiii\ \vt\ values and the \nii\ \vrad\ values gives a spatial velocity \Vomc~=~114 \kms\ with $\theta$=35\arcdeg. It appears that the East-Jet is crossed in the plane of the sky by a similar jet (the East-Second-Jet).

The -45 \kms~feature in the top part of the ellipse is probably associated with the 
expanding elliptical pattern that one sees in the \oiii\ tangential motions images.
This pattern on the north and south sides of the ellipse enclosing the East-Jet indicates that the geometric configuration is that the jet is breaking out from a source embedded within the Orion-S cloud.  The identification of the probable source is discussed in Section~\ref{sec:sources} and the possible relation to the large high ionization shocks composing HH~529 is discussed in Section~\ref{sec:HH529}.

\begin{figure}
\epsscale{0.5}
\plotone{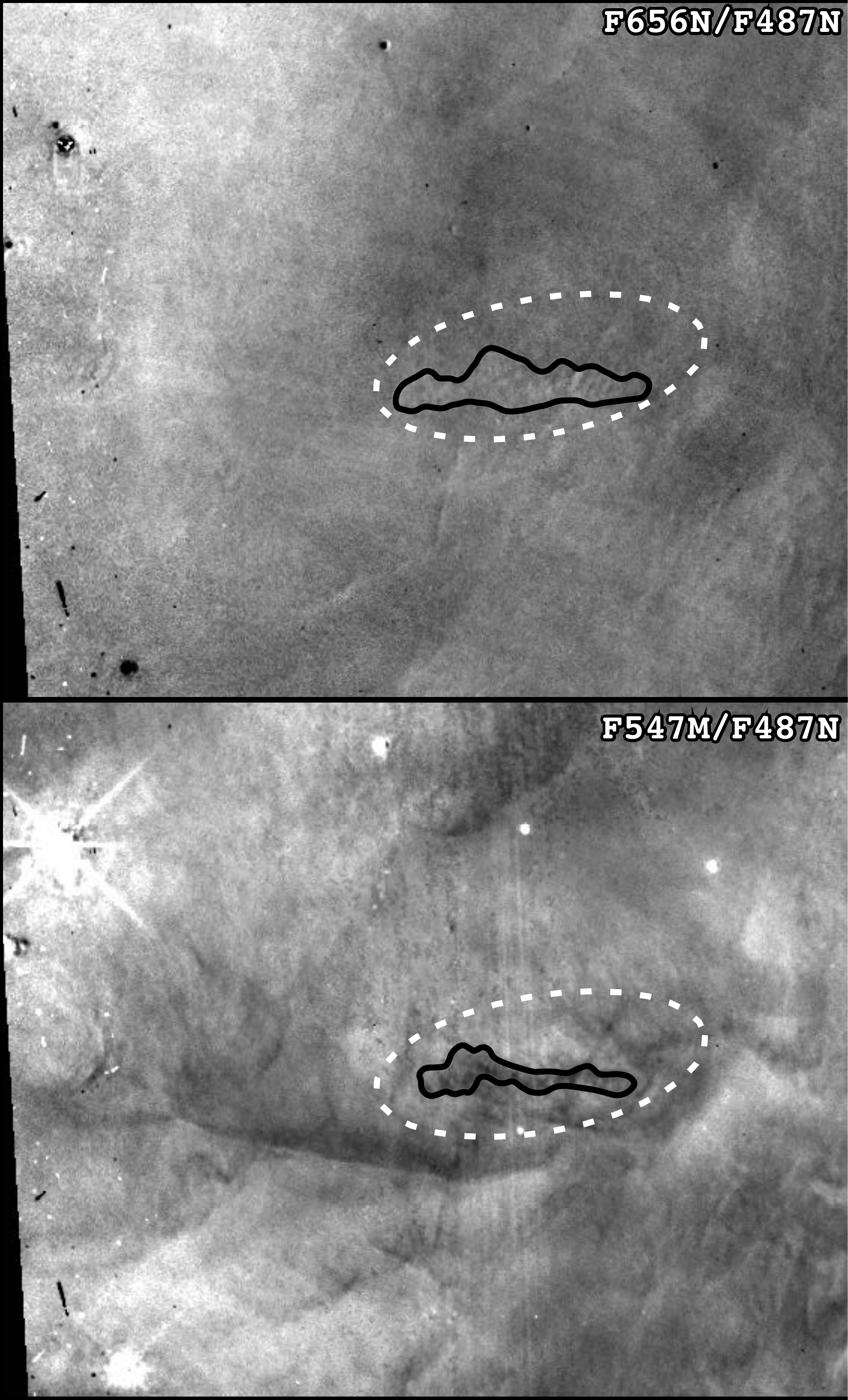}
\caption{Local variation is the F656N/F487N ratio-image is shown in the top panel and the F547M/F487N ratio-image in the lower panel. These are discussed in Section~\ref{sec:OOSeast}. The ellipse defines the same area as in Figure~\ref{fig:fig13} and Figure~~\ref{fig:fig14}. The irregular dark-line outline in the upper panel identifies the area of fine-scale variations in the F656N/F487N image. The similar outline in the lower panel indicates the areas of low and structured values of the F547M/F487N images.}
\label{fig:fig15}
\end{figure}

\placefigure{fig:fig15}

Physical conditions within the East-Jet are suggested by examination of Figure~\ref{fig:fig15}. The upper panel indicates that no big changes occur across the East-Jet's region. However one sees a pattern of sub-second of arc  changes within the ellipse enclosing the East-Jet. The axis of those fluctuations is a fewer degrees lower (104\arcdeg) and is similar to the PA suggested for the motions. These fine-scale changes in F656N/F487N may be caused by differences in \Te\ within the jet or small scale changes in the extinction (in which case this would mean that the jet lies closer to the observer than the material causing the surrounding extinction). In the lower panel one sees that the value of F547M/F487N is lower along the same  104\arcdeg\ region. This probably indicates that the  light from the jet is mostly gaseous material (which has a low F547M/F487N emission) superimposed  on the high scattered light continuum radiation coming from the background PDR.These combined considerations indicate that the changes in F656N/F487N are due to the temperature in the East-Jet being different from that of the nebular gas.

\subsubsection{Properties of the East and West regions}
\label{sec:properties}

Definitive discussions of the stellar sources of the flows are presented in Section~\ref{sec:fans}, Section~\ref{sec:269529} and Section~\ref{sec:EastAndSecond}.

In a parallel study led by co-author Henney we have isolated the auroral (575.5 nm) and nebular transitions (658.3 nm and 654.8 nm) of \nii. The ratio of these line intensities I(575.5 nm)/I(658.3 nm + 654.8 nm) is mostly electron temperature dependent, with larger values indicating higher temperatures. The FOV covered includes the East-Jet and the West-Jets and in both cases the jets have much larger values than the surrounding nebula. In the case of the East-Jet the high temperature features closely mimic the object as seen in F658N. In the case of the West-Jet the high temperature features resemble the West-Jet in F658N until about 1\arcsec\ west of the star 140.9-351.2. At that point the squiggly primary feature that defines the western end of the West-Jet is no longer a well-defined high 
temperature feature, rather, the high temperature region is extended with many high temperature knots. This region falls into a triangle defined by about 5:35:14.03 -5:23:51 on the east, 5:35:13.77 -5:23:49 on the north and 5:35:13.72 -5:23:55 on the south. The western end of the West-Jet lies along the middle of this extended region of high temperature.

\newpage
\subsection{The Eastern Systems}
\label{sec:Eastern}

This section presents and examines the results for the region immediately east of the Orion-S molecular cloud. It is shown as HH~529 in Figure~\ref{fig:fig2}. It includes the OOS region originally identified as the common centers of several outflows \citep{ode03a,ode03b}. Our new data has allowed a more accurate determination of tangential velocities and detection of patterns of motion. We begin with a discussion of 
HH~529, a well-known series of shocks moving east,  and a new  system designated as HH~1149. We then present the likely connection of this region to the large-scale objects HH~202, HH~203, HH204, and HH 528. 
Where possible the tangential velocities were determined from the combination of GO~5469 and GO~12543 images. East of the boundary passing through
about AC~Ori we used WFPC2 images from GO~5469 and GO~11038.

\subsubsection{HH~529}
\label{sec:HH529}

\begin{figure}
\epsscale{1.0}
\plotone{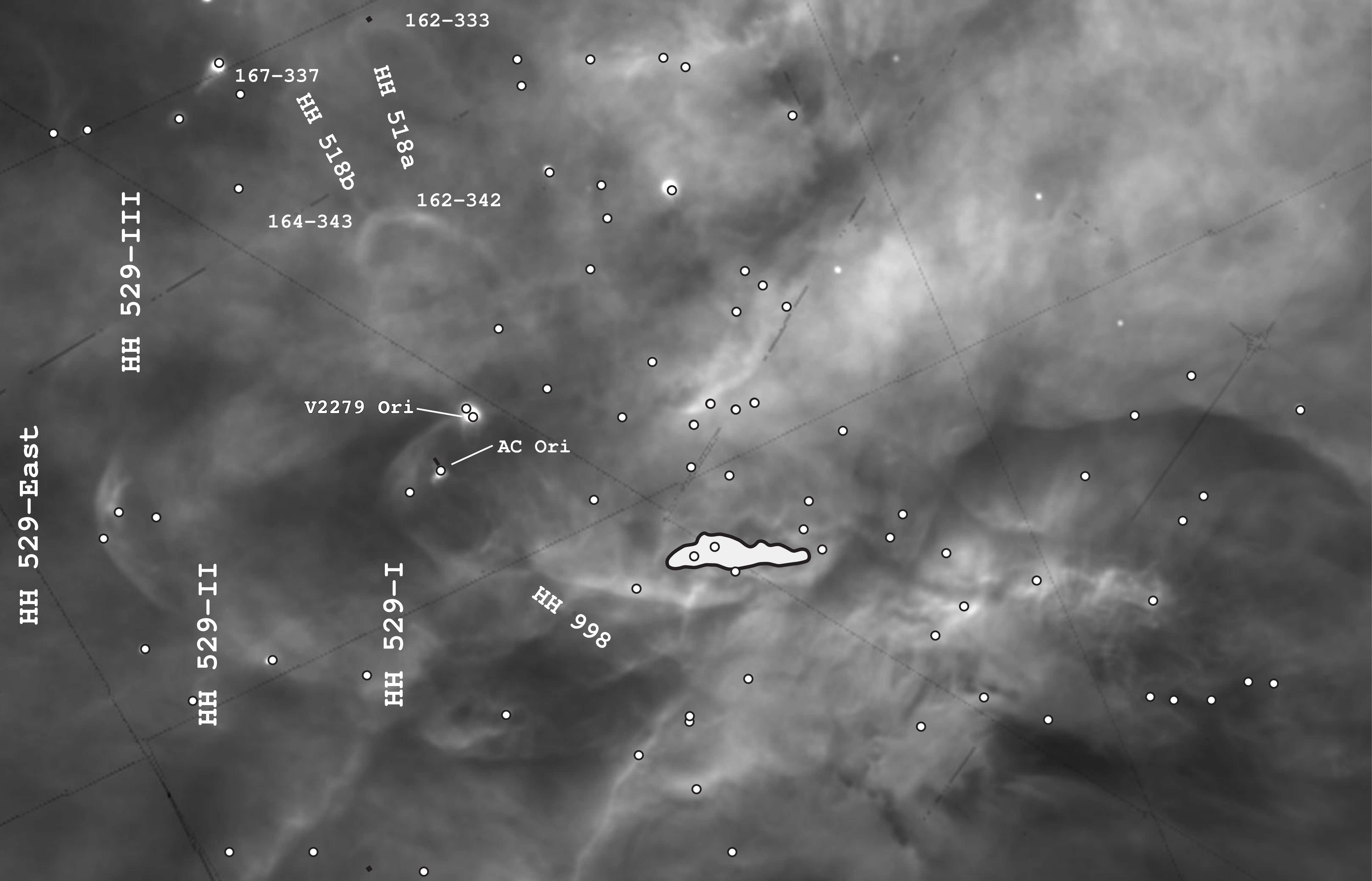}
\caption{This 64.6\arcsec $\times$41.5\arcsec FOV  is taken from the east boundary of the North FOV shown in Figure~\ref{fig:fig2}. It shows the F656N image. The breakdown of HH~529 into three regions \citep{ode08a} is labeled and a newly identified HH~529-East feature is added. The irregular closed bright line figure outlines the East-Jet region identified in Figure~\ref{fig:fig15} as having either fine scale F656N/F487N or F547M/F487N changes. Two key stars (V2279 Ori and AC Ori) are labeled. The filled circles are compact sources from the SIMBAD listing.}
\label{fig:fig16}
\end{figure}

\placefigure{fig:fig16}

\begin{figure}
\epsscale{1.0}
\plotone{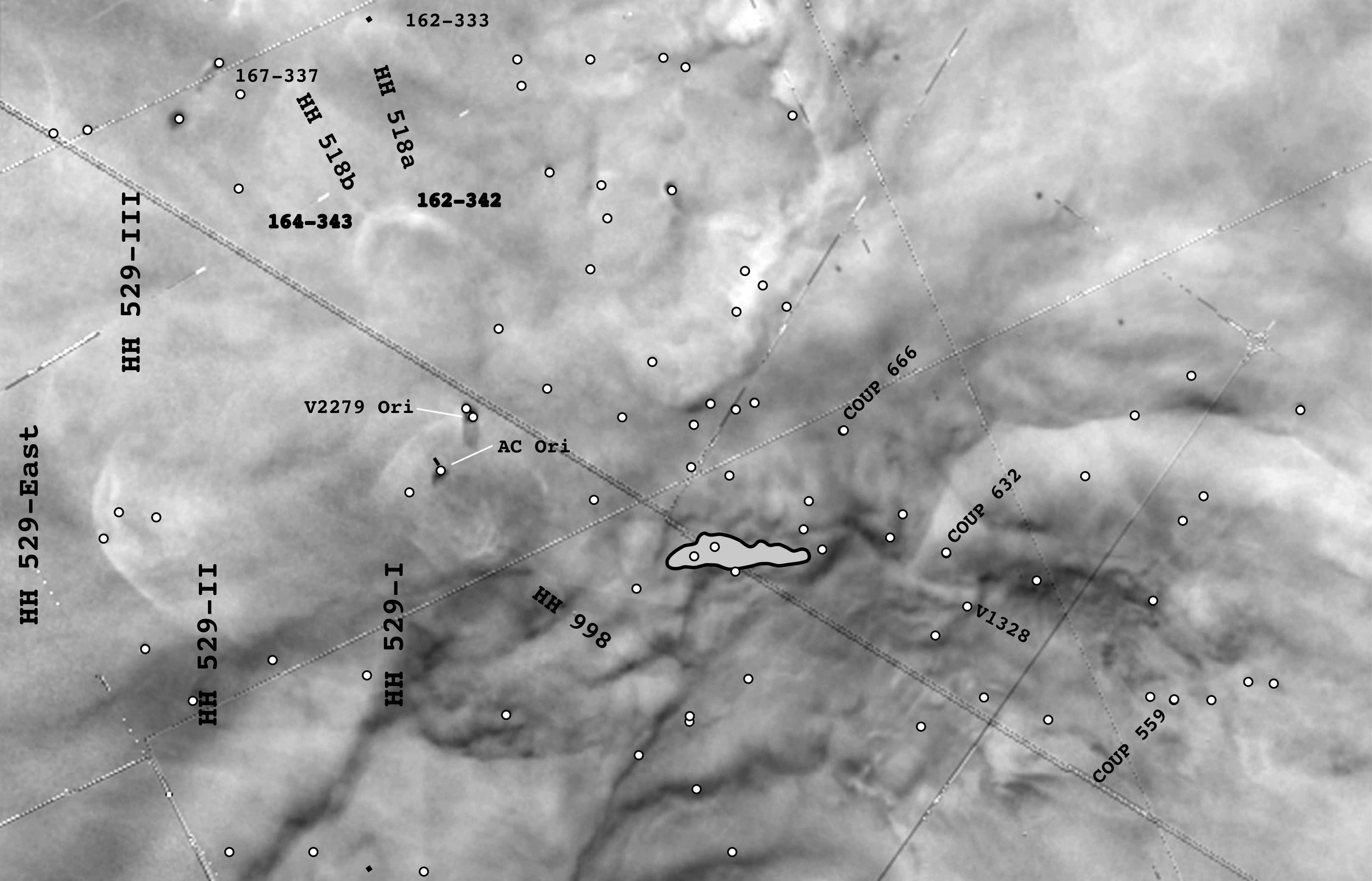}
\caption{Like Figure~\ref{fig:fig16} except that it shows the ratio image F502N/F656N.}
\label{fig:fig17}
\end{figure}

\placefigure{fig:fig17}

\begin{figure}
\epsscale{1.0}
\plotone{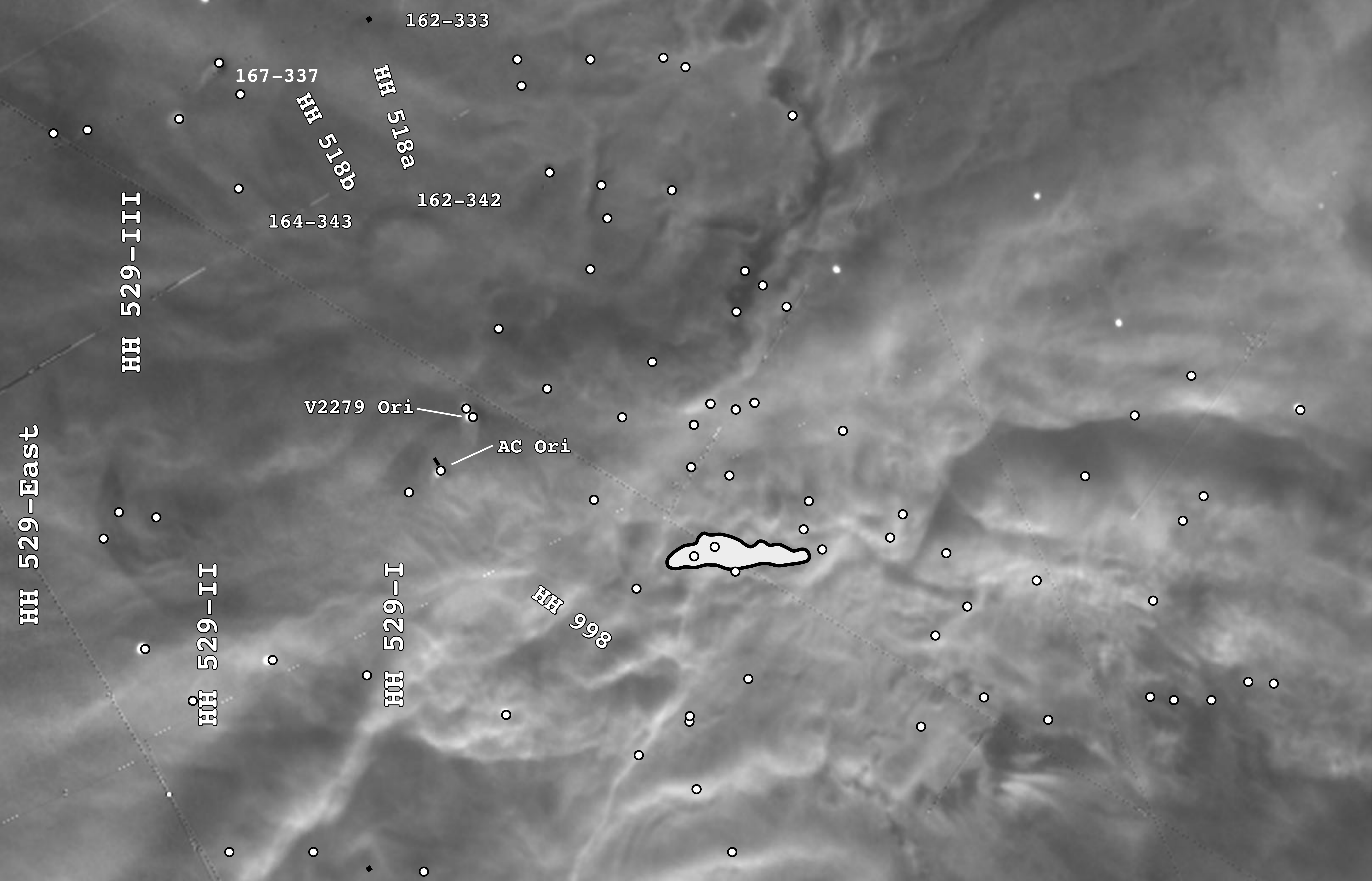}
\caption{Like Figure~\ref{fig:fig16} except that it shows the ratio image F658/F656N.}
\label{fig:fig18}
\end{figure}

\placefigure{fig:fig18}

\begin{figure}
\epsscale{01.0}
\plotone{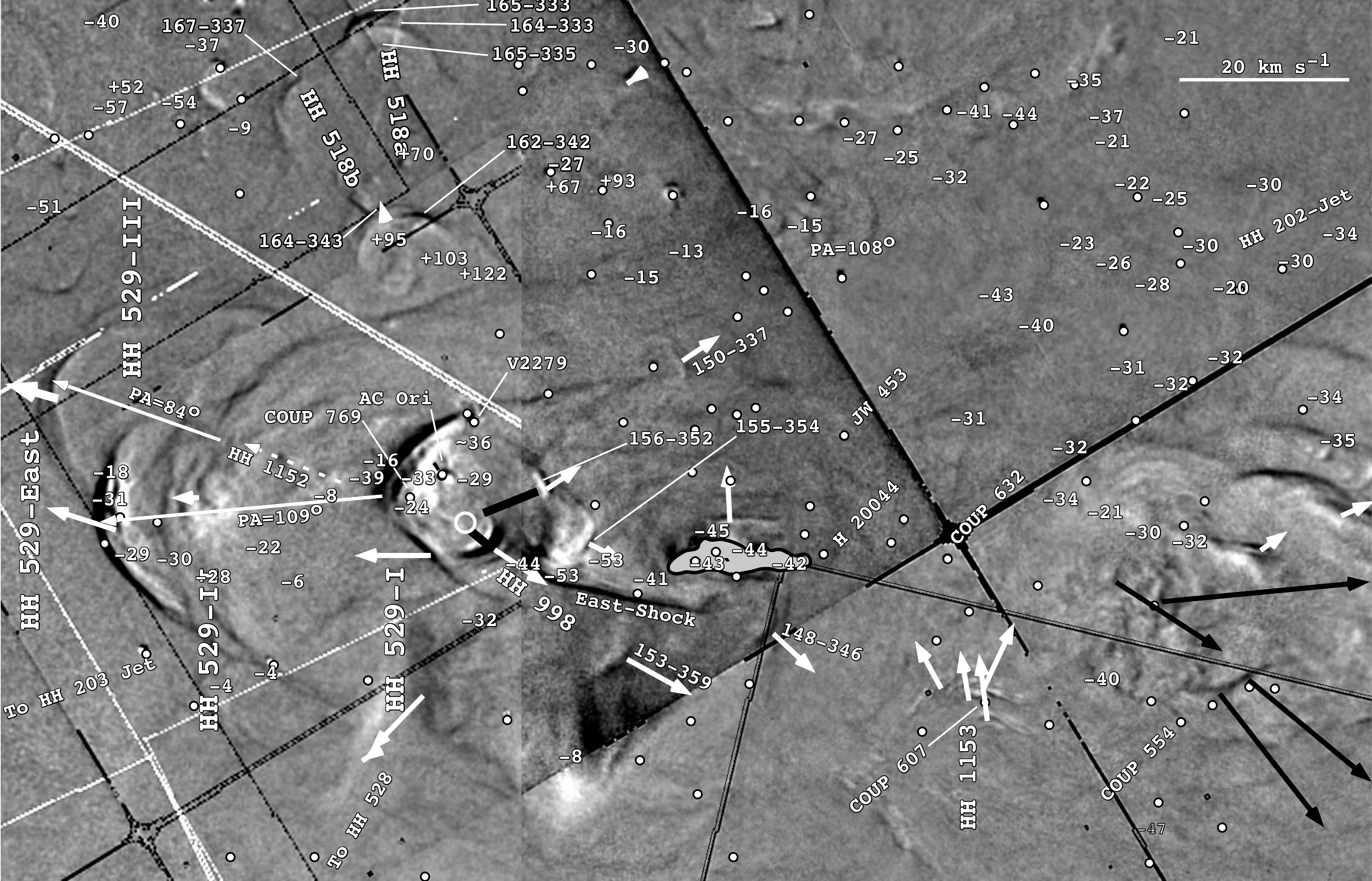}
\caption{Like Figure~\ref{fig:fig16}, but showing the motion image for F502N. Motion vectors for F502N are shown. The solid light white lines represent the several axes of the four HH~529 components, as discussed in Section~\ref{sec:HH529}. The heavy black lines to the SW of AC Ori lie along the symmetry axes of three corresponding shocks and the circle indicates the point of convergence of them as discussed in Section~\ref{sec:blanksAC}. The scale of the two rightmost velocity vectors is truncated in order to fit onto the figure. The numbers indicate the radial velocities in \oiii, although not all are drawn in crowded areas that are subjected to detailed examination. The dashed line west of COUP 769 is a series of moving knots discussed in Section~\ref{sec:HH998} and Section~\ref{sec:blanksAC}. The trapezoidal dashed boxes indicate the regions likely to contain the sources of HH~529-I--III and HH~529-East. }
\label{fig:fig19}
\end{figure}

\placefigure{fig:fig19}

\begin{figure}
\epsscale{1.0}
\plotone{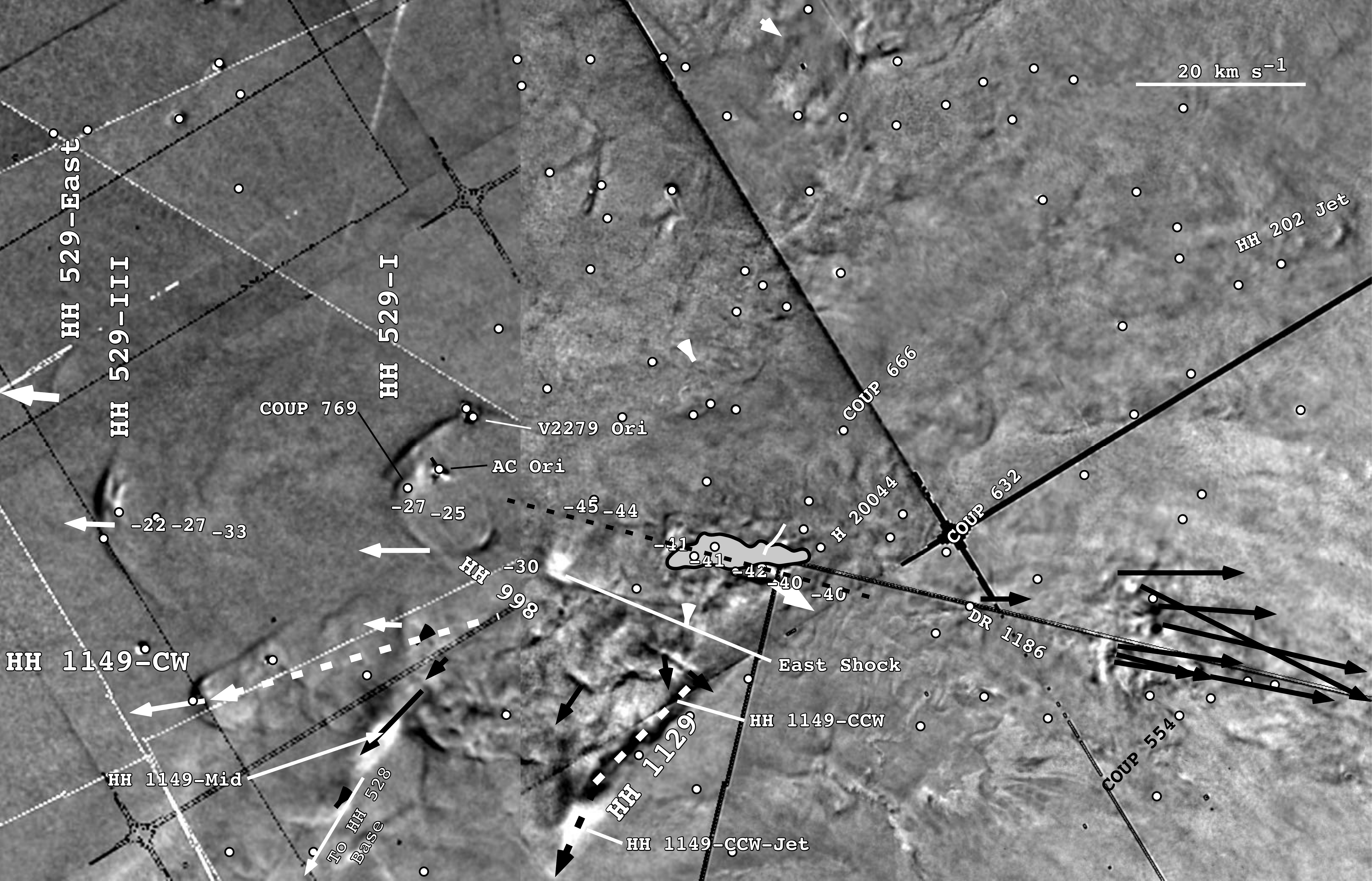}
\caption{Like Figure~\ref{fig:fig16}, but showing the motion image for F658N. Motion vectors for F658N are shown. 
The dashed lines represent the boundaries of HH~1149, as discussed in Section~\ref{sec:HH1149}. The dashed light black line shows the average PA for the East-Second-Jet (89\arcdeg). The scale of the rightmost three velocity vectors is truncated in order to fit onto the figure. The nearly linear white line crossing the East-Jet is the \nii\ strong feature seen in the lowest panel of Figure~\ref{fig:fig12}. The numbers indicate the radial velocities in \nii. 
}
\label{fig:fig20}
\end{figure}

\placefigure{fig:fig20}

The system called HH~529 was introduced in \citet{bal00} and was subsequently divided into three components (I, II, and III) by \citet{ode08a}. 

Our merged GO 12543 and GO 11038 images are shown in Figure~\ref{fig:fig16}, Figure~\ref{fig:fig17}, and Figure~\ref{fig:fig18}.  
Examination of these figures show that shocks HH~529-I, HH~529-II, and HH~529-III  share a very similar structure in F502N, but HH~529-II is much fainter, and in contrast, none of these three shocks are visible in F658N. 
The spectroscopic study of HH~529-III by \citet{bla06} concluded from its ionization balance that this was a shock formed in fully ionized gas, thus accounting for the object's lack of visibility in F658N.
The symmetry axis of HH~529-I and HH~529-III shocks is  109\arcdeg, which passes through the less well defined crest of the HH~529-II bowshock.

 In contrast with the single epoch filter images,  shocks HH~529-I and HH~529-III appear in both the F502N motion image (Figure~\ref{fig:fig19}) and the F658N motion image (Figure~\ref{fig:fig20}). 
 
\subsubsubsection{Features Near the Newly Identified HH~529-East Shock}
\label{sec:529east}

Our new analysis shows that there is an additional nearby shock that we designate as HH~529-East. The symmetry axis of it is about  84\arcdeg. 

The F502N motions image (Figure~\ref{fig:fig19}) also shows a faint series of moving knots centered 8.3\arcsec\ east  of COUP 769. The axis of these knots is about  82\arcdeg.

\subsubsubsection{Other Shocks Near HH~529}

There are other features that are associated with sources in this area, but they cannot be identified exactly.
The shock 150-337 lies at 12.2\arcsec\ from AC Ori towards  309\arcdeg. There are no intervening candidate sources, although COUP~769 lies only 1.8\arcsec\ beyond AC Ori. 

F502N/F656N  and F658N/F656N ratio images covering the entire \hr\ show that there is an additional large high ionization shock to the east of the OOS region (Figure~\ref{fig:fig24}). This may belong to one of the HH~529 outflows and is discussed in section~\ref{sec:SEbig}. 
  
\subsubsubsection{3-D Motion of HH~529-III}

Of the several components of HH~529, we have combined \vt\ and \vrad\ velocities for HH~529-III. Taking the average of the \vt\ values (7 \kms) and the average of the \vrad\  values (-28 \kms) yields \Vomc~=~54 \kms\ at $\theta$=83\arcdeg. 

\section{LARGE-SCALE FEATURES IN THE HUYGENS REGION}
\label{sec:connections}

In addition to the HH~269 and HH~529 systems there are several other large-scale flows that appear to originate in the same vicinity.
In the following sections we will present in the following order the results for HH~202, HH~203+HH204, and HH~528. We address their origins in Section~\ref{sec:fans}.

\subsection{HH~1149}
\label{sec:HH1149}
The feature that we have labeled as HH 1149-CW in Figure~\ref{fig:fig20} is a well-defined shock lying at the SE end of an extended region of moving 
shocks seen clearly only in F658N motions images. The symmetry axis of this easternmost shock is shown in Figure~\ref{fig:fig20} and is about 120\arcdeg. 
Although this shock lies near a line projected from the red-shifted SiO flow arising from COUP 554 \citep{zap06}, that flow has  PA=100\arcdeg\ and it cannot be driving this shock. 
There is a series of other shocks lying CCW from the HH~1149-CW feature that show a pattern of motion and symmetry with progressively larger PA and they end abruptly at a complex linear feature (at  154\arcdeg) in the lower-right of the pattern. The CCW boundary is ill defined and can be drawn as segmented with the upper boundary at 160\arcdeg\ and the lower at 168\arcdeg. We designate this series of shocks as HH~1149. 

The HH~1149 feature lies in a region of enhanced \nii\ emission (c.f. Figure~\ref{fig:fig43}). An enhanced F658N/F656N ratio this close to \tc\ is an indication of viewing an ionization front nearly edge-on \citep{md11}. This is why the eastern side of optical Orion-S images has high F658N/F656N and the west side does not.  The abrupt CCW end of the fan of HH~1149 shocks may indicate that they are overlaid by a step of material on the Orion-S east side. A problem with this interpretation is that one sees what may be an extension of the HH~1149 southern edge extending beyond the HH~1149-CCW boundary in the \nii\ component of Figure~\ref{fig:fig43}.

The northern end of the CCW boundary (marked as HH~1129) may be due to structure in the west side of Orion-S, but the lower part (marked as HH~1129-Jet) lies on an extension of a linear feature arising from the star COUP~666 at an angle of 160\arcdeg (Section~\ref{sec:final1149}). It is likely that the CCW boundary is part of a linear flow, hence the designation as HH~1129. There is a similar feature (marked HH~1149-Mid). Its axis is at 162\arcdeg. It may be unrelated to the fan of bow shocks defining HH~1149. 

In the material presented below we give the results of an analysis of the tangential velocities and in more detail the radial velocities. An  analysis similar to that of the radial velocities but of the relative line intensities is given in Appendix~\ref{AppendixD}. 

\subsubsection{Tangential Velocities In and Near HH~1149}
\label{sec:HH1149tan}

\begin{figure}
\epsscale{1.0}
\plotone{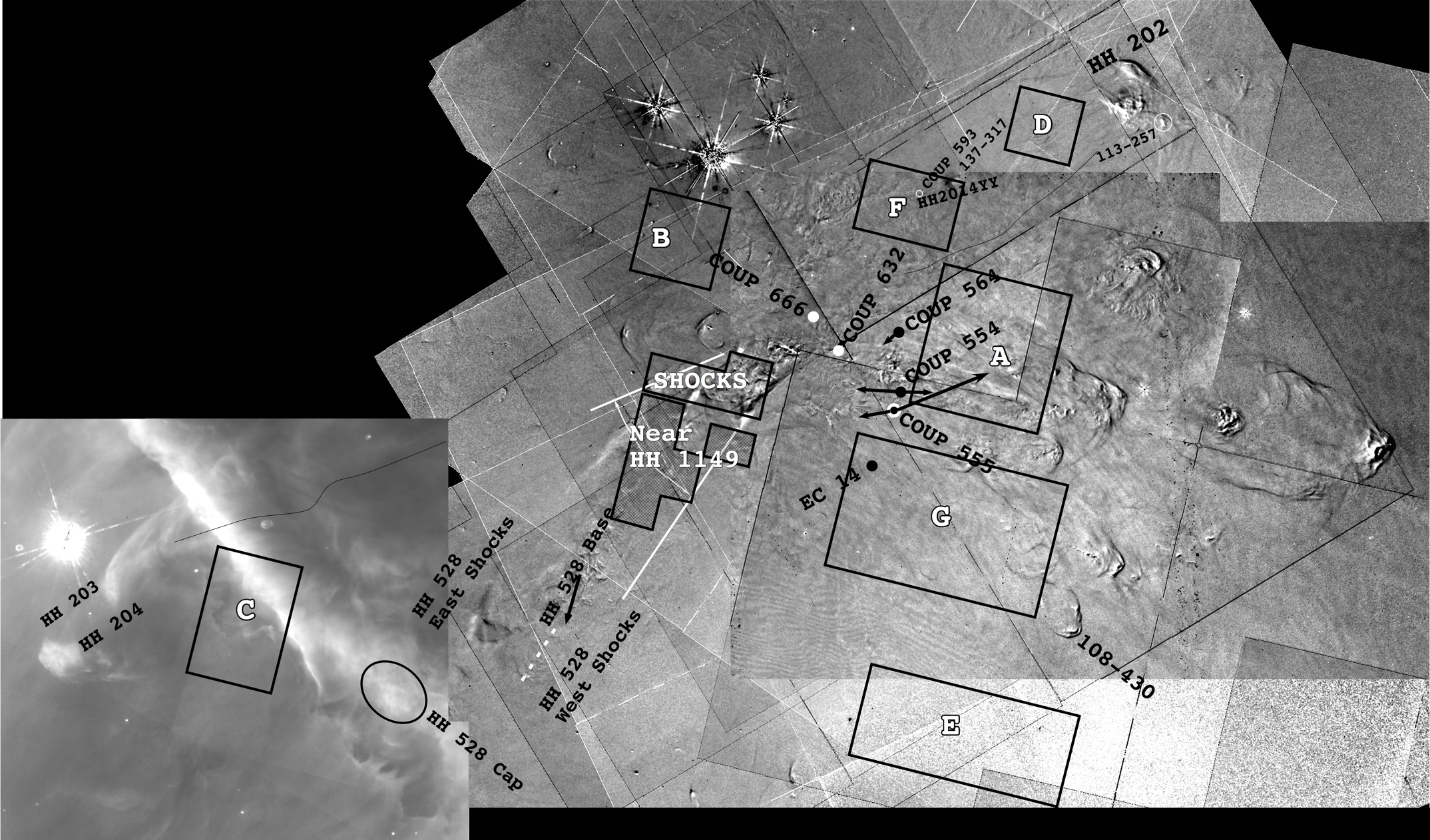}
\caption{This 260.1\arcsec $\times$120.5\arcsec\ FOV (c. f. Figure~\ref{fig:fig1} where this FOV is shown) shows the region sampled in F658N, numerous individual features, and area averaged spectra samples (A through G) used in Section
~\ref{sec:HH1149} for comparison with the irregular samples used for analysis of HH~1149. The 
"SHOCKS'' sample includes only an area within HH~1149 and the ''Near HH~1149'' sample includes an area immediately outside of HH~1149. The arrows indicate known molecular outflows. The filled white circles near the center represent the positions of the stars COUP 666 (upper left) and COUP 632 (lower right).}
\label{fig:fig21}
\end{figure}

\placefigure{fig:fig21}

HH~1149 appears only on the F658N motions images (Figure~\ref{fig:fig20}) owing to its low ionization and the presentation of multiple bow shocks.
The tangential velocity of most of the features composing HH~1149 are low. Changes in structure prevent determination of tangential motions from the study of large-scale samples, but a few objects do have individual determinations and these are never large.The five measured features have an average of only \vt~=~5 \kms\ and these are the larger motion features. The series of shocks is real, but the tangential motions are small.

\subsubsection{Radial Velocities In and Near HH~1149}
\label{sec:HH1149rad}

 No radial velocity features are seen in the Atlas, so that the radial velocities too must be small. Area averaged spectra across the entire feature have been used to increase the S/N ratio of the data. These area averaged spectra have been compared with other area averaged spectra within the \hr. Henceforth we will call the area average spectra ''samples''. The results of this comparison is presented later in this section. 
 
 \subsubsubsection{The Area Averaged Spectra and Selected Results}

We investigated the radial velocity properties in \oiii\ and \nii\ of the HH~1149 object and the region just beyond it, labeling them as SHOCKS and Near HH~1149 (henceforth "NEAR") respectively in Figure~\ref{fig:fig21}.  The SHOCKS sample was best fit into three components: a low velocity component (henceforth BLUE) described in most detail by \citet{npa06}, a component (henceforth MIF) identified with the MIF, and  a high velocity component (henceforth Red) due to backscattering from dust in the PDR behind the MIF \citep{ode92,hen94,doi04,npa06}. The results are shown in Table~ \ref{tab:HH1149}.

We also created seven large area averaged spectra as shown in Figure~\ref{fig:fig21} within the \hr\ for comparison with our SHOCKS and NEAR samples. The results for them are also given in Table~\ref{tab:HH1149}. Some results were very different from the others and were not used in deriving averages. The \oiii\ MIF velocity for Sample-A was 14.7 \kms\  and the \oiii\ MIF for Sample-G was 12.7 \kms\ (both of these are within the probable errors of the other samples, but as we'll see below the \oiii\ emissions in these regions are anomalous).  Sample-D was not used because its unusual value (7.8 \kms) for \oiii\ may be the result of blending with the blue-shifted flow that drives HH~202. 

The sample F \nii\ BLUE velocity was not used because of contamination of that sample by proplyds and shocks. For comparison, the most complete compilation of radial velocities in the \hr\ is that of \citet{gar08} who find average velocities of \vrad (\oiii)=16.3$\pm$2.8 \kms, \vrad (\Ha)=16.8$\pm$3.0 \kms, \vrad ([S III])=19.7$\pm$3.0 \kms, \vrad (\nii)=20.5$\pm$2.9 \kms, \vrad (\sii)=21.2$\pm$2.4 \kms, and \vrad (\oi)=25.7$\pm$3.4 \kms.

\subsubsubsection{The Most Important Radial Velocity Results}

The purpose of the \vrad\ study was to see if there were detectable velocity differences in HH~1149. Since this
object is only seen in \nii\ and any differences will be near the MIF values, it is these velocities that were compared. For \nii\ the large samples gave an average \vrad(MIF)=20.0$\pm$1.9 \kms, the NEAR sample gave \vrad(MIF)=21.0 \kms, and for the sample on the object \vrad(SHOCKS)=21.9 \kms. This means that the NEAR sample's velocity was well 
within the scatter of the large samples and the SHOCKS sample's velocity was only at one standard deviation.
The conclusion is that any systematic difference in radial velocity of HH~1149 must be very small (a few \kms); but, 
an alternative interpretation is that the HH~1149 shocks contribute little to the signal near the MIF velocities.
The latter interpretation would be consistent with the fact that we only see HH~1149 clearly in the motions \nii\ image and not the F658N image. In summary we can only say that we cannot determine HH~1149 \Vomc\ values. 

In the case of the analysis of the \oiii\ NEAR sample spectra the strongest velocity component (9.7 \kms) is similar to the typical blue component velocities, as discussed in Appendix~\ref{AppendixD}.

\begin{deluxetable}{lcccccc}
\tabletypesize{\tiny}
\tablecaption{Data for the HH~1149 Samples and the Large Samples*
\label{tab:HH1149}}
\tablewidth{0pt}
\tablehead{
\colhead{Sample Name} &
\colhead{Component} &
\colhead{\vrad ~\nii\ (\kms)} &
\colhead{\vrad ~\oiii (\kms)} &
\colhead{$\frac{S(MIF-\oiii)}{S(MIF-\nii)}$ } &
 \colhead{$\frac{S(BLUE-\nii)}{S(MIF-\nii)}$} & 
 \colhead{$\frac{S(BLUE-\oiii)}{S(MIF-\oiii)}$}
}
\startdata
Large Samples & MIF & 20.0$\pm$1.9~(7) & 15.5$\pm$2.6~(4) & 0.60$\pm$0.11  &  0.07$\pm$0.05(6) & 0.17$\pm$0.09(4)\\
-----                    & BLUE & 1.3$\pm$2.5~(6) & 2.6$\pm$2.2~(5) & --- & --- & ---\\
SHOCKS &MIF & 21.9       & 17.0  & 0.29  & 0.02 & 0.06\\
-----                      &BLUE & -4.7  & 1.1 & --- & --- & --- \\
NEAR      &MIF & 21.0  &  9.7/20.9 & ---  & 0.06  & --- \\
-----                      & BLUE & 2.9   &--- & --- & --- & ---\\
\enddata
\tablecomments{~*Numbers in parentheses indicate the number of samples used in deriving the averages.}
\end{deluxetable}
\newpage

\begin{figure}
\epsscale{1.0}
\plotone{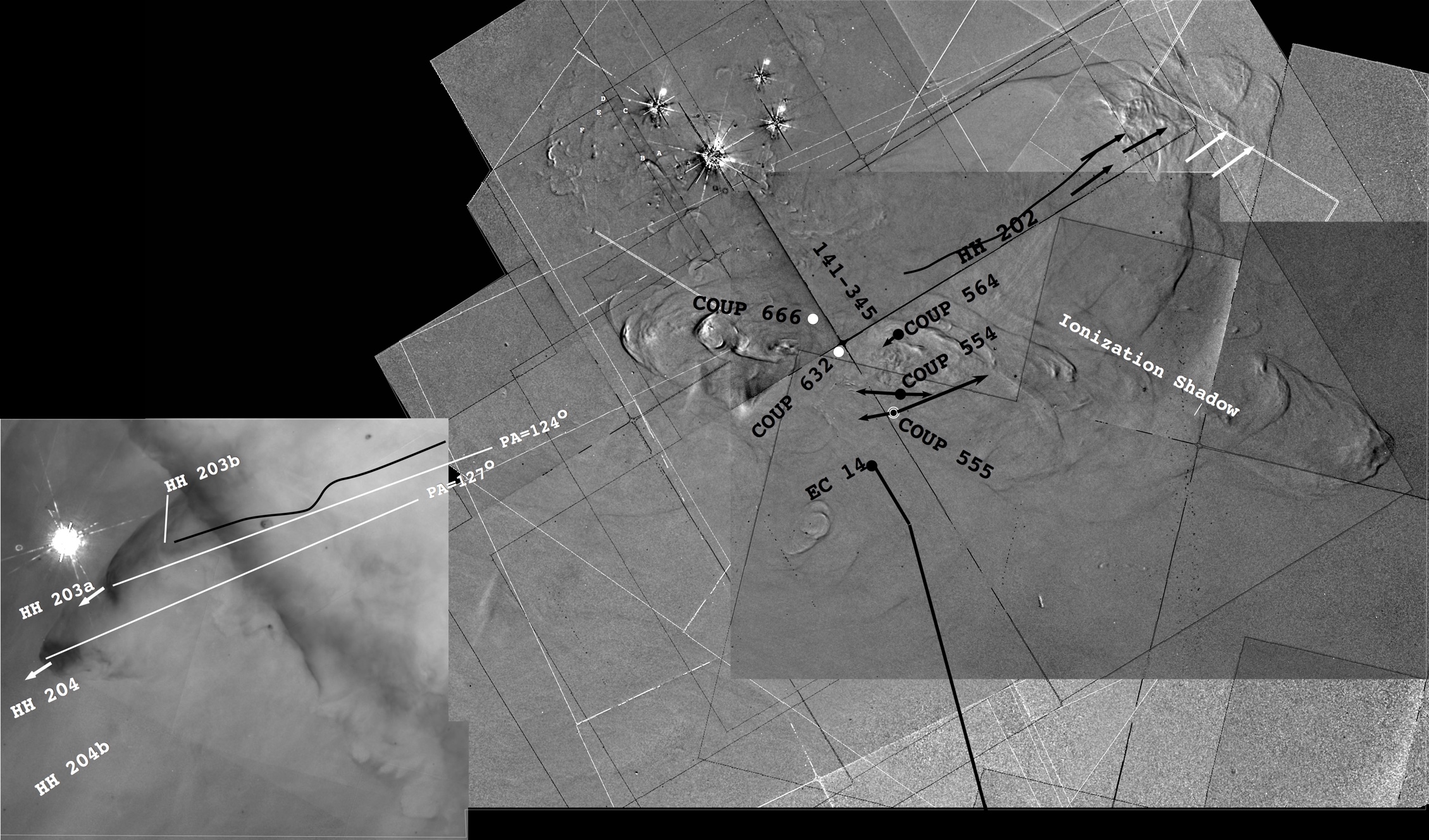}
\caption{This 260.1\arcsec $\times$120.5\arcsec\ FOV (c. f. Figure~\ref{fig:fig1} where this FOV is shown) shows a F502N motions image except for the inset at the lower left that is a ratio image of F502N/F656N. The irregular dark lines leading to HH~202 and HH~203 are the blue shifted jets reported in \citet{doi04}. Candidate stars for the major optical outflows (COUP 632) are shown with white filled circles and the candidate sources for the major molecular outflows (COUP 554, COUP 564, and EC 14) with black filled circles. The star COUP 555 that is the source of both a molecular outflow and the HH~625 optical HH object has a white circle.  The molecular outflows are shown with straight heavy vectors (except for EC 14, where it is known that the molecular jet \citep{zap10} changes PA near the source. The letters A--F refer to shocks that are part of the HH~518a and HH~518b flows (Section~\ref{sec:HH518}). The HH~1149 shock is too faint to be seen on this image, but its position is shown.}
\label{fig:fig22}
\end{figure}

\placefigure{fig:fig22}

\begin{figure}
\epsscale{1.0}
\plotone{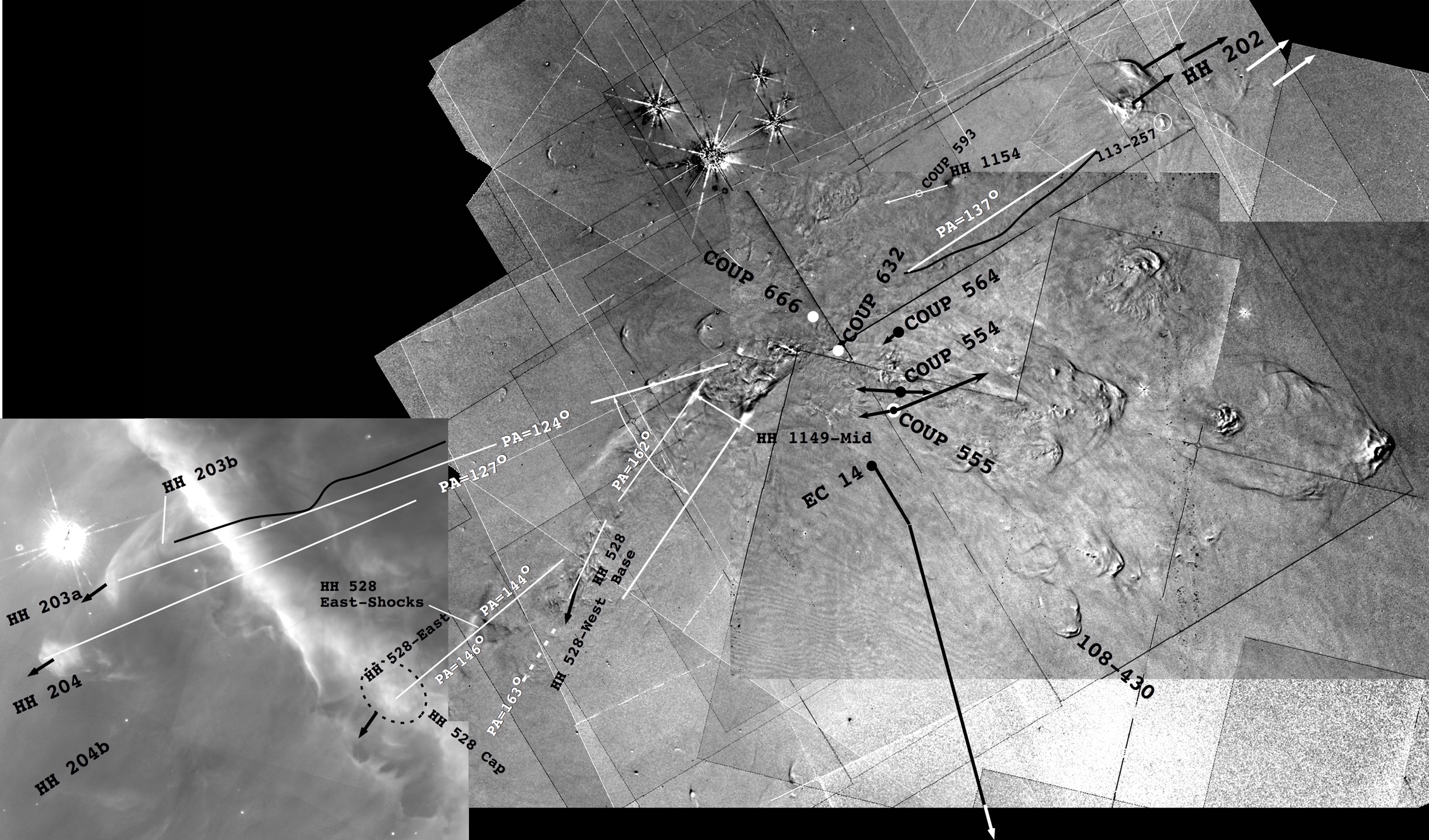}
\caption{The same as Figure~\ref{fig:fig22} except for depicting F658N data. The heavy white lines with an enclosed arc identify the boundaries of the series of low ionization shocks that include HH~203 and HH~204. 
The light white arrow shows the motion of object shock with a tip at 136.5-317.5 (designated as HH~1154) near the proplyd 139.2-320.3 (COUP 593). HH~1154 is discussed in Section~\ref{sec:HH1154}. 
The irregular dark lines aimed towards HH~202 and HH~203 are high velocity jets \citep{doi04}.
The white lines with PA values are reference lines discussed in the text. The HH204B shock is too faint to be seen on this image, but its position is shown. The white dashed ellipse shows the region of the intersection of the symmetry axes of HH~203a and HH~204a.}
\label{fig:fig23}
\end{figure}

\placefigure{fig:fig23}

\subsection{HH~ 202}
\label{sec:HH202}
The mixed ionization structure known as HH 202 was found by \citet{can80}. It has been the subject of many studies since then, starting with early HST imaging \citep{ode97b}. It has a detailed, multi-shock structure that has been the subject of several radial velocity, tangential velocity, and physical condition studies \citep{doi02,ros02,ode03a,doi04,gar08,ode08a,md09a,md09b,ode09}. The highly blue-shifted [O~III] jet delineated by \citet{doi04} also shows up in the HeI study of \citet{tak02}, although the source of the strength of the 1083.0 nm line was incorrectly explained there \citep{doi04}.  It has already been established that the apex region 
of HH~202 is a composite of multiple shocks. We see in Figure~\ref{fig:fig22} and Figure~\ref{fig:fig23}
that the jet is pointed at the middle of a number of shocks. There are several clearly defined shocks extending NNW beyond the group usually identified as HH~202. The average of the four major components \citep{ode08a} gives 68 \kms\ and $\theta$=51\arcdeg. 

The jet that currently drives one or more features of HH~202 curves and it is hard to extrapolate back accurately enough to identify the source.  
However, at an extrapolation of the jet's average path of  137\arcdeg\ 18\arcsec\ one finds COUP 666, which upon this criterium is a much more likely source than COUP 632. The rather approximate symmetry axes of the shocks 
slightly favor COUP 666.  The apexes of the several shocks fall over a range of PA values (310\arcdeg--319\arcdeg).  If the shoulders on the broadest shocks are taken as directions of other related shocks, the PA range increases (296\arcdeg--327\arcdeg). 

\subsection{HH~203 + HH~204}
\label{sec:HH2034}

HH~203 and HH~204 were found by \citet{mw62}. They appear as two overlapping but incomplete low-ionization parabolic shocks with quite different structures near their apexes. \citet{hen07} established from combinations of radial and tangential velocities that they have different (18\arcdeg) angles with respect to the line-of-sight (HH~203, \Vomc=104 \kms, $\theta$=45\arcdeg; HH~204, \Vomc=103 \kms, $\theta$=27\arcdeg). This pair of objects has had their structure, radial velocities and tangential motions well determined \citep{jw82,ode97a,ode97b,hu96,doi02,doi04,gar08,hen07}. HH~204 clearly shows that the volume behind the moving shock is photoionized by \tc. Their tangential motions are closely along their symmetry axes and those axes point back to the vicinity of the OOS, although the distance to that region does not allow an association with a specific source. As indicated before, an association with the Orion-S region was made quite early \citep{ode97a,ros02} and the case became more compelling with the identification of jets of high velocity material streaming towards them \citep{tak02,doi04}.  

In Figure~\ref{fig:fig22} and Figure~\ref{fig:fig23} we have drawn the high velocity jet determined in [O~III]  by \citet{doi04}, omitting the portions of high velocity material that must be part of the shocks. The [O~III] jet  (124\arcdeg) lies slightly CCW from a projection of the symmetry axis of the most CW shock belonging to HH~1149. \citet{doi04} also presented some evidence in [O~III] for a much shorter jet feeding into the apex of HH~204.

Study of HST images and the location of the [O~III] jet show that HH~203 is actually two parabolic shocks, with the brighter HH~203 shock (124\arcdeg) being a partial shock front and a fainter component being more complete. Whenever relevant, these components will be designated as HH~203a (the brighter, partial shock) and HH~203b (the more complete, fainter shock fed by the high ionization jet).
The [O~III] jet as delineated by \citet{doi04} does not exactly align with the brighter shock (HH~203), rather, it aims at a slightly more northerly position but with the same PA. Examination of Figure~\ref{fig:fig23} shows that there is a well-defined smaller shock at the end of the jet, this is HH~203b. 

HH~204 has a very different form \citep{ode97b} from that of HH~203, the apex appearing as a mottled grouping of shocks. However, its sides are well defined and the symmetry axis is  127\arcdeg.  

 \citet{hen07} discovered that there is a larger shock (now designated here as HH~204b) feature with a PA slightly greater than that of HH~204 and this fainter shock is large 
enough to include HH~203 within its CW boundary and its CCW boundary extends to almost the PA of HH~528's Cap (Section~\ref{sec:HH528}). Its boundaries are shown as a white dashed line in Figure~\ref{fig:fig24}. It is not clearly seen in any of our Figures 22-24, but is easily seen in Figure 7 of \citet{hen07}.

Figure~\ref{fig:fig23} shows an ellipse around the intersection of the axes of HH~203a and HH~204. This is in
the same region of an inferred source found previously for other sources.

\subsection{HH~528}
\label{sec:HH528}
HH~528 was discovered on early HST images \citep{bal00} and its radial and tangential velocities have been documented \citep{bal00,ode03a,hen07,ode08a}. It manifests itself initially by a series of small low ionization  (c. f. the [O~I] image in Figure~21 of Bally et al. (2000)), the [S~II] images in Figure~10  and Figure~11 of Doi et al. (2002), and lower resolution images in multiple ions in \citet{gar08}) shocks called in \citet{hen07} the jet. 

\cite{ode08a} point out that the original nomenclature for HH~528 presented by \citet{bal00} is better, where the broad array of small low ionization shocks are analogous to the ``base' of a mushroom and the broad shocks lying just inside the Bright Bar as the ``Cap'' of the mushroom. The ``Cap'' feature is difficult to see in the much higher surface brightness of the Bright Bar. \citet{ode08a} report that the ``Base'' is oriented towards  155\arcdeg\ (we judge it to be  169\arcdeg) and moving towards  178\arcdeg\ and the ``Cap'' is oriented approximately towards  146\arcdeg\ with its components moving towards  159\arcdeg.  Figure~\ref{fig:fig23} illustrates the direction but not magnitudes of the motions and some of the PA values for groups of features. \citet{hen07} determined that \Vomc=11$\pm$3 \kms\ and $\theta$=24$\pm$8\arcdeg for the feature that we call the HH~528 Base (illustrated well in \citet{ode03a}).

Figure~4 of \citet{ode08a} and our Figure~\ref{fig:fig23} show new features SE of the base that we now separate and identify as part of two different systems of flow.

SSE of the HH~528 ``Base'' shocks we see multiple larger shocks aligned at about  163\arcdeg\ and a dashed white line is shown connecting them. They are labeled as the HH~528-West shocks.
The location of the HH~528 West shocks with
respect to the HH~528 Base and their similar PA values indicate that they may be two parts of a common flow. The north end of the Base feature lies on the extension of the linear feature HH~1149-Mid (Section~\ref{sec:HH1149}.
To the east of the HH~528 West shocks  we see two well defined large shocks with a symmetry axis pointing both towards the ``cap'' feature and back towards the north end of the Base. The HH~528 East shocks and the ''cap'' feature are probably part of a common flow, which we'll call here HH~528-East.
 A line projecting NW from the HH~528-East flow passes (after crossing the HH~528 Base feature) several small moving features, then passes a few seconds of arc south of the CCW boundary of HH~1149. With a slow CW change of PA over this track it points to star COUP 632. 

\subsection{Large Scale Shocks Near and Beyond the Bright Bar}
\label{sec:SEbig}

\begin{figure}
\epsscale{1.0}
\plotone{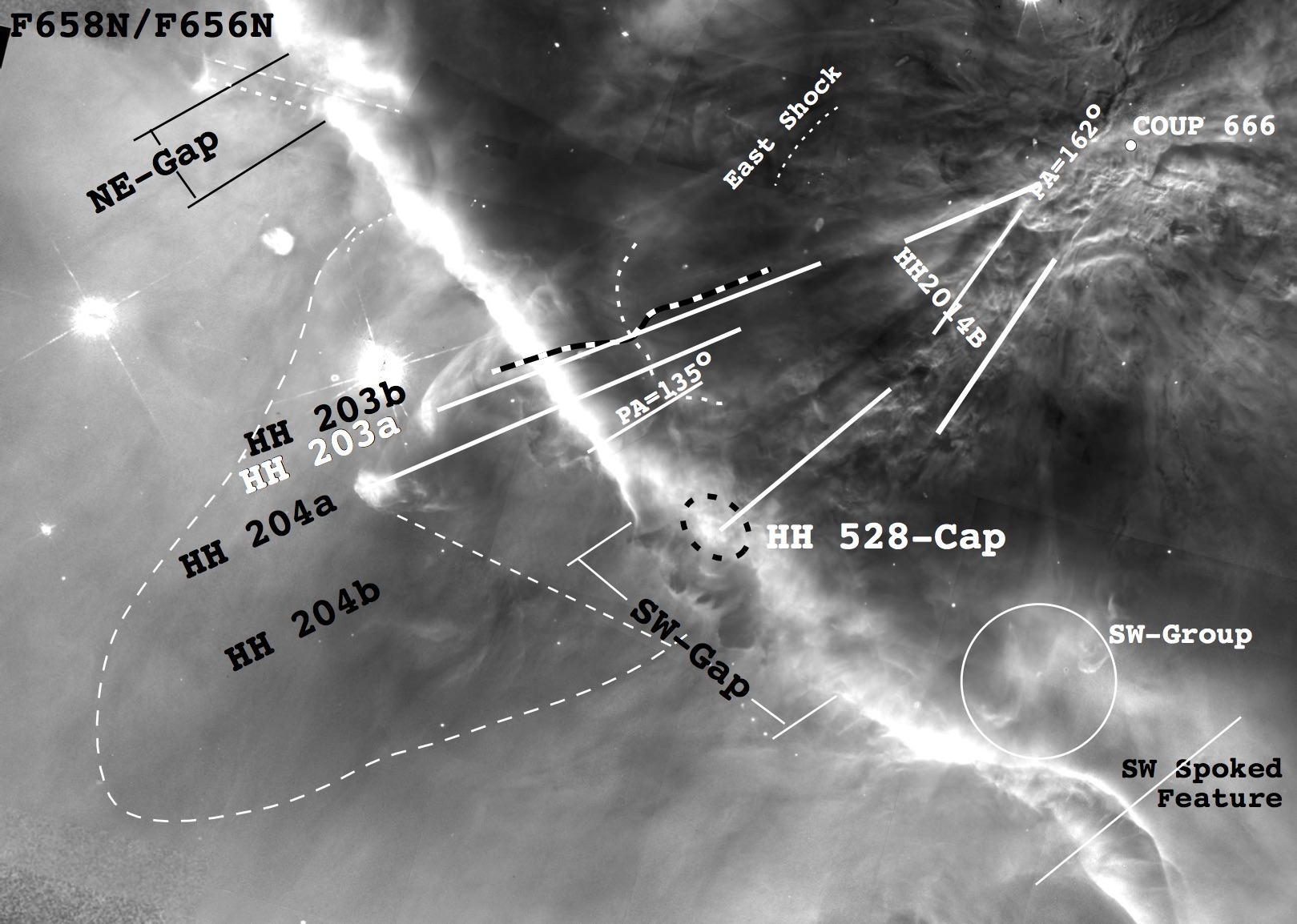}
\caption{This ratio image samples a 232\arcsec$\times$166\arcsec\ FOV shown in Figure~\ref{fig:fig1} and is taken from an early mosaic of the \hr\ using the HST WFPC2 \citep{ode96}. The two long-dashed straight lines indicate flaws at the intersection of two separate FOVs. The marked features are discussed in Section~\ref{sec:SEbig}. The position of COUP~666 is shown as a white filled circle in the upper-right of the image.
The alternating black-white line shows the position of the high ionization jet directed towards HH~203b.
 }
\label{fig:fig24}
\end{figure}

The Bright Bar is well established to be a tilted region in the MIF, where one is looking almost along the MIF. This explains that we see a narrow streak of low ionization (F502N weak) that is bright in F658N. Most of the Bright Bar is marked by very sharp ionization boundaries with dimensions of about 0.2\arcsec. Further from the Trapezium the MIF flattens out, but not so much that it is not illuminated directly by \tc. This means that the features we see beyond the Bright Bar are probably seen in projection on the MIF, rather than being formed in the MIF.

\subsubsection{The HH~203 and HH~204 Shocks}
 It has been argued that the HH~203 and HH~204 shocks are produced when collimated flows from the OOS region strike the upper part of the escarpment caused by the tilt of the MIF \citep{doi04}. The original interpretation was that it was the result of collimated flows striking the foreground Veil \citep{doi04}, which became unlikely when it was argued that the Veil was about 1 pc from \tc\ \citep{npa06,hen07}. Now the pendulum has swung and \citet{pvdw} have presented evidence that the Veil is sufficiently close that the shocks must arise there.
 
 There is a large irregular form (similar to two parallel but shifted parabolas) that is outlined with long-dashed lines encompassing HH~203 and HH~204 in Figure~\ref{fig:fig24}. It is of high ionization, being bright in F502/F656N and faint in F658N/F656N. There is a shorter and less clearly bounded adjacent region of high ionization to just beyond the CCW edge of the SW-Gap (c. f. Section~\ref{sec:crenellations}). The CW form agrees well in position along the Bright Bar with the  21-cm absorption feature L in the study of \citet{pvdw}, but the optical feature extends further away from the Bright Bar.

 At the CW edge of the above feature there is a well defined partial shock, shown as a short-dashes curved line in Figure~\ref{fig:fig24}. It is too incomplete to argue for an orientation except to say that it's axis points somewhere between the OOS and the Trapezium.
 
 \subsubsection{Crenellations}
 \label{sec:crenellations}
 
Beyond the Bright Bar (to the SSE) we see a number of crenellated structures that lack the bow shock form of HH~203b,  HH~203a, and HH204a, but are likely to be shocks driven by flows from the region of the OOS.  

The first group occurs just beyond the HH~528-Cap feature and is easily visible in both the F502N/F656N image and the F658N/F656N
image shown in Figure~\ref{fig:fig24}.These features occur in a region where the Bright Bar lacks a crisp SE boundary, a region labeled "SW-Gap" in Figure~\ref{fig:fig24}. The most
visible of these features occur between the projection of the CCW edge of HH~1149 (160\arcdeg) which points back towards the vicinity of COUP~666 and a wide high ionization linear feature (HH~1149-Mid) 
with 162\arcdeg. Given the wealth of shocks in HH~1149 within this PA range, it is very likely that these are bow shocks forming in the tilted portion of the Bright Bar or the foreground Veil.

Proceeding NE along the bright bar we see are series of similar features  that begin at  135\arcdeg\  and extend to the axis of HH~204b (127\arcdeg). Again, the Bright Bar lacks a crisp boundary along most of this region. 

There are a group of features similar to the crenellated features that are found inside the Bright Bar. These are designated as 
the SW-Group in Figure~\ref{fig:fig24}. Having individual designations of 147-532, 144-523, and 137-519. They lie at 184\arcdeg from COUP 666 as shown in Figure~\ref{fig:fig24}

\subsubsection{Features Inside the Bright Bar}
\label{sec:insideBB}
Inside the Bright Bar we see a thick and wide arc that is probably a bow shock. It appears to have two peaks, one in the direction of HH~203 and the other in the direction of HH~204.  It is almost certainly related to those shocks and the outflow in this direction arising from near the OOS. It is outlined with a heavy short-dash line in Figure~\ref{fig:fig24}.  

An isolated high ionization shock labeled East Shock in Figure~\ref{fig:fig24} may be related to the outflows associated with the HH~529 components, but it is impossible to determine its direction because only part of the Bow Shock is seen. The feature is on the east side of a broad bow shock formed feature that continues to the south, reaching an apex at about  80\arcdeg.  

\subsubsection{The SW Spoked Feature Pointing Inside the Bright Bar.}
\label{sec:SWSF}
Along the SW portion of the Bright Bar imaged in Figure~\ref{fig:fig24} is an unusual object that we label as the "SW Spoked Feature" (SWSF).
It is a region with a concave feature that resembles a composite of three bow shocks pointed NNW, and with a base of 49\arcsec. Within the concave feature there are five nearly linear features, best seen in the F658N/F656N panel of Figure~\ref{fig:fig24}.
These features converge on 5:35:13.7 -5:25:56 with an scatter of only a few seconds of arc. There are no features in SIMBAD there.  
The sharp boundary of the Bright Bar continues below the east portion of the SW Spoked Feature, but then has a gap of 
12\arcsec. It resumes and continues with decreasing brightness to the SE. Just beyond the west boundary of the SWSF a new sharp boundary begins (off the FOV in Figure~\ref{fig:fig24}, c.f. \citep{ode96}) and becomes the brighter sharp boundary during the region when the Bright Bar has two, nearly parallel sharp fronts.

Although it is tempting to assign the origin of the SWSF to the point of convergence of the projection of the eponymous spokes, the source may be more distant. There may be an association with the NNW components of HH~540. The NW center of the SWSF lies at 182\arcsec\ towards  325\arcdeg\ from the proplyd 181-826 \citep{bal05} that is the source of the HH~540 flows, and close to the symmetry axis of the SWSF shocks ( 323\arcdeg). 
\citet{bal05} assigned two shocks to the NNW of 181-826 as part of its system. The nearer shock (HH 540 N2)  is the better defined and lies at 46\arcsec\ and  337\arcdeg.  There are two shocks lying farther NNW from HH~540~N2 at smaller position angles (the closer called HH~540~N1).  These angles are very similar to the symmetry axis (about 341\arcdeg) of the curved rings that give 181-826 the descriptive nickname (if one recalls mid-20th century cartoons) the Beehive \citep{bal05}.
\citet{bal06} demonstrate that shocks from long collimated flows commonly are curved away from the Trapezium and/or the center of the \hr. If the position of the HH~540~N2 and its further companion are at smaller PA values because of curvature, there is no relation of the SW Spoked Feature and 181-826. Of course the association may still be valid if the SW Spoked Feature is the result of a different period of collimated outflow under conditions where the curvature is less (e.g. higher velocity collimated flow from 181-826).

There may be another feature that is related to NNW outflow in HH~540. It clearly appears in the first HST (WFPC2) mosaic of \hr\ images \citep{ode96} and the more recent ACS mosaic \citep{hen07} as a parabolic high extinction area lying to the NNW from the SWSF. It stands out in the high resolution extinction map of \citet{ode00b}, where it was first called the SW Cloud. The tip of its parabolic form is at 112-436 and the axis of symmetry is about 343\arcdeg. The tip lies at 480\arcsec\ and 348\arcdeg\ from the HH~540 source 181-826. Although its PA and symmetry axis are not as close to the HH~540 features as SWSF's, this may be due to the fact that the velocity and angle of the 181-826 outflows have changed with time.

A caveat arguing against the association of the SW Cloud and HH~540 is that one sees within it a series of narrow arcs of high extinction over 40\arcsec\  along  277\arcdeg\ and ending at 102-453. These are similar to features seen within and at the west end of the Dark Bay  high extinction feature east of the Trapezium \citep{ode00b}. 

\subsubsection{Opposite Flow Through a Gap in the Bright Bar.}
There is an additional gap in the sharp boundary of the Bright Bar that is labeled in Figure~~\ref{fig:fig24} as the NE-Gap. There is a 
set of shocks nearby whose alignment is indicated by a small-dash straight line (unfortunately there is scar from combining the 
early WFC CCD's nearby and it is indicated by a long-dash straight line).  These shocks together with several less obvious and smaller shocks between the two ends indicate motion away from a source at  87\arcdeg. There are no obviously good candidate sources nearby in this direction.
The diffuse feature shaped like a seahorse (looking left in Figure~\ref{fig:fig24}) is low ionization and is not in alignment with the NE-Gap shock system, however, it is likely that this feature is caused by the same flows producing the NE-Gap shocks. 

The NE-Gap  does not appear upon a projection of the axis Identified for HH~529-East or for HH~529-III (Section~\ref{sec:HH529}) and the form of its features all indicate a motion in the opposite direction (west).  

van der Werf's (2013) 21-cm absorption feature D is a C shaped configuration oriented towards the WNW. The optical features of the NE-Gap system fall into the middle of this structure. It appears that the optical parts are impinging on neutral material in the foreground Veil, producing the high density, high negative radial velocity component observed at 21-cm. 

\subsection{HH~625}
\label{sec:HH625}

\begin{figure}
\epsscale{0.8}
\plotone{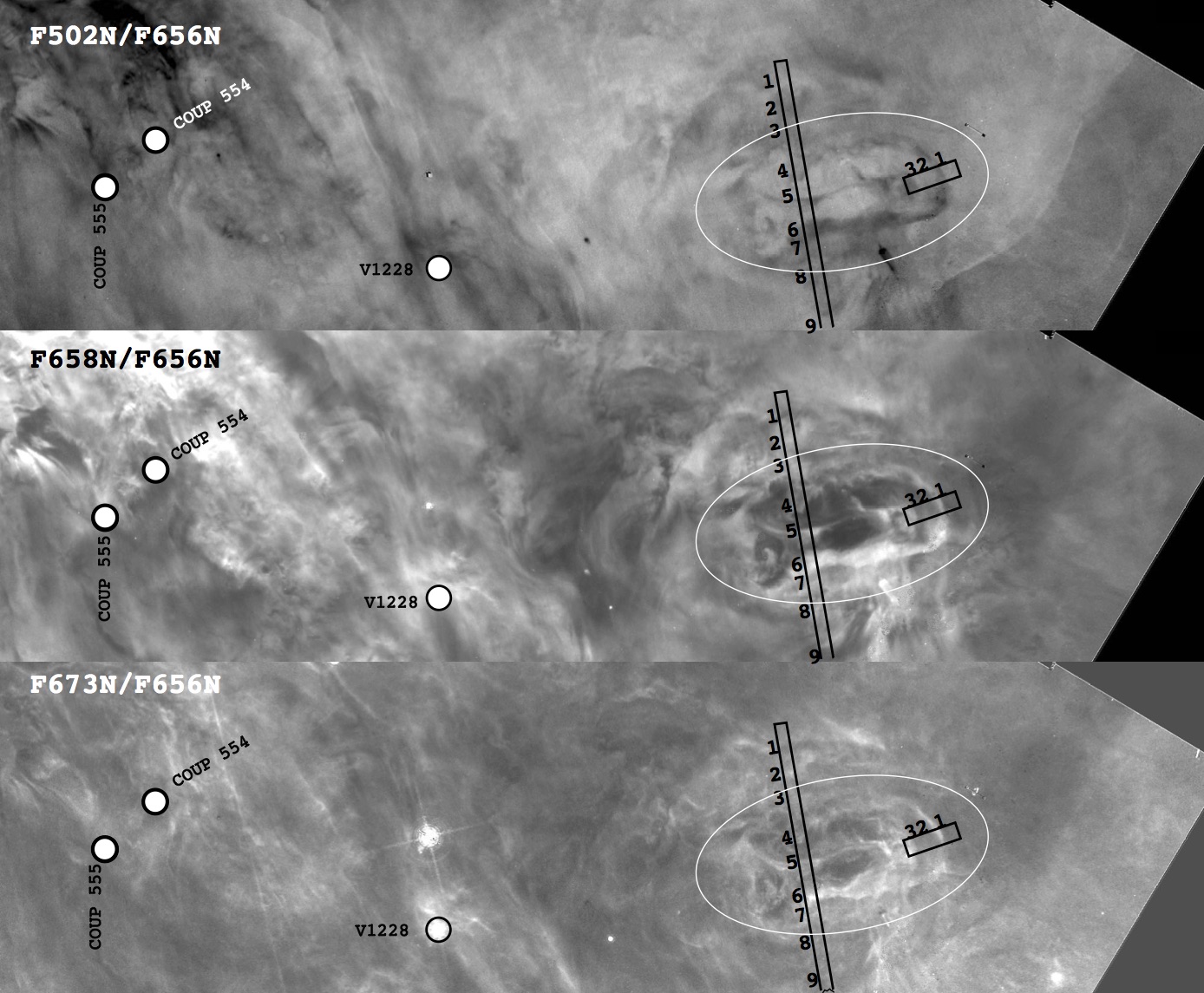}
\caption{These three ratio images show the ionization structure of HH~625 (designated with a white ellipse here and in the next figure). The irregular FOV is shown in Figure~\ref{fig:fig1} and is 80\arcsec\ wide and 22\arcsec\ high, with the positive vertical axis pointed towards 45\arcdeg. The F656N/F487N (\Ha/\Hb) is constant over this FOV to a few percent, indicating that there is little effect of location wavelength dependent extinction when using the higher S/N ratio F502N/F656N images as an indicator of low ionization regions. The positions of the three candidate sources for HH~625 are shown as filled circles. The numbered rectangles shown are where the profile samples discussed in Section~\ref{sec:HH625structure} were taken.}
\label{fig:fig25}
\end{figure}

\begin{figure}
\epsscale{1.0}
\plotone{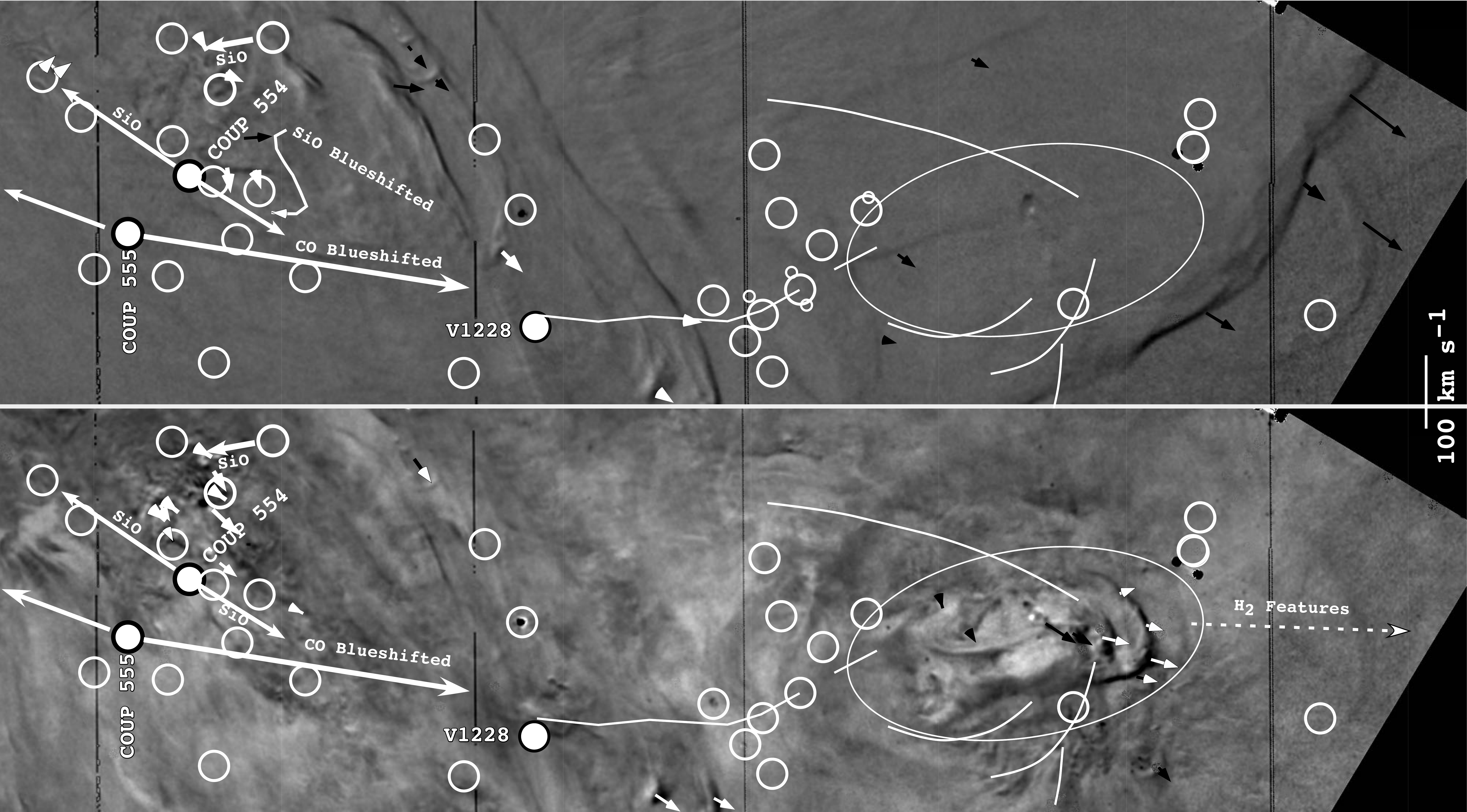}
\caption{The same FOV as Figure~\ref{fig:fig25} except that now motion images are shown with F502N images in the top panel and F658N images in the lower. Velocity vectors most likely connected with HH~510 (c.f. Section~\ref{sec:HH510}) have not been shown. Compact sources included in the SIMAD data base are shown as circles or filled circles. The candidate sources (two with molecular outflows) are shown as filled circles and have the extent of any outflows shown as heavy straight lines \citep{zap05,zap06}. The small open circles are compact  \htwo\ from \citet{kai00}.
Extended \htwo\ features in Kaifu et al.'s images are also shown as thin white lines. The dashed line marked as "\htwo\ features " indicates the direction to the nearest \htwo\ feature in the chain of knots discovered by \citet{sta02}.}
\label{fig:fig26}
\end{figure}

HH~625 was identified as a structured, moving feature that could be identified as an HH object by \citet{ode03a}, who first measured its tangential motion (as 26 \kms) at a point on the furthest extend of the leading edge. Our new observations (Figure~\ref{fig:fig25} and Figure~\ref{fig:fig26}) show that the symmetry axis is towards  324\arcdeg$\pm$3\arcdeg\ and that the tangential velocity is 30$\pm$9 \kms\ towards  311\arcdeg$\pm$21\arcdeg\ as measured at the leading three points. In Figure~\ref{fig:fig1} we see that there is a high velocity He~I feature at the leading edge of the object. \citet{ode08a} give \vrad~=~-8 \kms. Combining this radial velocity with our new tangential velocity gives \Vomc~=~45 \kms\ and $\theta$~=~49\arcdeg.
 
  \citet{ode03a} made an association of the brightest portion with \htwo\ features  to the southeast \citep{kai00} (Figure~\ref{fig:fig26}) where we have added outlying curved features in the \citet{kai00} images) and the series of \htwo\ knots identified by \citet{sta02}. The Stanke and McCaughrean knots begin at a direction of 313\arcdeg\ and curve slightly CW out to a distance of 204\arcsec.  

\placefigure{fig:fig25}

\placefigure{fig:fig26}

\subsubsection{Source}
\label{sec:HH625source}

\citet{ode08a} argued that the bipolar CO flow from the embedded source COUP 555 (136.0-359.0) is the most likely driving source for HH~625. 
In the light of the new data, it is worth examining if there are alternative driving sources. The strength of the aligned \htwo\ features and the richness of low ionization features in HH~625 itself argues that the object is breaking through an ionization front (c.f. Section~\ref{sec:physics}) and this favors an embedded source, but does not exclude an optical source since the optical extinction in a column sufficient to form an ionization front is not great. The first two candidates discussed below must be embedded as they have no detected optical counterparts.

Associating HH~625 with COUP 555 requires that the CO flow curves, a not unusual feature in long outflows and one that can be explained by passage of the collimated outflow through ambient material of differing density \citep{can96}. 
 As mentioned above, \citet{ode08a} used the radial velocity data of \citep{doi04} to determine that the radial velocity of HH~625 was -8 \kms. In contrast, the CO blue shifted flow \citep{zap05} from COUP 555 is traced out to \vrad= -36 \kms. If the shocks of HH~625 are driven by COUP 555, then the slower velocities in the shock would require that the HH~625 features are mass-loaded shocks.  

COUP 554's blue shifted outflows have heliocentric radial velocities of -62 to -2 \kms, which makes it a good alternative source for HH~625, but that identification would require an even greater deviation in direction than COUP 555 before reaching HH~625.  An association is unlikely.

V1228 lies immediately at the end of the \htwo\ features that appear to be associated with HH~625.  This object is unresolved on our HST images and there is no indication of a microjet. It has an optical image (V1228 Ori) and falls exactly at the end of a curving \htwo\ feature that leads to HH~625.  

The selection between V1228 and COUP 555 depends upon one favoring a source (COUP 555) with a known blue shifted molecular flow (but requiring a deflection of the flow) or a closer fainter star (V1228) linked to HH~625 by an elongated \htwo\ feature. The deciding factor is the fact that HH 625 appears to be breaking through an ionization front, which favors the embedded star COUP 555 as the source.

The tangential motions of HH~625's leading edge are not moving along the line of symmetry of the feature. As one examines the shocks in a CW progression around the leading edge, one sees a progression to smaller PA Values and the feature in the expected position of a Mach disk  and another to its east share the direction of motion of the CW portion of the leading edge. At 11.5\arcsec\ to the northwest from the leading edge there is a large shock, seen in both the F502N and F658N motion images (Figure~\ref{fig:fig26}). Its axis of symmetry falls between that of the overall HH~625 feature and the CW components of the leading edge. It may be the product of two flows.

\subsubsection{Structure}
\label{sec:HH625structure}

We see in Figure~\ref{fig:fig25} that HH~625 lies within a region of high ionization (high F502N/F656N, low F658N/F656N). However its immediate surroundings ( a circle of about 20\arcsec\ diameter centered about the middle of the long profile in Figure~\ref{fig:fig25}) is of lower ionization. This indicates that HH~625 is affecting its surroundings. Since it is of low velocity, shock heating on a large scale is unlikely to be important and it is more likely that the local gas density and the form of the ionization front has been modified by the penetration of the jet's gas column.

\begin{figure}
\epsscale{0.50}
\plotone{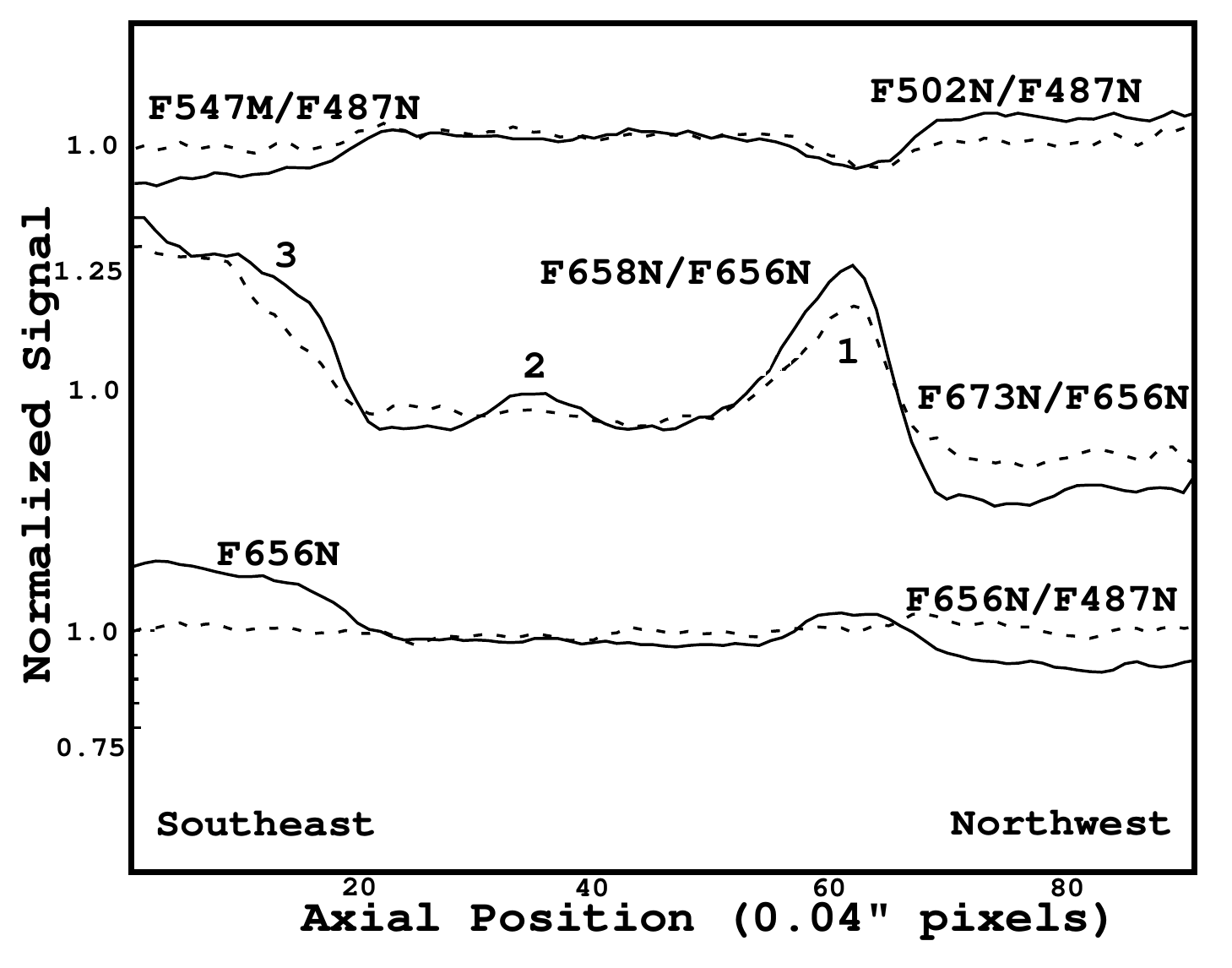}
\caption{Normalized profiles of 20 pixel (0.8\arcsec) width along the short (``axial'') line indicated in Figure~\ref{fig:fig25}  are presented, with features of particular  interest as discussed in Section~\ref{sec:HH625structure}, numbered. Ratios of images are shown except for the lowest line, where only the F656N signal is shown.}
\label{fig:fig27}
\end{figure}

\begin{figure}
\epsscale{0.5}
\plotone{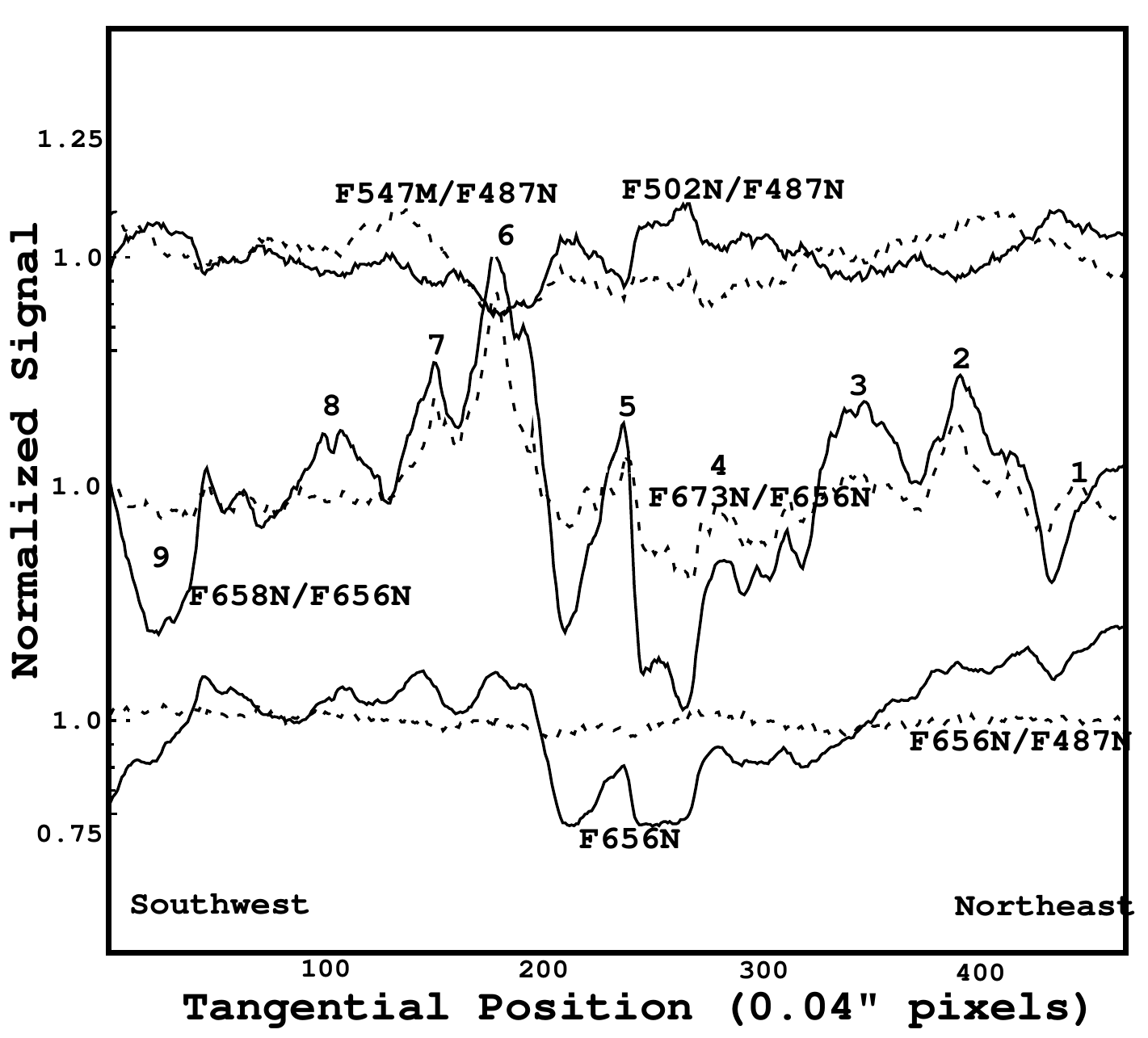}
\caption{Like Figure~\ref{fig:fig27} except for showing results along the long (``orthogonal'') line.}
\label{fig:fig28}
\end{figure}

Examination of an axial profile of the F656N brightness and various line and continuum ratios, as shown in Figure~\ref{fig:fig27} and Figure~\ref{fig:fig28} are useful in determining the structure of HH~625. In this analysis it must be kept in mind that much of the radiation along a line of sight occurs in the foreground or background, so that small scale changes probably represent large scale changes in the object.
In Figure~\ref{fig:fig27} feature 1 occurs at the bright leading edge of the object. The profiles show that F656N increases there, while the high ionization tracer (F502N/F487N) drops and the low ionization tracers (F658N/F656N and F673N/F656N) increase. This is the behavior that one expects from a tilted ionization front. The measure of the relative strength of the continuum as compared with the \Hb\ recombination line (F547M/F487N) drops at feature 1. If the continuum was dominated by atomic emission, the recombination coefficients for \Hb\ and the optical region emissivities of \citet{ost06} would mean that the electron temperature at this ionization front has decreased, which is the opposite to what is expected from photoionization models that incorporate radiation hardening near the front itself. However it is well established \citep{ode01} that the continuum in the Huygens region is dominated by scattered light coming from the dense gas and dust layer lying on the neutral side of the ionization front. In a region where one is looking nearly along an ionization front, one is looking at emission primarily from the ionized gas. This means that a trace across such a tilted ionization front would have dust scattered light plus atomic continuum (hence a high F547M/F487N ratio on both sides of the front, but in the immediate direction of the front the continuum would be primarily from ions and the F547M/F487N ratio would drop. This is what is observed.

In Figure~\ref{fig:fig27} there are no big changes in the ratios at feature 2, except for a small rise in F658N/F656N, which means that it is probably a small tilted ionization front. At feature 3 we see the same signatures as at feature 1. However, the characteristics continue off the plot and examination of Figure~\ref{fig:fig25} shows that this continues until reaching yet another clearly defined tilted ionization front.

Figure~\ref{fig:fig28} shows a similar profile plot, but in this case giving the profile completely crossing HH~625 perpendicular to the symmetry axis. These profiles are used below to understand features within and near HH~625.

The most striking feature along the symmetry axis within the boundary of HH~525 is a nearly linear feature that begins about 1\arcsec\ to the southeast of the long sample and extends almost to the southeast end of the axial sample.  It is feature 5 in Figure~\ref{fig:fig28}. It is clearly not an extinction feature (the F656N/F487N ratio is essentially constant on both profiles).  With a dip in F502N/F487N and peaks in F658N/F656N and F673N/F656N it has the characteristic of a tilted ionization front. But, there is no dip in F547M/F487M, so that it must be very thin. If it is very thin one does not get an isolated region of solely atomic emission. 

Feature 6 has the full set of characteristics of a thicker tilted ionization front viewed nearly edge-on and
with less definition, this pattern is continued to the southwest. This means that there must be a series of ledges. 

When reaching feature 9 there is a reversal in the pattern (F502/F487N rises, F658N/F656N drops, and F656N drops). F673N/F656N remains constant at feature 9, unlike the other regions-indicating that all of the peaks in this ratio are due to local features, with the background F673N being nearly constant. The drop in F656N indicates that the local front is tilted away from the photoionizing star (\tc). the strong rise in the low ionization tracers at feature 6 indicates that this is a tilted front that is narrow as compared with the local \Ha\ emitting region. 

Given that the curving driving jet, as traced by \htwo\ emission, feeds directly into the symmetry axis where feature 5 lies, it is likely that this causes a deformation in the ionization front.  There is no evidence that the jet has emerged, although it must be close to the surface of the Orion-S neutral cloud. Since there are a series of discrete \htwo\ features along the line of symmetry of HH~625, this information combines to indicate that the molecular outflow from COUP 555 must be intermittent. 

\placefigure{fig:fig27}

\section{SMALLER HERBIG-HARO SYSTEMS}
\label{sec:smallHH}

The \hr\ is the host to a large number of HH objects, some identified by jets coming from stars, others from shock forms, and others from two-epoch imaging. In this section we report on a number of previously known HH objects and several new ones discovered on the motions images. We do not report the measurements of the LL~1 object located in the SW corner of the GO~12543 FOV as those results have already been published \citep{hen13}.

\subsection{HH~510}
\label{sec:HH510}

This young-star jet driven outflow was first described by \citet{bal00} using an HST-WFPC2 wide-band filter (F606W) image.
The source of the outflow is the edge-on disk proplyd d109.4-326.7 (d119-327, \citep{bal00}; MLLA 379, \citep{mlla}; HC 286, \citep{hil00}.
The \citet{bal00} image revealed a monopolar east pointing microjet beyond which there were two identifiable knots lying on a projection of the jet.
\citet{ric08} has published images made with the HST-ACS camera using the F435W, F555W, F658N, F775W, and F850LP filters, most of which are most sensitive to continuum radiation, and the narrow-band filter F658N of the ACS  (which is approximately the same sensitivity at both the \Ha\ 656.3 nm and \nii\ 658.3 nm emission lines). 
The most useful earlier work is that of \citet{ode08a} who imaged the system in the HST-ACS  \oiii\ F502N, F658N, and \nii\ F660N filters. These images determined that the jet was extended and knotty and the dominant feature (called s2 by Bally et al. 2000) was rapidly moving, with a dynamical age of about 120 years.

\subsubsection{The Structure of the Proplyd and Its Jets}
\label{sec:HH510proplyd}

\begin{figure}
\epsscale{0.50}
\plotone{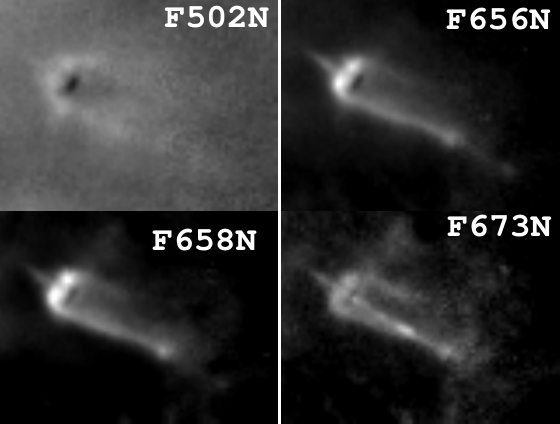}
\caption{Four 3.20\arcsec $\times$ 2.44\arcsec\ images are presented, each with the vertical axis pointed toward position angle  14\arcdeg.}
\label{fig:fig29}
\end{figure}

Our new high quality WFC3 images have better imaging sampling due to the 0.04\arcsec\ pixels and better isolation of emission lines than the earlier studies. Through the availability of AstroDrizzle images of the earlier WFPC2 images there is the possibility of obtaining improved measurements of tangential motions and changes in the structure. Figure~\ref{fig:fig29} shows a mosaic of four monochromatic images. The F502N, F656N, and F658N images show best the nearly edge-on optically thick circumstellar disk of about 0.3\arcsec\ width and 0.1\arcsec\ thickness. Examination of our F547M continuum filter image shows glow of scattered starlight on the west side of the disk, indicating that the near side of the disk is tilted slightly to the east. In this sense it is like the largest of the pure silhouette proplyds 114-426 \citep{mcc96,shu03} in that the central star is revealed only by scattered light. 

\begin{figure}
\epsscale{0.5 }
\plotone{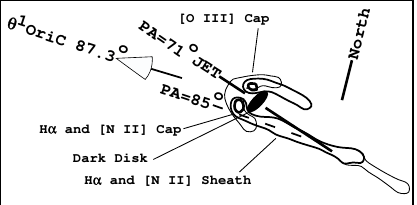}
\caption{This drawing is based on the images shown in Figure~\ref{fig:fig29} and illustrates the primary features of the proplyd d109.4-326.7 that drives the HH~510 outflows.}
\label{fig:fig30}
\end{figure}

\begin{figure}
\epsscale{1.0}
\plotone{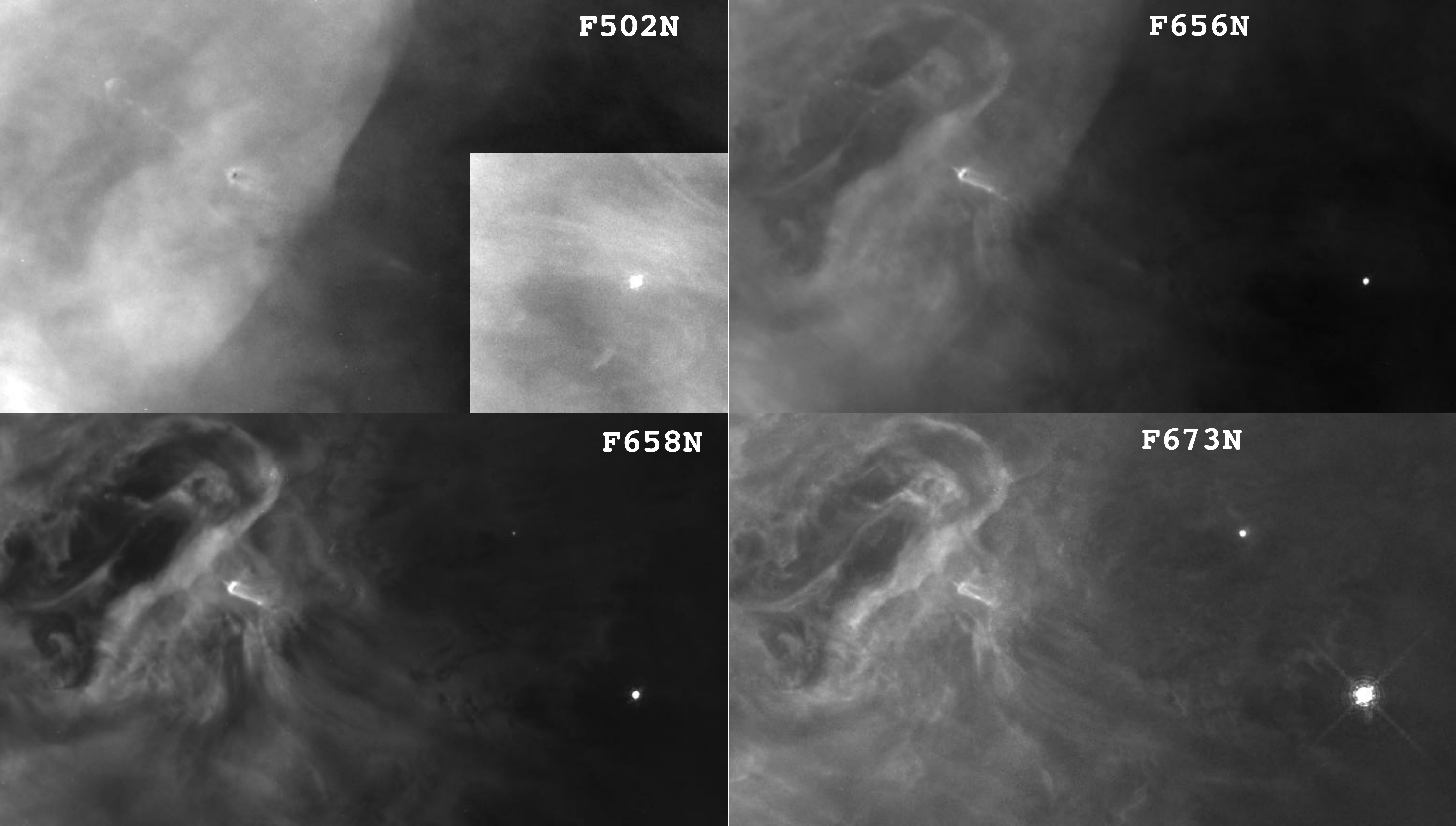}
\caption{Four 31.1\arcsec $\times$ 17.6\arcsec\ monochromatic emission line images  of the the vicinity of proplyd d109.4-326.7. The proplyd is at the left center in each and the large feature to the left in each is the optical feature HH~625 discussed in \ref{sec:HH625}.
 The vertical axis is pointed toward position angle  14\arcdeg. The 11.0\arcsec\ square inset in the lower right of the F502N image is simply a rescaling of the display range in this area. It is contiguous with the remainder of that image.}
\label{fig:fig31}
\end{figure}

\begin{figure}
\epsscale{1.0 }
\plotone{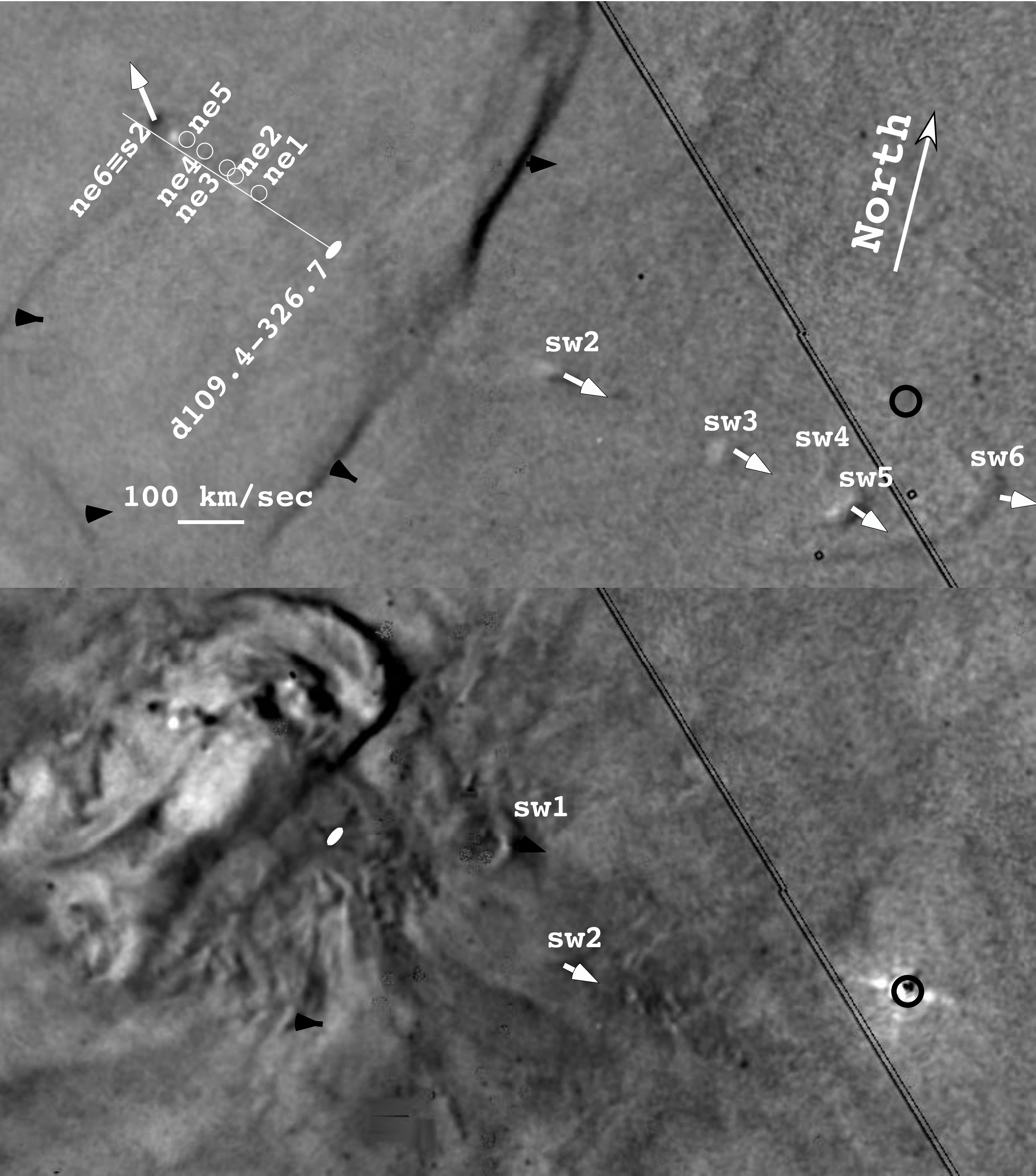}
\caption{The same field of view as in Figure~\ref{fig:fig31} is shown as the ratio of first epoch (GO~5469) over second epoch (GO~12543) images in the two filters capturing the high ionization (\oiii\ with F502N) and low ionization (\nii\ with F658N) emission from the nebula.The added details are explained in Section~\ref{sec:HH510Shocks}. The thin white line indicates the direction ( 71\arcdeg) of the microjet extending east from d109.4-326.7. The star 102.7-316.4 has been indicated by a heavy white circle in the lower-right.}
\label{fig:fig32}
\end{figure}

\begin{figure}
\epsscale{0.6}
\plotone{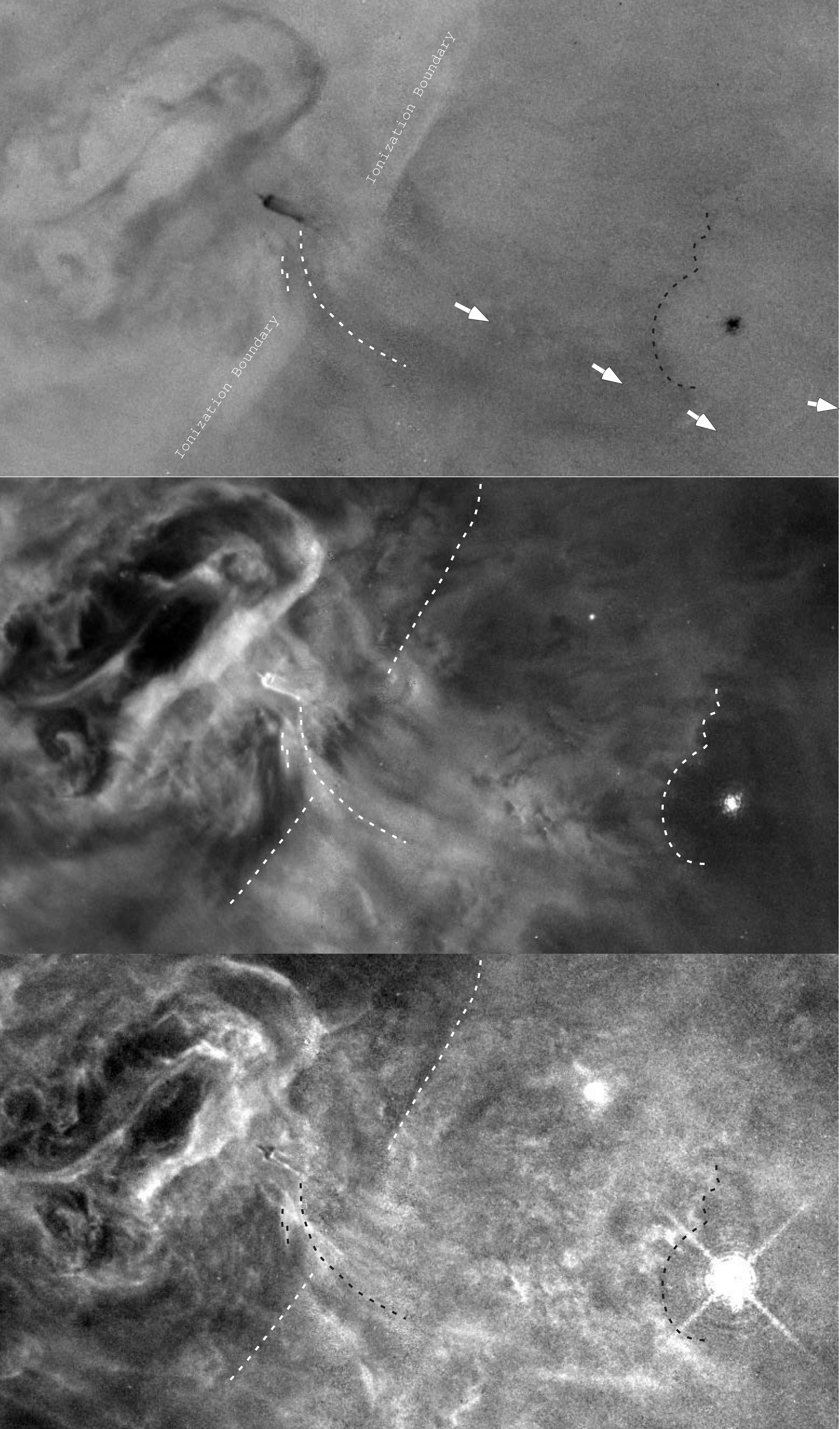}
\caption{The same field of view as in Figure~\ref{fig:fig31} is shown as the ratio of the extinction insensitive ratios of F502N/F487N (top)
F658N/F656N (middle), and F673N/F656N. The rapidly moving \oiii\ features sw2, sw3, sw4, and sw5 are shown as arrows. The features marked ``Ionization Boundary'' in the top panel is shown as a dashed white line in the lower two panels. An apparent ionization boundary east of star 102.7-316.4 is outlined as a dashed white line in the top two panels and a dashed black line in the lowest panel. An open partial shock lying to the west of the  proplyd d109.4-326.7 is  shown as dashed lines (white in the top two panels, black in the lowest panel).}
\label{fig:fig33}
\end{figure}

\newpage

We note that the \oiii\ emission is found on the outer part of the photoionized cap on the side of the disk facing the ionizing star \tc, which lies in the direction  87.3\arcdeg. The location of the other emission lines are unresolvable from one another, but are certainly closer to the disk than the \oiii\ emission. There is a gap in the ionized cap emission where the east jet passes through it. The \sii\ (F673N) images reveal the west pointing counter microjet. Both jets are perpendicular to the plane of the dark disk. 

The primary shape of the tail of a photoionized proplyd is determined by photoionization of material on the side of the proplyd not facing the ionizing star, i. e. the shadowed region. Material there has recently been expelled from the photoionized cap and accelerated away from \tc\ or was expelled from the molecular disk as part of the large-angle flow from the disk. When the density is high an ionization front is formed at the outer edges of the shadowed region and the sides of the tails are parallel, as in HH~510. When the density in the shadowed region is low, an ionization front is formed that is in equilibrium with the diffuse Lyman Continuum (LyC) produced by the surrounding nebula and the tail shape converges to a point. This shape is often called that of a tadpole. 

The boundary of the shadowed zone, which in the case of HH~510 is an ionized sheath,  should point back to the ionizing star, which it does, as shown in the drawing that summarizes the primary features of this proplyd (Figure ~\ref{fig:fig30}). The difference in direction of the ionization boundary (85\arcdeg $\pm$5\arcdeg) and jet (71\arcdeg $\pm$2\arcdeg) means that the east jet passes through the boundary and one sees an enhancement of the boundary there and that this boundary curves outward along the track of the jet.  

The west microjet is only seen in \sii\ emission because it lies within the ionization shadow of the bright cap. The ionization structure of the tail indicates that the tail is ionization bounded. Any illumination within the tail would arise from LyC  photons generated as the surrounding photoionized hydrogen recombines. The physics of such objects have been addressed by \citet{can98}.  The collisionally excited \sii\ doublet emission that we see in the F673N filter requires two major components, singly ionized sulphur and electrons having a few electron volts energy. Singly ionized sulphur is formed by 10.4 eV and higher energy photons produced by scattered light of the nebula and its main ionization front, which freely pass through the tail's ionization boundary. Photons of at least 23.3 eV are necessary to cause ionization to S$^{++}$ and won't pass the layer of neutral helium that must lie outside of the tail's ionization boundary. This means that S$^{+}$ is the preferred state of the jet within the shadowed region. If the jet material has a similar velocity to that of the shocks that it drives (about 100 \kms ), then electrons within the jet have energies of about 52 eV, sufficient to ionize neutral sulphur and to excite the levels producing the observed \sii\ doublet. 

\subsubsection{Shocks Driven by the HH~510 Proplyd}
\label{sec:HH510Shocks}

The HH~510 features are shown in Figure~\ref{fig:fig31}. The previously known east microjet arising from proplyd d109.4-326.7
 is seen in \oiii\ to be a series of clumps starting at about 2.4\arcsec\ from the proplyd and ending with a bow-shock feature of about 0.8\arcsec\ width at 6.6\arcsec\ from the proplyd. The knotty outer portions of the jet are most easily seen in \oiii\ and \Ha, being weak or absent in \nii\ and \sii. The shock is clearly seen in all filters except \sii\ and its faint appearance in the F673N may be due to the larger width of that filter, which admits much more continuum than the other emission line filters. 

The newly discovered west microjet is not traced beyond where the \sii\ jet intersects and distorts the west ionization boundary of the tail. 
Only a few hints of structure are present, once in \oiii\ and once in \nii. These features are described below in the discussion of the motions. 

We present in Figure~\ref{fig:fig32} the results of two epoch ratio images in F502N and F658N. We have drawn in a bright ellipse over the position of the dark disk of d109.4-326.7. This overlying feature is twice the dimension of the true disk. Arrows represent the velocities measured either by the ZSQ method or from direct measurement of the features.  One sees clearly the effects of the collimated jet interacting with ambient material along the projection of both the east and west jets. The direction of the jet's flow ( 71\arcdeg) is shown as a light white line in Figure~\ref{fig:fig32}.
 In the upper (F502N) panel we have added the positions of ten knots and and in the lower (F658N) image we have added two features, one of which (sw2) is common to both filters.  \citep{ode08a}. 
 The tangential velocities are presented in Table~\ref{tab:HH510}. Using the tangential angular velocity ($\mu$) and the distance of the feature from the proplyd ($\theta$) we have derived a characteristic age (T) from the relation t=$\theta$/$\mu$. It appears that increases in the jet outflows have been occurring at intervals of about 150 years. Such episodic outflows are not unprecedented \citep{rei86} and have been recently discussed by \citet{rag13} in the context of HH~34.  
 
 We do not have \vt\ and \vrad\ values in the same ions for either of the series of shocks associated with HH~510.
 However, for \oiii\ the \vt\ for the ne6 feature is 96 \kms\ and the average \vrad\ in \nii\ for the other ne shocks is -6 \kms.
 Within the assumption that these velocities are related, then shock ne is moving at  \Vomc\ = 101 \kms with $\theta$ = 18\arcdeg.

It is likely that the feature sw1, which is seen only in \nii\ is unrelated to the shocks driven by d109.4-326.7's jet. This is because the feature does not fall along the west jet's axis, it is clearly low ionization (whereas the other features are high ionization), and the tangential velocity is much lower than the other HH~510 features. 

\begin{deluxetable}{lcccc}
\tabletypesize{\scriptsize}
\tablecaption{Tangential Motions of Objects near HH~510\label{tab:HH510info}}
\label{tab:HH510}
\tablewidth{0pt}
\tablehead{
\colhead{Object Name} &
\colhead{$\mu$ (\arcsec / yr)} &
\colhead{Tangential Velocity (\kms)} &
\colhead{$\theta$ (\arcsec)} &
\colhead{Age (yrs)}}
\startdata
ne2 & 0.047 & 97 & 6.56 & 140\\
sw1* & 0.013 & 26 & 5.54 & 440\\
sw2 & 0.032 & 67 & 7.87 & 250\\
sw3 & 0.033 & 68 & 13.44 & 410\\
sw5 & 0.031 & 65 & 17.38 & 560\\
sw6 & 0.029 & 59 & 21.2 & 750\\
\enddata
\tablecomments{~*Object sw1 is unlikely to be associated with HH~510 in spite of its proximity to 
proplyd 109.8-325.3.}
\end{deluxetable}

\subsubsection{The 3-D Location of HH~510}
\label{sec:HH510location} 

The visibility of HH~510 is made difficult because of the presence of other structures. However, we can use them to determine their relative positions along a line of sight into this part of the \hr. There is no indication of interaction of the east jet and the associated shocks with HH~625. HH~510 must lie in the foreground relative to that object.

About 3\arcsec\ west of d109.4-326.7 the jet crosses a long feature  shown in each of the panels of Figure~\ref{fig:fig33} and most visible in the F502N panel (top) of Figure~\ref{fig:fig32}, which shows its motion. 
The ratio images show that the long feature is an ionization boundary, with the ionization preceding from \oiii\ to \nii\ to \sii\ going from east to west. 

This boundary is the northwest portion of a large area of high ionization northwest of the Trapezium grouping of massive young stars. It resembles several large parabolas  pointing northwest with centers displaced about 30\arcsec\  and extending back to the Orion-S region. The largest and easternmost parabola points toward the HH~202 shock complex and the shortest and westernmost is about the same orientation as HH~625.  They both are moving to the northwest \citep{ode08a} and their axes are separated by about 30\arcsec. These large structures were discussed in a study of the 3-D structure of the inner Orion Nebula by \citet{ode09}. 
It is likely that the easternmost parabola is associated with the HH~202 shocks and the westernmost parabola is associated with HH~625.

The west jet from d109.4-326.7 passes through a break in this ionization front. This argues that the break is caused by the jet and that d109.4-326.7 lies within the high ionization central cavity of the nebula. 
A caveat to drawing that conclusion is that we see a half shock in all the emission line filters. When the image of this half shock is flipped on itself, the axis of this large shock is the same as the d109.4-326.7 jet and shocks. The mirror image of the half shock indicates that the full width of the parabola is about the same size as the gap in the ionization boundary feature, suggesting that this large shock is the cause of the interruption, rather than the west jet. Of course there may be a physical relation of the west jet and this half shock feature.

It is of interest that near the star 102.7-316.4 there is an incomplete circle of high ionization and immediately outside that circle there are multiple irregular features seen in \nii\ and \sii. Certainly this low mass and low temperature star is not the source of ionization that produces the \oiii\ emission. However, its large-angle stellar wind could have excavated a local region that has allowed photoionizing radiation to enter.

The high ionization seen in both the east and west outflows from d109.4-326.7 may be due to collisional effects, rather than photoionization.
The new shock is moving at about 100 \kms, which corresponds to 52 eV of energy,
plenty to doubly ionize oxygen (removing the second electron takes 35.1 eV photons.

\subsection{HH~518}
\label{sec:HH518}

Various features have been assigned to HH~518  since its first designation in \citet{bal00}, where it had only one feature, a single shock at 162-342. \citet{ode08a} argued that it was composed of the shocks
 163-342, 165-334, 172-327 and 177-322. Our superior determination of the motions (seen only in F502N) shows that there are actually two separable flows. We have designated them as HH~518a (162-342, 165-335, 164-333 and 165-333) and HH~518b (164-343 and 167-337), as shown in Figure~\ref{fig:fig19}. 
 
 Further west along a slightly CW curving of the axes of HH~518a and HH~518b we see additional large moving shocks that are probably part of these flows. In Figure~\ref{fig:fig22} we see shocks A (170-325) and  B (172-327) WSW of \tc\ and shocks C (177-320), D (178-318), E (178-321) and F (180-325) WSW of $\theta^1$~Ori~D. Inclusion of these shocks makes the HH~518 flows among the largest in the \hr.
 
 The average axis of these two flows passes through the bright variable V2279 Ori, which is probably the source of the optically invisible jet driving both HH~518a and HH~518b. There are no shocks identifiable with material streaming in the opposite direction from V2279 Ori.
 
 Doi's (2004) velocity mapping showed that the shocks identified with HH~518 are  redshifted. This is confirmed in our analysis.
 These flows are among the few in the \hr\ that show positive values of \vrad, indicating motion towards the OMC. For shock 164-343 \vrad~=~+95 \kms\ and \vt~= 13 \kms. These yield \Vomc~=~70 \kms\ and $\theta$~=~-79\arcdeg. There are two high positive velocity shocks (+102 \kms\ and +122 \kms) to the EWE from 164-343 that are probably associated with the material driving HH~518a and HH~518b.

 \subsection{HH~530 Redefined}
\label{sec:HH530redefined} 

\begin{figure}
\epsscale{0.6}
\plotone{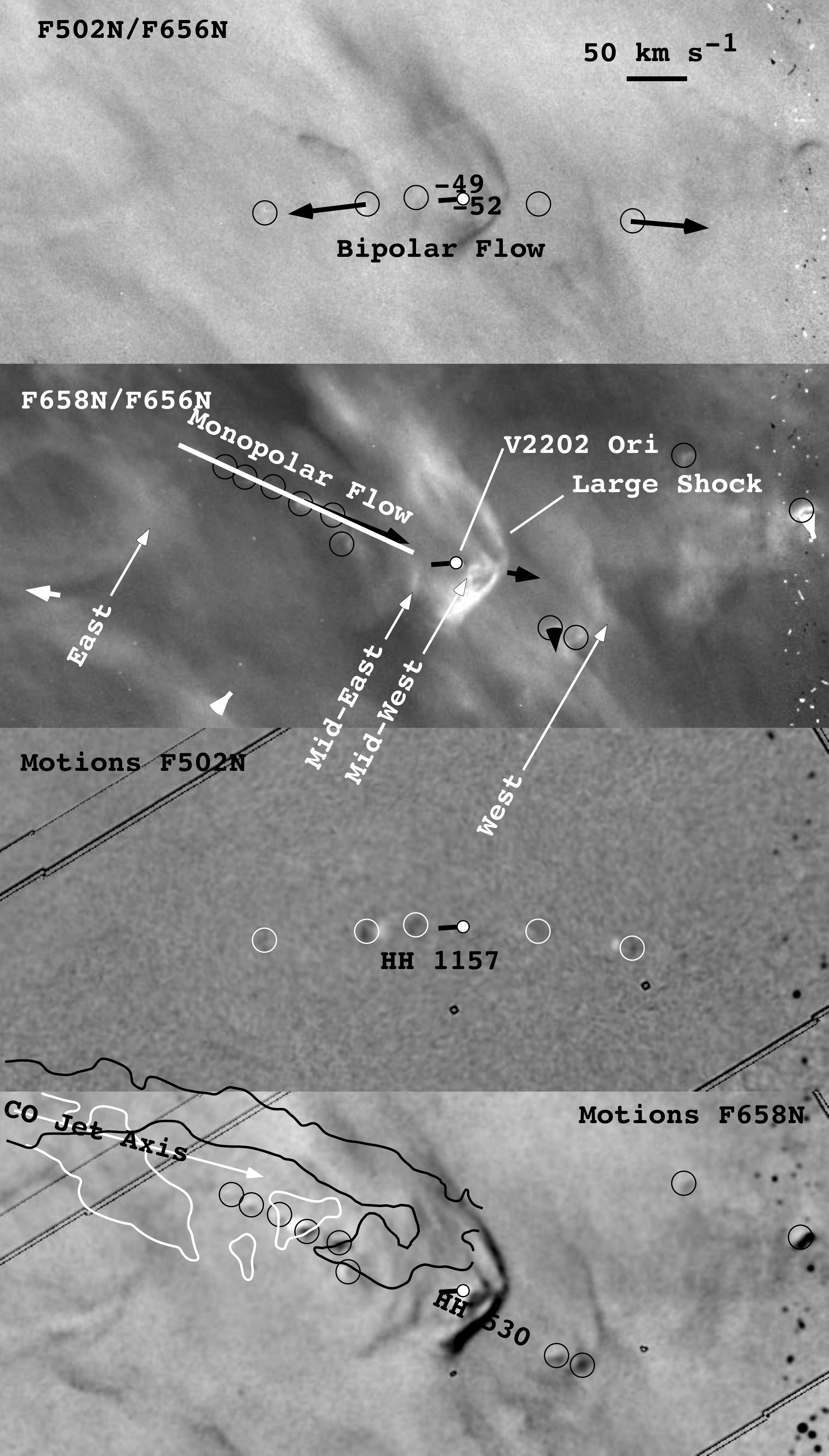}
\caption{This figure of four identical FOV  27.5\arcsec $\times$12.1\arcsec\ images centered on HH~530 show both ratio images (top two panels) and motions (lowest two panels.The FOV is shown in Figure~\ref{fig:fig2}.
The open circles indicate the position of components of the two flow systems (HH~530 and HH~1157), some of which could not be measured quantitatively. The only star in this FOV is V2202 Ori, the apparent source of the HH~1157 Bipolar flow. The numbers indicate the \oiii\ \vrad\ values. These lie at the extreme west boundary of the spectral Atlas. The irregular lines in the lowest panel represent the outer contours of blueshifted (white, 20.1--24.0 \kms) and redshifted (black, 28.6--32.4 \kms) CO emission reported by \citet{bal15}. The white arrow designates the direction (PA = 269\arcdeg) of the highest intensity portions of this molecular jet.} 
\label{fig:fig34}
\end{figure}

\placefigure{fig:fig34}

The object HH~530 has been mentioned in multiple studies since its discovery by \citet{bal00}. As shown in Figure~\ref{fig:fig34} the region is complex and in fact it must be
a composite of unrelated features, as first suggested by \citet{ode03a}. The dominant feature is a large, low ionization bowshock facing west with numerous nearby small moving features. We designate the large low ionization shock as
the Large Shock in Figure~\ref{fig:fig34} and its axis of symmetry is about  281\arcdeg.  The conditions in this region are illuminated and somewhat confounded by recent CO studies with ALMA \citep{bal15}.

\subsubsection{HH~1157}
\label{sec:HH1157}

In the F502N ratio and motion images (Figure~\ref{fig:fig34}) we see that there are a series of high ionization
shocks extending from both sides of V2202 Ori. We designate these as the Bipolar Flow feature. This flow is slightly curved and points towards  108\arcdeg\ to the east and  276\arcdeg\ to the west. A close examination of our images of V2202 Ori in the F502N and F656N filters reveals an unresolved width microjet about 0.2\arcsec\ 
long pointed to about  108\arcdeg\ (with an uncertainty of about $\pm$10\arcdeg. Within the uncertainty of this angle, it must be the driving source for the east moving series of shocks. We must be seeing shocks formed by an episodic bipolar flow that is curved due to the southward moving ambient high ionization gas. This bipolar flow is designated as HH 1157. The spectral Atlas coverage ends a few seconds of arc west of V2202 Ori, but, it does show high radial velocities (-49 \kms\ and -52 \kms) in \oiii\ coincident with that star and they  probably arise from its microjet.

The similarity of the Large Shock's orientation of about 281\arcdeg\ and the direction of the western HH~1157 flow (276\arcdeg) is likely to be coincidental. There is a gap in the well-defined Large~Shock where the west moving high ionization flow from V2202 crosses it in the plane of the sky.  This argues that this aligned high ionization HH 1157 flow is penetrating and disrupting the large low ionization Large~Shock. They are probably unrelated and the Large Shock is driven by another collimated outflow.



\subsubsection{HH~530}
\label{sec:HH530}

In the F658N ratio and motion images we see a series of low ionization shocks moving towards  259\arcdeg. We designate these as the Monopolar Flow feature.  Although it nearly aligns with V2202 Ori, it is moving towards that star. These westward moving shocks and V2202 must be unrelated and the origin  of the Monopolar Flow feature must lie to the east. If one includes the two low ionization shocks to the west of the Large Shock as part of the Monopolar Flow, it is seen that clearly the flow passes south of V2202.

Within the arc of the Large~Shock is a low ionization moving feature designated as Mid-West. This feature is near where one sees similar features in HH~269-West and HH~625, but not quite at the same position as this shock lies below the symmetry axis of the Large~Shock. It is close to the axis of the Monopolar Flow, but slightly below. The small low ionization moving knot designated as Mid-East falls below the projection of the Monopolar Flow but otherwise has the same characteristics of the Monopolar Flow. There are two low ionization knots lying just below a projection of the Monopolar Flow and they lie immediately to the east of the larger West shock. It appears that the Monopolar Flow curves slightly south and may produce the larger, low ionization West shock.

With the identification of the bipolar flow from V2202 as HH 1157, HH 530 can be said to now be composed of the Monopolar Flow, the Large Shock, the shock we label as West, and two small shocks (Mid-East and Mid-West). However, this grouping into one flow (HH 530) is probably too simplistic.

\subsubsubsection{HH~530 is Complex}

In a recent study of the nearby silhouette proplyd 114-426 , \citet{bal15} discovered CO emission velocity shifted with respect to the OMC. The west end of these features is shown in Figure~\ref{fig:fig34} and the east in Figure~\ref{fig:fig42}. The linear blueshifted CO axis is oriented PA=269\arcdeg, with a spur on the east end (Section~\ref{sec:HH1148}) and another pointing SE on the west end. The blue shifted axis orientation is 10\arcdeg\ larger than the Monopolar flow axis, but, they may be linked 
if the CO molecular flow is redirected in the plane of the sky as it passes through the MIF.

Because of their proximity, it is tempting to associate the redshifted and blueshifted CO flows. However, a link is difficult to understand unless a bipolar stellar outflow has suddenly reversed direction or the the elongated redshifted component is only a portion of an expanding envelope surrounding the blueshifted linear flow, which seems unlikely.

The redshifted flow has the fish-hook feature of an incomplete bowshock, with its apex within the optical Large Shock feature. 
However, the apex of the CO feature lies NE of the apex of the Large Shock. The average of its flow is 272\arcdeg, different from the 281\arcdeg\ symmetry axis of the Large Shock. A link of the Large Shock to either the redshifted or blue shifted components could be established by determination of the Large Shock's radial velocity. Unfortunately, the data base of velocities that we used  ends just east of the Large Shock.

\subsubsubsection{The Origin of the HH~530 Complex}




The reciprocal of the several directions associated with the HH~530 features all point in the direction of a concentration of imbedded young stars and known molecular outflows called the Ori-S6 region, whose most prominent source is EC 14 \citep{schm90,zap10}. EC 14 lies 49\arcsec\ at 82\arcdeg\ from the apex of the Large Shock.
82\arcdeg\ is similar to the reciprocal  of the HH~530 features (Large Shock symmetry axis, 101\arcdeg; Monopolar Flow, 79\arcdeg; CO blueshifted jet, 89\arcdeg).  If one begins the search for a source or sources at the east end of the CO jet, there are no candidate stars until reaching the region of Ori-S6, then there are many (including COUP 496,132-413, EC 12, H20097, EC 113, EC 14,  COUP 582, COUP 615). EC 14 is highly unlikely to be the source because it begins its outflow at 225\arcdeg\ before being deflected towards 209\arcdeg.

\subsection{HH~626}
\label{sec:HH626}

\begin{figure}
\plotone{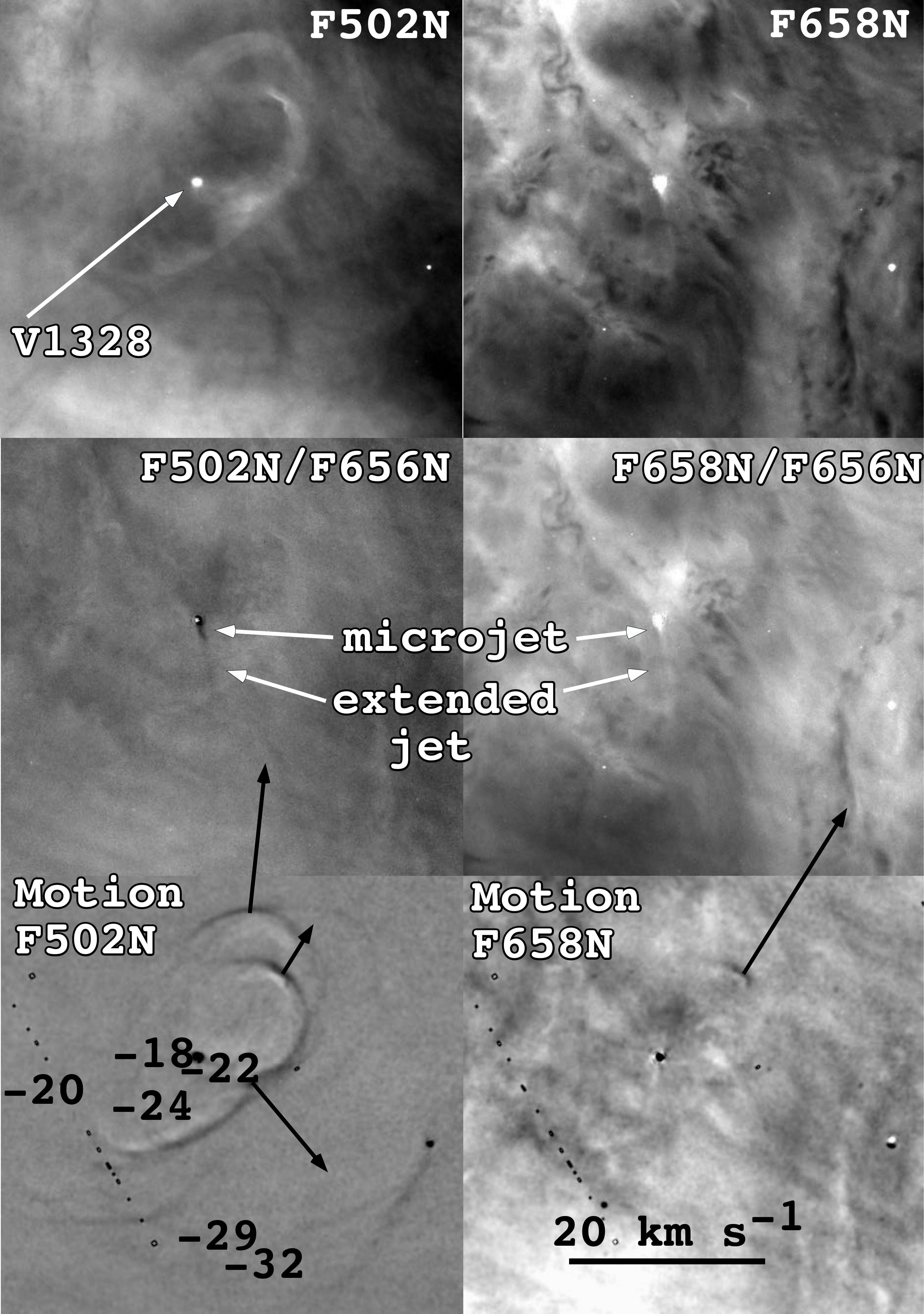}
\epsscale{0.8}
\caption{This mosaic is composed of 18.9\arcsec$\times$17.9\arcsec\ images of the components of HH~626. This FOV is labeled HH626 in Figure~\ref{fig:fig2}. The central star is V1328. The microjet, the extended jet, and the measured velocity vectors are shown. \vrad\ values in \oiii\ are inserted. }
\label{fig:fig35}
\end{figure}

\placefigure{fig:fig35}

The central star of this outflow is 
V1328 Ori (142.8-424.6). It was first detected outside the optical region in the radio study of \citet{fel93} and has now been observed through the infrared and into X-ray wavelengths. \citet{ode96} noted it as a a round proplyd with a tail, reporting that its size was 0.6\arcsec$\times$0.73\arcsec. Examination of our new images show that the star is embedded in a concentrated core of emission and that after deconvolution is 2.8$\pm$0.3 WFC3 pixels, or 0.1$\pm$0.01\arcsec\ Full Width Half Maximum. There is a low ionization microjet traceable to 0.88\arcsec\ from the star in the direction  224\arcdeg.  This is probably what led \citet{ode96} to conclude that the proplyd was elongated.  There is a much fainter linear feature at  214\arcdeg\ that extends to 3.0\arcsec\ from the star and is labeled the extended jet. V1328 Ori lies at  208\arcdeg\ with respect to \tc. This means that the microjet and the linear feature are not inline or in the direction expected for the tail of a proplyd and the features are part of a collimated outflow and not simply the tail formed by gas being accelerated away from V1328.

HH~626 was initially reported by \citet{ode03a} as an incomplete double ellipse of high ionization, being visible only in F502N and F656N images. Along a line with  149\arcdeg\ there are two shocks (as shown in Figure~\ref{fig:fig35}. The southeast component is 5.0\arcsec\ from the central star and a blunter shock is in the northwest direction at 4.8\arcdeg.
F658N emission is seen clearly only in the northwest blunt shock. None of these features align with the microjet and its possible extension. 

There are a number of additional shocks probably belonging to HH~626. At  354\arcdeg and 6.0\arcsec\ there is a bright high ionization shock and there are several fainter shocks to the west through southeast.  The dynamical age of the three measured components (from nearest to farthest) is 350 yrs, 800 yrs, and 1700 yrs.

As noted by \citet{ode03a}, HH~626 appears to be shocks associated with either a large-angle wind from the proplyd or photo evaporative flow from a disk of gas and dust close to the central star and unresolved in our images. The multiplicity of the shocks indicates that they might be formed by different flows in nearly orthogonal directions. The star is probably located within the high ionization, low density gas that exists between the MIF and the foreground Veil.  The lower brightness of the features in the northeast probably indicates that the slow wind from the star is moving into lower density ambient gas in that direction.

Only in the case of the feature 2\arcsec\ west of V1328 Ori are there both radial and tangential velocities. Using \vt=12 \kms\ and \vrad=-22 \kms\ gives \Vomc=49 \kms\ and $\theta$=66\arcdeg.

\subsection{HH~998}
\label{sec:HH998}

\citet{ode08a} designated the shock 158-355 as HH~998. This well defined shock is more visible in F502N, but is clear in both the F502N and F658N motion images (Figure~\ref{fig:fig19} and Figure~\ref{fig:fig20}). Its symmetry axis is towards  244\arcdeg\ and moves along that axis. There is a fainter shock of the same  form and direction of motion about 0.6\arcsec\ outside of it. Much further out (8.0\arcsec) there is the large and less well defined shock (153-359) that aligns suggestively near the axis of the HH~998 shock and may be the result of an earlier outflow from a common source. 

\citet{ode08a} discovered a microjet extending 0.4\arcsec\ from AC Ori at  47\arcdeg\ and could see no shocks identifiable with it in that direction. The microjet is shown as a short dark line in Figures~\ref{fig:fig19} and \ref{fig:fig20}.  

HH~998 lies at 4.1\arcsec\ at  230\arcdeg\ from AC Ori and  \citet{ode08a} argued that it was produced by a counter-jet ( 227\arcdeg) from that star. However, those directions do not agree with the symmetry axis of HH~998 ( 244\arcdeg). This calls into question the conclusion that AC Ori is the source of HH~998. There are two other possible sources, COUP~769 and and a currently undetected source (Blank-East). 
 
The highly obscured star COUP 769 is the first alternative source. The HH~998 shock lies 3.9\arcsec\ at  254\arcdeg\ from it. This means that neither AC Ori or COUP 769 lie exactly along the symmetry line of the HH~998 
shock, with the difference being +14\arcdeg\ for AC Ori and -10\arcdeg\ for COUP 769. 
No clear association is possible, with the known microjet in AC Ori favoring that star while the closer PA alignment favors COUP 769. 

If the driving source of the HH~998 shock is AC Ori, then the shock arose 330 years ago. If the driving source is COUP 769, then the shock arose 355 years ago.
However, it is most likely that the source lies in the Blank-East region. In this case the age of the HH~998 shock would be even shorter.

We have both radial and tangential velocities for only one of these features. The derived 3-D motion for 155-354 (\vrad=-53 \kms, \vt=4.6 \kms) is \Vomc=79 \kms, $\theta$=87\arcdeg,

\subsection{HH~1127}
\label{sec:1127}

\begin{figure}
\epsscale{0.8}
\plotone{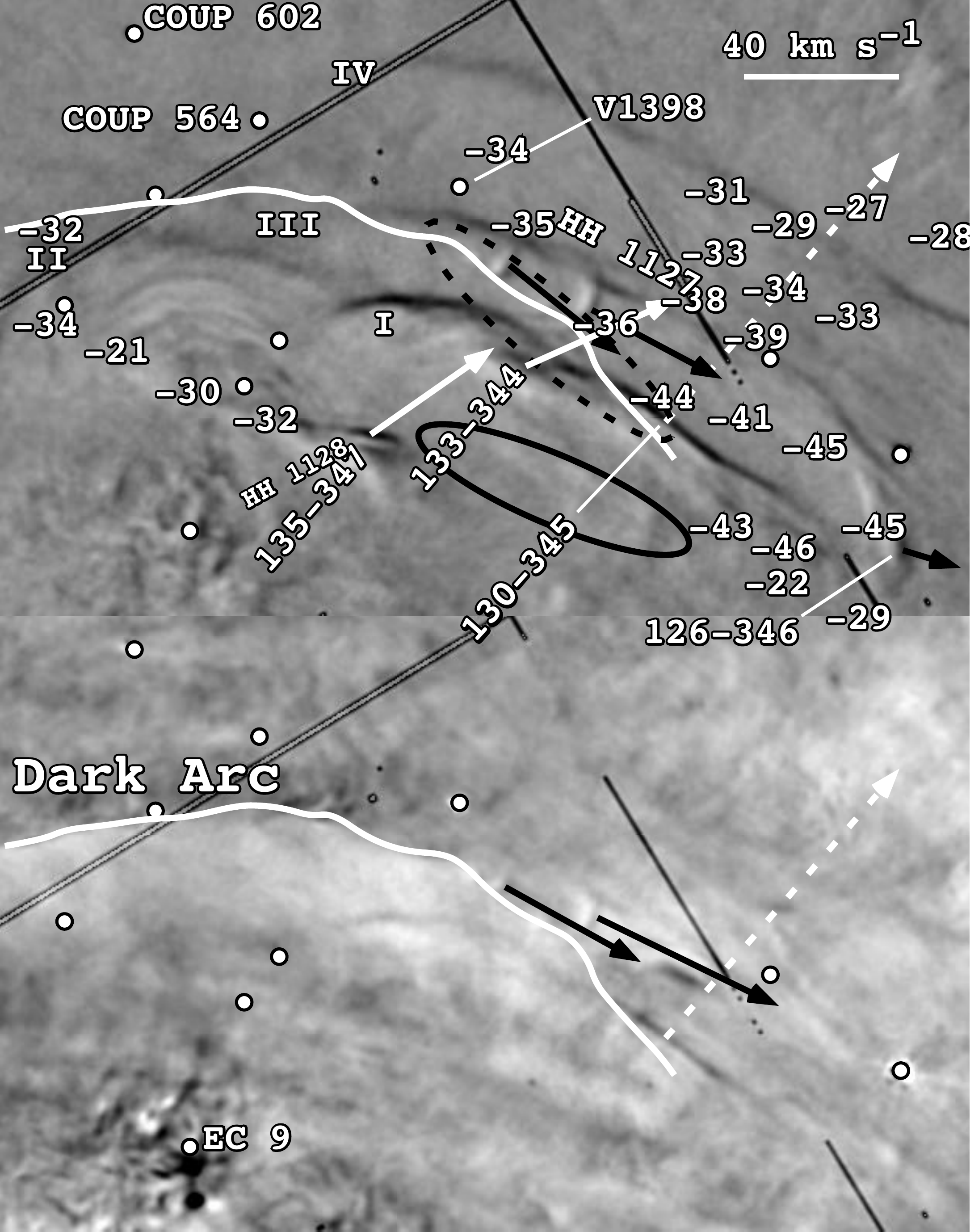}
\caption{This pair of  25.0\arcsec $\times$ 15.9\arcsec\ motion images centered on the northwest corner of Figure~\ref{fig:fig10} shows the tangential velocities measured in this study for the parabolic arcs  and the knot within HH~1127 (black arrows), the average (as a dashed white line) of the motions in \Ha\ and \nii\ measured by \citet{ode03a} for the feature they call 130-345, and the newly measured motions for features 133-344 and 135-347, which are not part of HH~1127. The velocity vector for 126-346 in the F502N (upper) panel is at 40\%\  the scale of the other vectors. The irregular curved white line delineates the sharp north boundary of the Dark Arc feature.The ellipses  and Roman numerals designate objects discussed in Section~\ref{sec:1127} and Section~\ref{sec:MotionsDarkArc}.  The negative numbers indicate the high velocity \oiii\ features. The white lines without arrowheads indicate the names of features in crowded fields.}
\label{fig:fig36}
\end{figure}

\placefigure{fig:fig36}

\begin{figure}
\epsscale{1.0}
\plotone{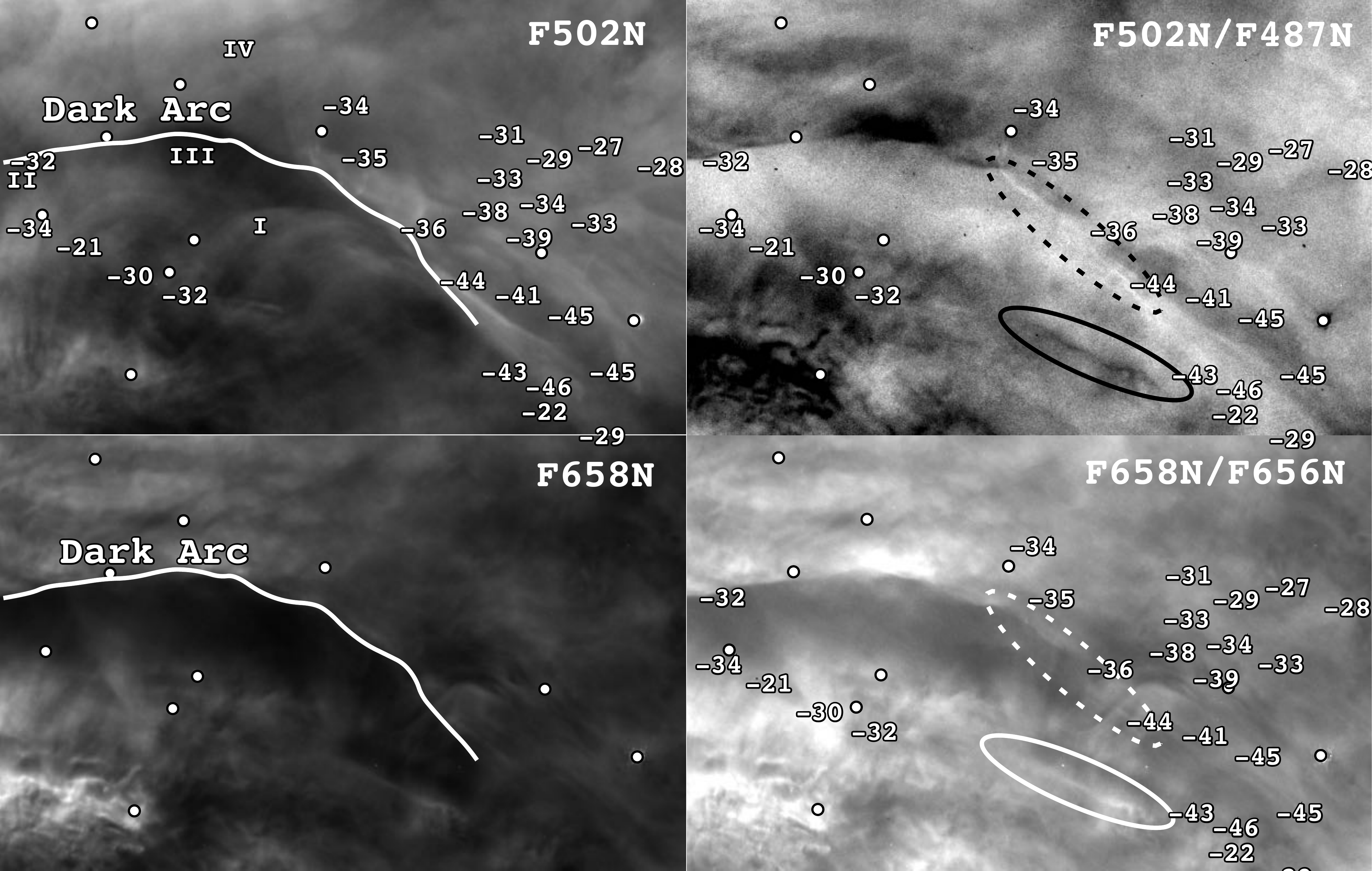}
\caption{The same FOV as Figure~\ref{fig:fig36} is shown in single filter and ratio images.
The boundary of the Dark Arc is highlighted by a wavy line. The numbers in the panels indicate high velocity  \oiii\ features.}
\label{fig:fig37}
\end{figure}

\placefigure{fig:fig37}

The isolated flows near the Dark Arc first seen by \citet{ode03a} are now shown clearly in our motion images and designated as HH1127.  The motion and emission line images of this object are shown in Figure~\ref{fig:fig36} and Figure~\ref{fig:fig37}.

The HH~1127 system is defined by two parabolic shocks of increasing size aligned and moving towards  255\arcdeg\ and three small shocks at  252\arcdeg. As indicated previously \citep{ode03a} they head away from star MAX 46, 
which is seen only in infrared wavelengths. However, their alignment is equally good with the more distant star COUP 602, which is seen in X-rays, optical, and the infrared. We can rule out an association with the nearest star (V1398) because the velocity vectors and the alignments pass south of that star. 
There are no nearby candidate stars beyond COUP 602.
The closer shock has a kinematic age of 400 yrs and the further 700 yrs if the source is MAX 46 and 540 yrs and 640 yrs respectively if the source is actually COUP 602.

Only the westernmost shock of HH~1127 may have an associated radial velocity (the \oiii\ \vrad = -36 \kms, \oiii\ \vt=38 \kms). The derived spatial motions is \Vomc=73 \kms\ and $\theta$=58\arcdeg. We note that there are similar unrelated  \oiii\ radial velocity features of about the same value in the vicinity of the HH~1127 shock. It may be that we cannot actually derive a \Vomc\ for HH~1127.

There are no shocks that clearly lie westward along the axis of the HH~1127 flow.
It is not clear if the high-ionization knot (126-346) shown at the bottom-right of the F502N panel (top) of Figure~\ref{fig:fig36} is part of the 
HH~1127 system. Its \vt\  (39 \kms) is comparable to that of the two parabolic shocks (average 37 \kms). Its direction of motion ( 267\arcdeg) is slightly CCW from the direction of motion of the HH~1127 shocks ( 255\arcdeg) and its position direction is slightly CW ( 249\arcdeg) with respect to the HH~1127 shocks. It may be part of the HH~1127 system, but this is unlikely.

There is a narrow, strong, irregular, nearly linear \oiii\ feature within the dashed ellipse in the top right panel of Figure~\ref{fig:fig37}. The orientation is towards  240\arcdeg\ and it extends for 7.7\arcsec. The position and orientation indicates that it is not a jet associated with
the HH~1127 flow. 
Its NE end is coincident with a linear F658N feature and no F658N feature is seen on the SW end. In F502N it lies along the edge of the Dark Arc feature.
If it is an edge-on ionization front, it is one of the most narrow that has been found.

The many \oiii\ radial velocity features in the west side of Figure~\ref{fig:fig37} show no obvious correlation with features seen in the motions images (Figure~\ref{fig:fig36}), with the exception of one feature possibly associated with the westernmost HH~1127 shock. One group with an average of \vrad=-33$\pm$4 \kms\ lies in the NNW area, another five features with an average of \vrad = -44$\pm$2 \kms\ lie in the region between HH~1127 and shock 126-346, and two other velocity features (average \vrad =26 \kms) are found south of shock 126-346. Obviously some types of medium-scale outflows are occurring in this area, but it is not clear what they are.

\subsection{Motions Within and Near the Dark Arc}
\label{sec:MotionsDarkArc}

There are important structures within and near the Dark Arc. One series of shocks has been reported before and the remainder were discovered in our new images.

We see in the F502N motion image (Figure~\ref{fig:fig36}, top panel) and the F502N image (Figure~\ref{fig:fig37}, top left panel) that there is a series of three large high ionization incomplete parabolic shocks directed approximately towards  68\arcdeg. The westernmost shock is the most visible of the three in Figure~\ref{fig:fig36} and is labeled I. 
The NW boundary of the westernmost shock is about 
parallel to the HH~1127 flow (there is probably no association) and about 1.6\arcsec\ south.  These shocks were initially reported by \citet{ode03a} on the images then available, where they reported the orientation as  85\arcdeg\ and they measured one boundary, calling it 130-345.

We also see in the F502N motion image (Figure~\ref{fig:fig36}, top panel) and the F502N image (Figure~\ref{fig:fig37}, top left panel) three other even larger high ionization shock features, labeled in Figure~\ref{fig:fig36} as II, III, and IV. The NW boundary of III passes almost parallel to and about  0.8\arcsec\ north of HH~1127. These objects appear as northern portions of very large incomplete parabolic shocks that are moving towards about the same direction of I (68\arcdeg). There is evidence in both \oiii\ and \nii\ for an even fainter
 large partial shock with an apex about 3\arcsec\ west of feature IV. 
 There is a linear sequence including velocities of -32 \kms, -30 \kms, -21 \kms, -34 \kms, and -32 \kms, as shown in the figures. Another feature at -31 \kms\ lies on an extension of this series. Their PA is about 74\arcdeg, close to the approximate  68\arcdeg\ of the large partial shocks. Although there may be a relation of this sequence to the large shocks, their positions are too far east to be the source of even the easternmost shock II. There are no candidate stellar sources for this series of high radial velocity features. The size and incompleteness of shocks I through IV renders it imprudent to comment on their sources, but these must lie at a considerable distance to the WSW. Shock II can also be designated as 141-345 and is discussed further in Section~\ref{sec:HH1148}. There is a strong resemblance to the wide shocks lying along the axis of the HH~202 flow (Section~\ref{sec:HH202} and Section~\ref{sec:HH510location}).

There is a narrow \nii\ strong linear feature within the ellipse in the lower right panel of 
Figure~\ref{fig:fig37} that resembles a jet. This feature is undetectable in \oiii\ and appears dark in a F502N/F487N image. This indicates that it is either a low ionization jet or
a local ionization front seen almost edge-on.  If it is a jet, the orientation of  81\arcdeg\ is 13\arcdeg\ less that the orientation of shock II, which is the best candidate for an association with a visible shock. We note that this linear feature points close to the series of high \vrad\ \oiii\ features discussed in the preceding paragraph. 
The difference in PA of the linear feature (81\arcdeg) and the series of \oiii\ high velocity features (74\arcdeg) argues against an association. The linear feature is probably an edge-on ionization front.

There are two features moving along the same direction lying to the NW ( 312\arcdeg) from star EC 9 (137.2-350.6 ).  Shock 135-347 lies at 5.2\arcsec\ from the star and shock 133-344 at 9.5\arcsec. The orientation, motion, and proximity argue that 
these are shocks driven by an optically invisible jet arising from star EC 9. This object is only seen at millimeter wavelengths and
is object 24 in the catalog of \citet{eis06}. This source and pair of shocks is designated as HH~2034AA.

Along the same line passing through shocks 133-344, 135-347, and star EC 9 is a series of shocks to the south near the star COUP 607, as seen in Figure~\ref{fig:fig19}.
The motions of these shocks indicate that there is no association with the EC 9 driven shocks. The former are discussed in Section~\ref{sec:GroupShocks}.

\subsection{Complex Shocks Forming HH~1153.}
\label{sec:GroupShocks} 

There is a group of high ionization shocks near the star COUP 607. On a motions image such as Figure~\ref{fig:fig19} the 
movement appears to be to the south. However, this is a result of the complexity of the field, which has both bright and dark features. When
the individual features are measured, a movement towards  19\arcdeg\ is apparent, a conclusion supported by the symmetry axes of the shocks.
We designate this group of shocks as HH~1153.

There are no nearby stars towards the reciprocal heading and the Ori-S6 region \citep{schm90,zap10} is the first group of sources encountered. The source EC 14 that is identified by \citet{zap10} as the source of a deflected SO outflow lies at  196\arcdeg\ with a distance of 12.9\arcsec. The direction of  EC 14 is close to the reciprocal of the motions of the COUP 607 shocks but it is impossible to link the shocks to the molecular outflows since the initial flow is towards  245\arcdeg\ and after deflection it is  238\arcdeg.  

\subsection{HH~1154}
\label{sec:HH1154}

\begin{figure}
\epsscale{0.6}
\plotone{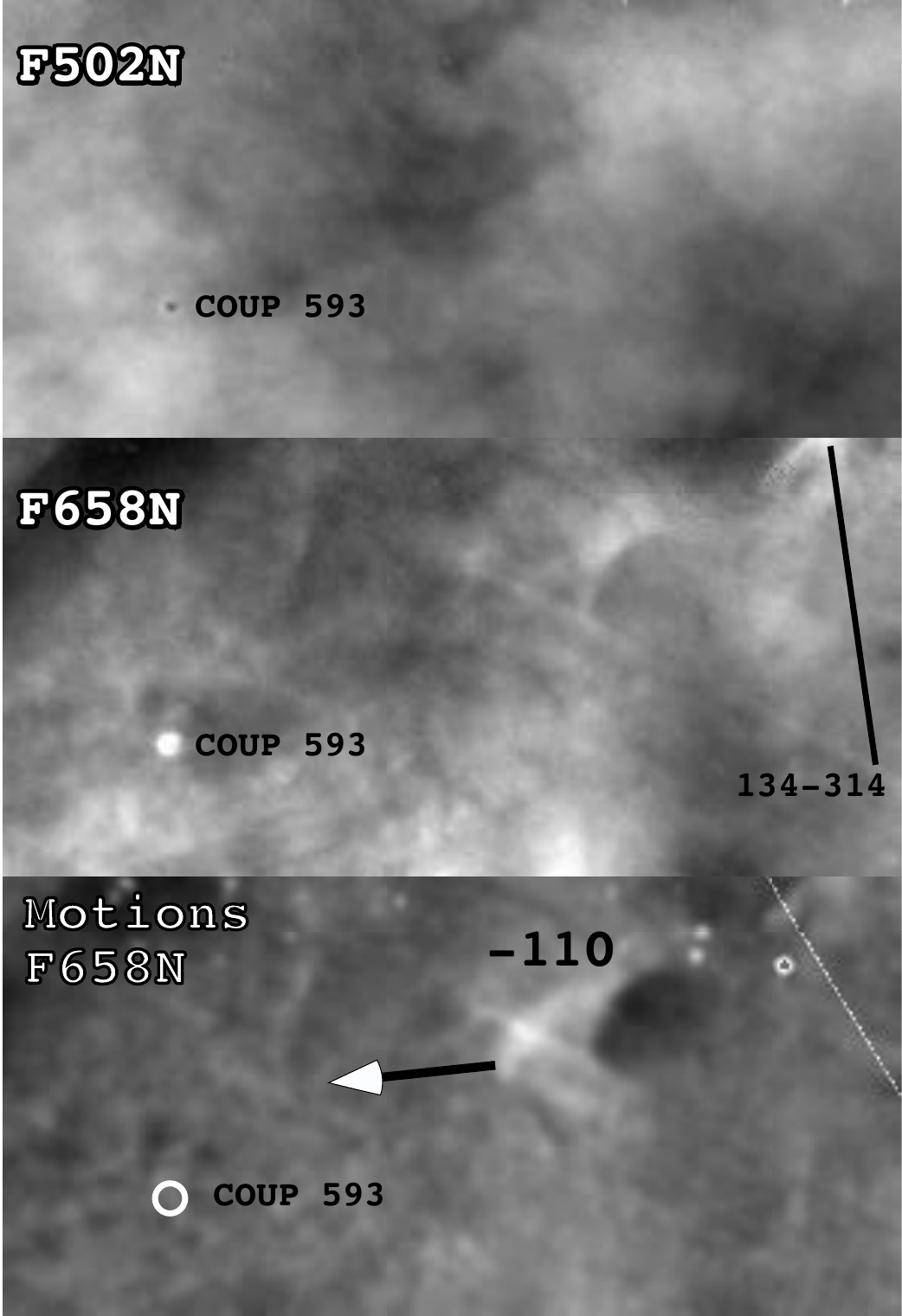}
\caption{This set of 12.0\arcsec $\times$ 5.8\arcsec\  FOV images show the region around HH~1154. The top panel shows the GO 12543 F502N images and the middle panel the GO 12543 F658N images.  The lowest panel shows the motions image displayed so that the position of objects on the second epoch images are brighter. The FOV is designated as HH~1154 in Figure~\ref{fig:fig2}. The only star in this region is shown. The stationary shock feature (134-314) is pointed out in the middle panel with a black line. In the Motions image the only high radial velocity feature (\nii) in this region is shown. The cluster of star-like features in the upper right of the bottom panel are from cosmic ray events not edited out of the second epoch F658N images.}
\label{fig:fig38}
\end{figure}

\placefigure{fig:fig38}

\citet{ode08a} noted an unusual feature, best seen in F658N, and designated it as 137-317. They interpreted it as a bright object moving at  155 \kms\ towards  113\arcdeg. This may not be the case if the apparent motion was caused by the changing structure of the feature.  
The object is very low contrast with the background, even in its stronger F658N line.  It appears as a crescent, oriented to the NE and is absent in F502N. The earlier epoch images of this region (GO 5469 and GO10921) are of course lower resolution than the GO 12543 images, but, the change of appearance is much more than can be attributed to improved resolution.  If the bright feature in the lowest panel is interpreted as a change of position, the motion would be 146 \kms\ towards  120\arcdeg, similar to the results of \citet{ode08a}. The tip can be designated as 
136.5-317.5.  The presence of an isolated high radial velocity feature in \nii\ at -110 \kms\ argues that the correct interpretation is that it is a moving feature (the -110 \kms\ feature is only 0.0028 the strength of the emission from the MIF, but it is well resolved). We designate the apparent high velocity shock as HH~1154. 

There is a single nearby star. The proplyd 139.2-320.3 (COUP 593) was discovered by \citet{ode94} and was designated there as 139-320. The proplyd has been detected in x-rays, in optical-wavelengths, and in infrared wavelengths. In
our images we see an unresolved dark disk in F502N that is surrounded by a ring of \oiii\ emission that is 0.3\arcsec\ in diameter.  In F658N it is a partially resolved star-like source of FWHM=0.2\arcsec.

If the derived \vt (146 \kms) and \vrad (-110 \kms) arise from the same feature, then \Vomc  =200 \kms\ and $\theta$=43\arcdeg. HH~1154 is left here as a curiosity, one that begs for resolution of its motions through additional radial velocity spectra and WFC3 images in F658N.

We see a curious linear feature that tangentially grazes the low dark feature (seen best in Figure~\ref{fig:fig38}, bottom panel). The PA is 257\arcdeg\ and it starts before the dark feature and extends after it.
This eliminates it as an ionization shadow cast by HH~1154. A projection towards 77\arcdeg\ points towards 
$\theta^1$~Ori~B, which is much cooler and much lower in LyC luminosity than \tc.  

\subsection{Outflows in the Region Immediately SW of the Trapezium.}
\label{sec:nearSW}

The region immediately SW of the Trapezium contains several previously unrecognized  high velocity flows. We characterize four and when possible identify the best candidates for their sources.

\begin{figure}
\epsscale{1.0}
\plotone{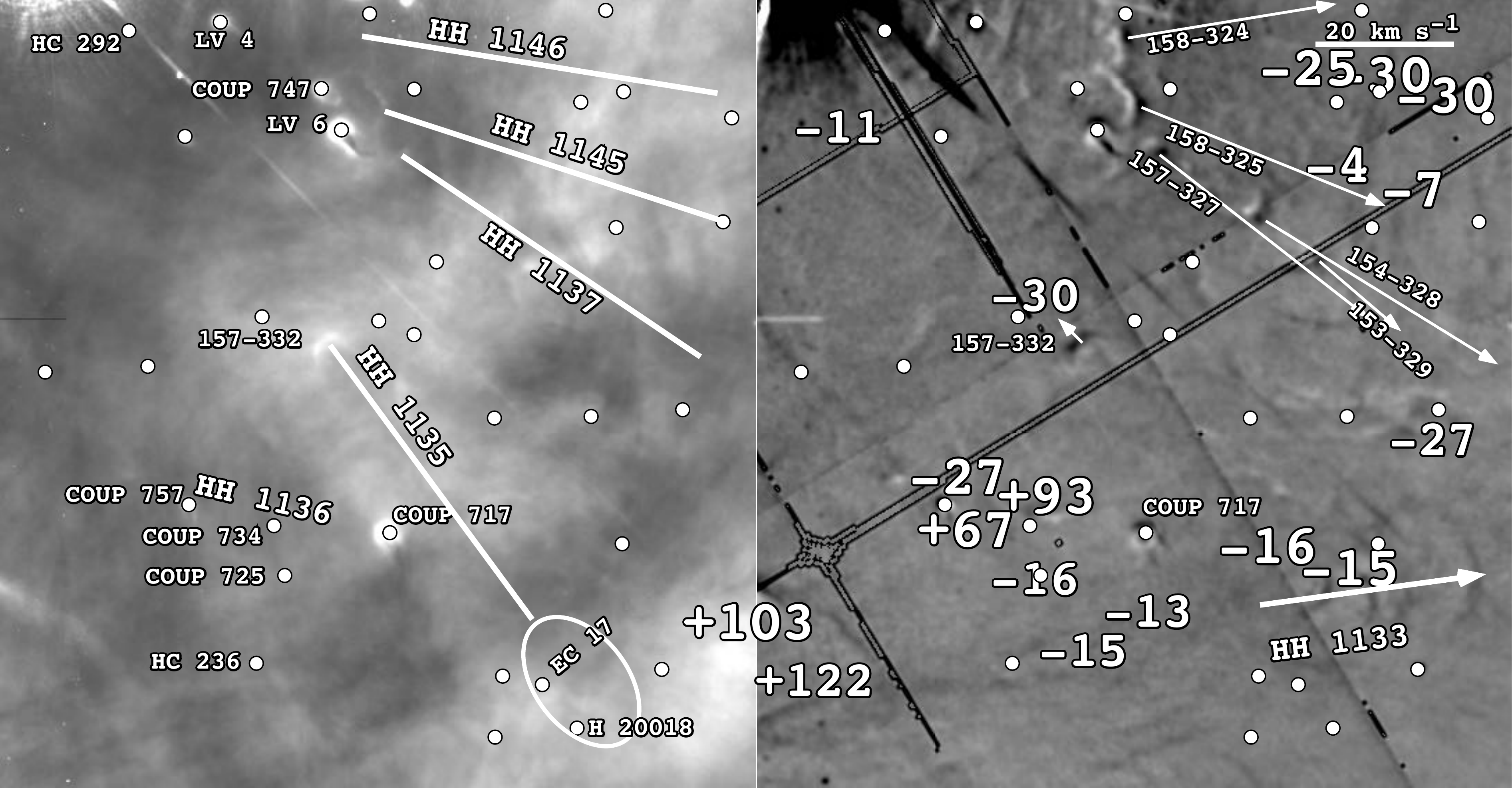}
\caption{This pair of 21.8\arcsec $\times$ 22.7\arcsec\  images presents the F502N image (left panel) and motions (right panel) of a FOV designated as SouthWest in Figure~\ref{fig:fig2}. The velocity vector for shock 158-324 is shown at one-half the scale of the others. The filled circles represent SIMBAD sources in this region. The inserted numbers are values of the \oiii\ \vrad. The other features are discussed in the text.}
\label{fig:fig39}
\end{figure}

There are two well defined large shocks (158-324 and 158-325) lying about 9\arcsec\ west of \tc. Their motions are 
shown in the right hand panel of Figure~\ref{fig:fig39}. There is no obvious low mass star that can be associated with them and they are discussed in Section~\ref{sec:theta1C}.
There are three high velocity \oiii\ features (-25 \kms, -30 \kms, and -30 \kms) in the upper right-hand of the right panel of Figure~\ref{fig:fig39} that appear to be unrelated to any of the flows discussed here. They are also
discussed in Section~\ref{sec:theta1C}.

\newpage

\subsubsection{HH~1137}
\label{sec:HH1137}
Three aligned shocks have been discovered in our new F502N images (Figure~\ref{fig:fig39}) that we designate as HH~1137. In addition to 157-327, 154-328, and 153-329, there is an additional, barely detectable, larger shock to the SW, along the same axis (250\arcdeg) as the others. A backwards projection of their axis passes through the proplyd COUP 747 and at a greater distance proplyd LV~4. It is impossible to judge definitively between the two as the source of HH~1137as neither shows a jet on our images. The axis  of HH~1137 passes about 0.5\arcsec\ north of \tc\ (Section~\ref{sec:theta1C}), well within the uncertainty of the projection of its axis.

There are no radial velocity features that would help in determining their 3-D motion. Their axis also passes close to \tc\ (Section~\ref{sec:theta1C}).

\subsubsection{HH~1145}
\label{sec:HH1145}
The partial bow shock labeled 158-325 is moving towards about 262\arcdeg, which is compatible with its symmetry axis. Further out in this direction are two \oiii\ radial velocity features of -4 \kms\ and -7 \kms. 
Within the assumption that these are part of the same flow that produces the shock, the axis is 276\arcdeg\ 
and this points exactly at the nearby proplyd COUP 747 and at a much greater distance HC 292. 

\subsubsection{HH~1146}
\label{sec:HH1146}
The moving knot at 158-324 would not ordinarily be assigned to a designated flow. However, there is a
series of high velocity \oiii\ features (-25 \kms, -30 km/s, and -30 \kms) lying beyond it. If these are associated with knot 158-324, then this forms HH~1146. The flow is then at 275\arcdeg. The axis points back to LV ~4 and the more distant proplyd HC 292. 

\subsubsubsection{Summary of Preceding Three Sections}
\label{sec:SumOfThree}
In the previous three sections we identified three flows that all arise from a small region to the west of \tc.
There are four candidate proplyd sources there (COUP 747, LV 4, LV 6, HC 292). Of these only LV~6 fails to align with any of the flows. However, each of the flows aligns with two possible sources. The well defined flow, HH~1137, aligns with the nearby COUP 747 and the more distant LV 4. The less well defined flow,
HH~1145 aligns with the nearby COUP 747 and the more distant HC 292. The other less well defined flow, HH~1146 aligns with the similarly distant LV 4 and HC 292. It is possible that two of the stars produce Multi-Direction flows. However, if all the sources produce monopolar flows, then the most likely pairings are HH 1137 with LV 4, HH 1145 with COUP 747, and HH~1146 with HC 292.

\subsubsection{HH~1133}
\label{sec:HH1133}

The series of shocks labeled as HH~1133 (centered at 5:35:15.1 -5:23:37) align along a vector that passes  between the two candidate stars HC 236 and COUP 725 (156.8-339.0). The latter
star is listed in SIMBAD as COUP 725 (5:35:15.68 -5:23:39.0) and \citet{lad00} 49 (5:35:15.69 -5:23:39.5), which are probably the same source, with the COUP position being used here. The alignment of the HH~1133 shock's symmetry favors HC 236. 
 The \vt\ velocity vector shown in Figure~\ref{fig:fig39} is the average result from the group of shocks.
  The \oiii\ \vrad\ values (-16 \kms\ and -15 \kms) are almost certainly associated with HH~1133, yielding \Vomc =53 \kms\ and $\theta$=52\arcdeg. 
  The similarity of magnitude of the  \oiii\ \vt\ values to those lying to the east (-16 \kms, -15 \kms, and -13 \kms) and their alignment along the symmetry axis of HH~1133 argue that these five radial velocity features are all part of the HH~1133 system. This means that its source lies in the direction of star HC 236, which is seen in optical through radio wavelengths. 

\subsubsection{HH~1136}
\label{sec:HH1136}
The very high and positive \oiii\ \vt\ (+103 \kms\ and +122 \kms) features in the lower left of the right panel of Figure~\ref{fig:fig39} are part of the flow producing HH~518a and HH~518b, as discussed in Section \ref{sec:HH518}, while the two other high velocity features (+ 67 \kms/ and +93 \kms) are out of the line of progression of the two HH~518 flows (Section~\ref{sec:HH518} are probably not part of them.

This region also contains the proplyd COUP 717 [d155-338 \citep{bal00}, more accurately d155.1-337.2]. This object is seen
from X-rays through the radio continuum. It has the peculiar feature of a series of equally spaced bright rings of emission in F502N, extending out to 4.2\arcsec\ towards 38\arcdeg.
Unlike the concentric rings in the Dark Arc (Section~\ref{sec:OOSwest} and Section~\ref{sec:blanksDarkArc}) these have no detectable tangential motion.

There are two knots of high tangential velocity lying at 91\arcdeg\ from COUP 717, the closer at 3.3\arcsec\ and the further at 7.6\arcsec. The high \oiii\ radial velocity features -27 \kms, +67 \kms, and +93 \kms\  lie between these moving objects. The moving knots and the 
+67 \kms\ and +93 \kms\ are likely to be an outflow from COUP 717, which we designate as HH~1136.

The nearby -27 \kms\ feature is probably associated with the all wavelengths source COUP 734 (proplyd 157.3-337.9).

\subsubsection{HH~1135.}
\label{sec:HH1135}

There is a conspicuous shock seen in F502N in Figure~\ref{fig:fig39}  whose motion is shown in the motions (right hand) panel and labeled
157-332. The symmetry axis of the shock is towards  50\arcdeg\ and is shown as a line in the left-hand panel of Figure~\ref{fig:fig39}, whereas the less accurately determined vector of the motion 
is  59\arcdeg.  Both the symmetry axis and the vector direction argue against an association with the nearby COUP 717.  
There are two stars that fall within the range of uncertainty of the symmetry axis, these are 151.6-340.6 at  226\arcdeg,
and H 20018 at  227\arcdeg. They are enclosed within an open ellipse in the lower left hand panel of Figure~\ref{fig:fig39}. 
The closer object (EC 17) has only been seen as 
a millimeter thermal object \citep{eis06} and the further object H 20018 has only been seen in the near infrared (NIR)\citep{hil97}, therefore neither are a promising source. 
The shock 157-332 and the possible sources are designated as HH~1135 in Figure~\ref{fig:fig39}.

\subsection{Shocks Near the Trapezium Stars}
\label{sec:theta1C}

The Trapezium is at the core of the optical components of the Orion Nebula Cluster. On the average, the low mass stars forming near this line of sight will be subject to a more intense ionizing radiation flux from \tc\ than stars further away in the plane of the sky. This made it particularly worthwhile to study the flows in this region.

\begin{figure}
\epsscale{1.0}
\plotone{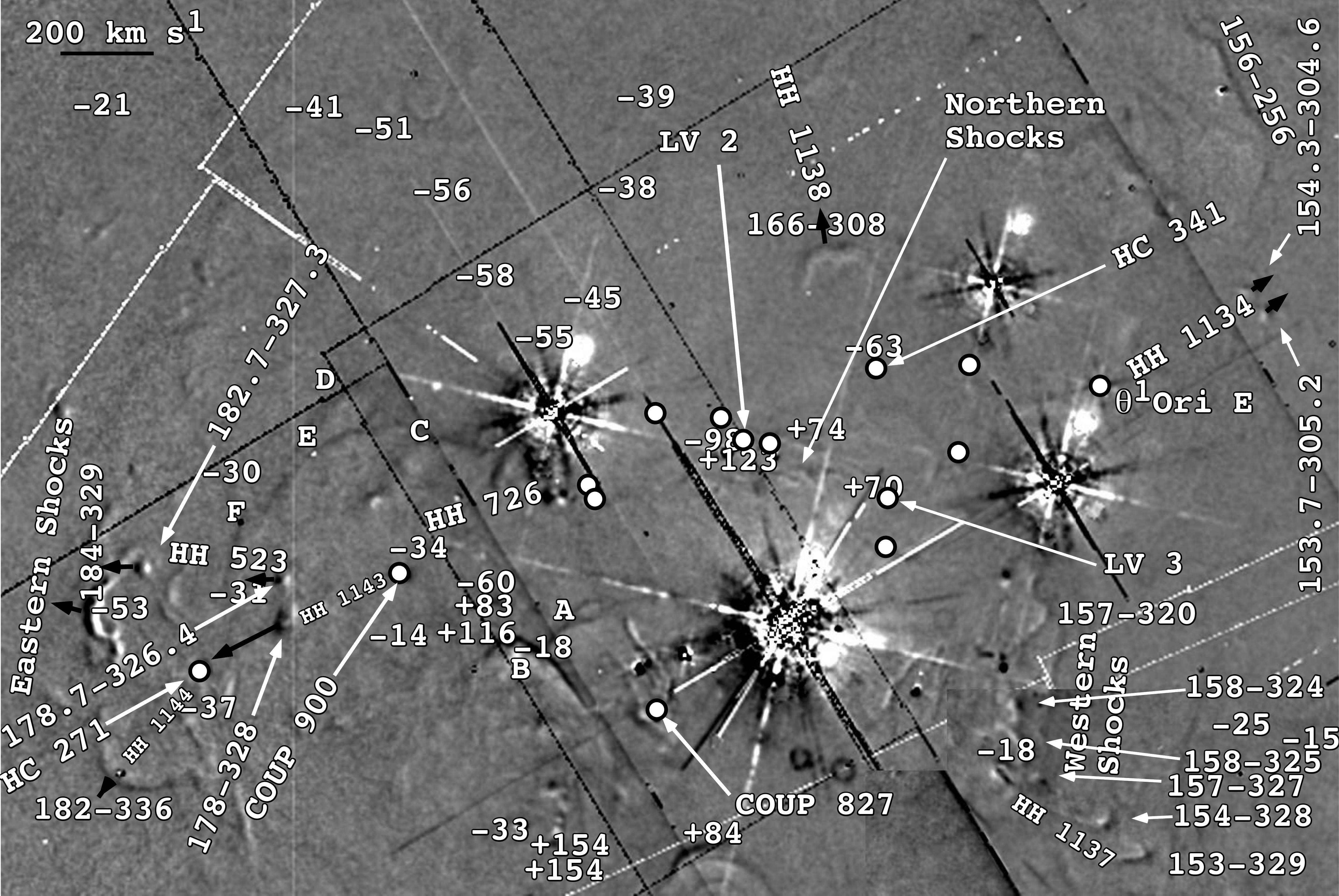}
\caption{This 56.8\arcsec $\times$ 38.0\arcsec\  image presents the F502N motions image derived from non-WFC3 observations. The letters A-F designate features in the northern portions of HH~518a and HH518b (Section~\ref{sec:HH518}). Radial velocities in \oiii\ are from \citet{doi04}. Features discussed in Section~\ref{sec:theta1C} are labeled. Not all stars in this FOV are shown, but all source in SIMBAD that fall within the four Trapezium stars are. Features with measured tangential velocities are shown with black arrows.}
\label{fig:fig40}
\end{figure}

\subsubsubsection{Shocks to the West of \tc}

To the west of \tc\ we see (Figure~\ref{fig:fig40}) a series of nearly N-S shocks (about 10.5\arcsec\ and 259\arcdeg) that we call the Western Shocks. The individual components of this series are oriented towards \tc\ or slightly south of it, with the exception of the 157-327, 158-324, and 158-325 features that have been assigned to specific HH flows (\ref{sec:nearSW}).

\subsubsubsection{Shocks to the East of \tc}
\label{sec:eastoftc}

Similar to the situation to the west of \tc, there is an extended series of shocks to the east that we call the Eastern Shocks. Most of these were identified in \citet{bal00} and assigned to HH~523. However, we are now able to assign them to the redshifted lobe of the bipolar flow that originates in the well studied proplyd LV~2. 
The brightest proplyd is LV~2  ($\theta^1$~Ori~G, COUP 826). It's position is under the -98 \kms\ entry in Figure~\ref{fig:fig40}.
Among the many studies of this bright proplyd, \citet{hen02} established that it has a jet extending 2\arcsec\ towards  120\arcdeg\ and a fainter counter-jet extending 2.5\arcsec\ towards  300\arcdeg. They found that the SE jet has a velocity of +126 \kms, in agreement with our value of +123 \kms. The counter-jet probably accounts for the coincident -98 \kms\ feature. \citet{doi04} created a detailed radial velocity map in \oiii\ of this region and showed that the redshifted flow could be traced to the east boundary of the Trapezium and the blueshifted flow extends to the west boundary. These flows follow the 120\arcdeg\ and 300\arcdeg\ orientation of the jet and counter-jet of LV~2.  They were given the designation HH~726.

\citet{bal00} defined HH~523 as shocks at 184-330, 183-327, and 179-326. We more accurately define these individual features as 184-329, 182.7-327.3, and 178.7-326.4. They lie at 120\arcdeg , 117\arcdeg , and 122\arcdeg\ from LV~2 and almost certainly belong to HH~726. Were it not for the known flow from LV~2 one might assign as their source COUP 900, but the fit with HH~726 is excellent. Moreover, there is a high velocity (\vt\ = 160 \kms ) shock moving away from COUP 900 and we define that flow as HH 1143. We have found that COUP 900 (175.7-324.7) lies indistinguishably close to the proplyd  175.3-324.7 (176-325 in Henney $\&$ Arthur 1998). 
This proplyd has a cusp in F502N and an 0.8\arcsec\ long tail pointed about 102\arcdeg. \tc\~ lies at 106\arcdeg , and this tail is probably determined by the ablation of material by \tc .

The large shock at 182-336 is symmetric along a line to the star HC 271 and the measured motion shares this direction. We define this newly discovered flow as HH~1144. In summary we can say that all of the Eastern Shocks can be assigned to known or new HH objects and HH~523 should now be considered non-existing.

Most of the multiple high velocity features about 10\arcsec\ east of \tc\ are probably associated with the extended
sequence of shocks assigned to the two branches of HH~518. Those may restrict identification of radial velocity features. 

\subsubsubsection{Shocks to the North of \tc}
\label{sec:NorthOfTc}

North of \tc\ at 7\arcsec\ are a large number of outward moving shocks lying nearly along an E-W line over a spread of PA values of 67\arcdeg. They are less well defined than those in the Western Shocks series, but again they are oriented away from \tc. We call these the Northern Shocks. There are multiple proplyds and several measured high radial velocity features in the vicinity of the Northern Shocks. However, the stars and the high velocity features lie further from \tc\ than the Northern Shocks.

Further north of the Trapezium, at 15.9\arcsec\ and  8\arcdeg\ from \tc,  lies the double shock 166-308, the west component of which approximately aligns with \tc\ (there are no intervening candidate sources) and the other towards the proplyd 164-316 (HC 341, which lies near the -63 radial velocity number).  The tangential velocity vector indicates motion of the shock 166-308 away from HC 341.   We designate proplyd HC 341 and 166-308-East as HH~1138. 

There is a similar double shock form at 156-256 symmetrically oriented towards 
$\theta^1$~Ori~D. It lies at 309\arcdeg\ and 34.5\arcsec\ from $\theta^1$~Ori~D. There is a more symmetric 
shock with the same orientation 3\arcsec\ beyond it. However, an association of these shocks with $\theta^1$~Ori~D
is uncertain because there are numerous intervening other candidates.  

Near the right hand edge of Figure~\ref{fig:fig40} we see a pair of NW moving knots (154.3-304.6 and 
153.7-305.2). Their positions straddle the projection ($\pm$1\arcdeg) of the HH~726 axis. However, their motions indicate that they arise from the nearby binary star $\theta^1$~Ori~E and we designate this flow as HH~1134.
\tE\ is a remarkable star as it is a spectroscopic binary, both components of which are mid-G-Giants \citep{her06,cos08}. Most unusual is that its systemic radial velocity is 8.3 \kms\ greater than the ONC.
This means that the star is not dynamically part of the Trapezium. The orientation of HH1134 is 309\arcdeg. It is interesting that a line projected back 7.7\arcsec\ from \tE passes through a small bow shock of the same orientation.
Unfortunately, the proper motion study of \citet{oli13} does not give a useful result for the tangential motion of \tE\ with respect to the Trapezium stars considered as a group.

The broader context of the \tc\ features is discussed in Section~\ref{sec:NonHH}.

\begin{figure}
\epsscale{1.0}
\plotone{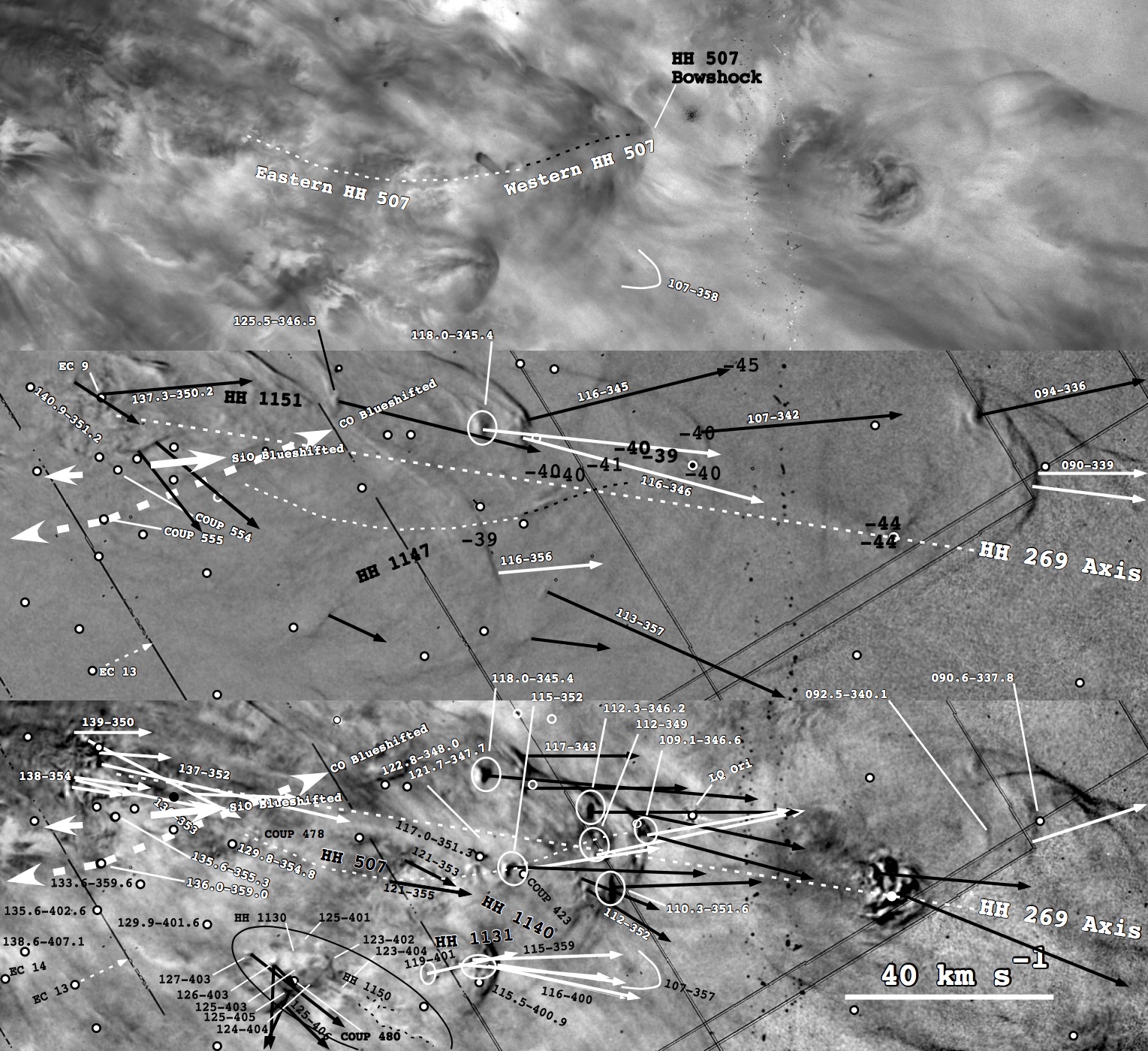}
\caption{The FOV is designated as LargeWest in Figure~\ref{fig:fig2} is shown in this set of 89.1\arcsec $\times$ 26.9\arcsec\  images. The F502N ratio image on the top, with the motion images in the the middle and lower panels. The velocity scale shown applies for all velocity vectors except those labeled 094-336 and 124-404, where the lines are at one-half the scale of the others. The curved lines shown crossing near star COUP 423 are faint patterns in the ratio and motion images. White and black vectors are used to distinguish different flow patterns, although the same color is used for different groups when it is clear that one is referring to different flows. The dashed line emanating from EC 13 indicates the direction of the blueshifted SiO outflow \citep{zap06}. The heavy black letters in the middle panel indicate \oiii~ radial velocities taken from Figure~\ref{fig:vradOIII}. The truncated black ellipse indicates the complex of objects called 126-406 in Section~\ref{sec:1150} and Section~\ref{sec:HH1148}. The open circle on the ellipse designating 109.1-346.6 marks the positions of the shock 109.7-346.8 discussed in Section~\ref{sec:others}.}
\label{fig:fig41}
\end{figure}

\newpage
\placefigure{fig:fig41}

\newpage
\subsection{Outflows Near the HH~269 Axis}
\label{sec:LargeWest}

 In our examination of the westward flows that constitute HH~269 we saw that there were numerous shocks that moved at similar PA values and had locations near the HH~269 axis. The optically dominant feature is the group of shocks lying to the east of the HH~269-East feature. A careful examination of these objects indicate that almost none can be identified with the HH~269 flow. 

\subsubsection{HH~507}
\label{sec:HH507}

This object is part of a complex system of overlapping shocks. This accounts for the assignment of membership and the 
accompanying features in the literature changing  with time. Our new images have allowed a more accurate identification of the flows involved.

HH~507 was first designated in \citet{bal00}.  They identified the shock 109-347 (designated here as 109.1-346.6) as being of a different orientation (315\arcdeg) than the other
shocks upon which it is superimposed and different from the HH~269 axis (which we establish in this paper as  275\arcdeg-276\arcdeg). 
They reported a 10\arcsec\ linear feature nearly along the symmetry axis of the shock and noted that this pointed south of the nearby proplyd 
d117-352 (117.0-351.3) and  towards the complex
of molecular outflows to the east. \citet{ode03a} assigned the adjacent shocks at 111-352 (designated as 110.3-351.6 in this paper)  and 113-345 to HH~507 and cast doubt on the association of HH~507 with the molecular outflows. Improved values of positions, designations, and angles are given below.

We note that on our F658N image the proplyd d117.0-351.3 (COUP 443) has a previously unreported microjet extending to 1.9\arcsec\ from the central star at  56\arcdeg,
There is a low ionization moving shock about 0.3\arcsec\ from the star that is probably driven by this microjet.  

There are multiple unassociated small shocks lying within the HH~507 bowshock. Their motions are usually along or near the HH~269 axis and several are probably part of that system of flows. 

Figure~\ref{fig:fig41} shows that the form and measured motions present a more complex picture than previously realized and that there are at least two overlapping flows. The axis of symmetry of the 109.1-346.6 shock that defines HH~507 (300\arcdeg) lies along its tangential motion.
The putative 10\arcsec\ linear feature in Bally et al.'s (2000) study lies 
along this symmetry axis. It is a low ionization feature extending east only 8.3\arcsec\ (this feature is shown as a black  dashed line in the middle panel of Figure ~\ref{fig:fig41}). We will call this the western HH~507 jet. 
The shock 112-349 lies along and is symmetric with this linear feature and is probably part of the HH~507 flow.  Immediately beyond the east extension of the western HH~507 jet is the shock 115-352. It has a small shock 1.0\arcsec\  west that is certainly part of western HH~507. The position of this shock is less that 1\arcsec\ from the star COUP 423, but since the shock lies 0.3\arcsec\ north of the star and is moving to the NW, COUP 423 cannot be the source of HH~507. 

The more pronounced 115-352 shock is the westernmost feature of a series of shocks (8.7\arcsec\ length and  291\arcdeg) and are shown as white dashed lines in the middle panel of Figure ~\ref{fig:fig41} and we call here the eastern HH~507 jet. 
The axis of the eastern HH~507 jet curves more rapidly at the small shock 121-355, remaining populated by small 
shocks and extending to 129-354 with a final orientation of  261\arcdeg. 

The combined western HH~507 jet and the eastern HH~507 jet are marked by a continuous non-monotonic change in 
direction. This combined jet extends over 31\arcsec. There is no obvious source, but there are many candidate stars starting at about 5\arcsec. 
  
 \subsubsection{Shocks Near but Probably Unrelated to HH~507}
\label{sec:others}
 
Unlike the other shocks in the HH~507 region 118.0-345.4 and the shocks to the west of it  (117-345 and 116-345 116-346) appear in both F502N and F658N.  Their motion and symmetry indicate an axis of
 284\arcdeg, in agreement with the orientation of the small arc that constitutes 118.0-345.5. 
This orientation is close to the angle of the blueshifted SiO from COUP 554, but is displaced 2\arcsec\ from that 
star. A link to the SiO flow would require a rotation by a few degrees CW beyond the region 
traced by the SiO emission. The PA of this grouping is sufficiently uncertain that it is impossible to establish a link to the SiO outflow.  However, their motion is very nearly parallel to the HH~269 axis, arguing for an association with HH~269 in spite of the displacement,

Features closer to the bow shock of HH~507 are probably also associated with HH~269. 
Shocks 109.7-346.8 (which falls 0.7\arcsec\ east of the large bow shock 109.1-346.6), 110.3-351.6,  and 112.3-346.2 are not symmetric about the symmetry axis of 109-347, rather, their symmetry and motion is at about  284\arcdeg. This is closer to the the HH~269 axis (275\arcdeg)
than to the HH~507 axis (300\arcdeg), arguing that these are associated with the HH~269 flow. Although their average  \nii\ \vt\ 
is large (32 \kms), there are no detected radial velocities for them. This indicates motion almost in the plane of the sky.
  
The moving knots a few arc seconds east of 112-352 appear to be independent of either HH~269 or HH~507.

\subsubsection{A New Outflow Source South of HH~507}
\label{sec:nearHH507}

There is a faint  large bow shock seen in the F658N motions image oriented towards the west and its apex is at 107-357. 
There is a diffuse moving shock within the arc of the bowshock about 3\arcsec\ from the peak. There are no obviously related other features, although the axis of symmetry pass close to the shock 121-353, whose motion is towards 107-357. The separation of these features is 21\arcsec. The axis of these features pass through the star COUP 478, which is their likely source. This star is seen in X-rays, our F547M image,  and the infrared. We designate this flow as HH~1140.
  
\subsubsection{A Possible Link of Large Shocks by Radial Velocity Data to the EC 13 Outflow Complex}
\label{sec:13574082objects}

There are a series of large shocks and high velocity features that may be related to a compact region of multiple outflows. In the F502N motions panel of Figure~\ref{fig:fig41} we see that the measured shock at 107-342 shares a similar orientation (304\arcdeg) and tangential motion with the large shock containing 116-356. 
Shock 107-342 is seen best in F502N and is not the same as the shock at 109.1-346.6 that is the leading bowshock of HH~507.  

Possible additional evidence lies 
in the  \oiii~radial velocities shown in Figure~\ref{fig:vradOIII} and Figure~\ref{fig:fig41}. All have velocities of about -40 \kms~and an axis of about PA~=~295\arcdeg, about the same as the axis of 107-342 and 116-356 shocks (304\arcdeg). Although parallel to the position of the west HH~507 shocks, they are displaced to the north by well more than the uncertainty of position of the radial velocities.
The easternmost of this set lies on the large shock containing 116-356 and the westernmost is 107-342.  It appears that these large shocks (116-356 and 107-342 are driven by a source at PA~=~124\arcdeg. 
This direction passes close to the source 137-408 \citep{zap06} which lies within a complex group of sources \citep{zap10} and has a bipolar SiO outflow \citep{zap06}. In its discovery paper \citep{zap06} it was designated as 137-408, but later studies \citep{nut07} show that it should be called EC 13. The blue shifted flow (the reciprocal of the better defined red shifted flow) has  PA=129\arcdeg~ as shown in Figure~\ref{fig:fig41}. This suggests a causal relation.

A caveat upon using the \oiii~radial velocities for identification of a flow is in the two nearby features with -44 \kms\ lying exactly on the HH~269 axis.  Although their position is exactly on HH~269-East, their velocities are very different from HH~269-East, as noted in Section~\ref{sec:HH269}.  Clearly these features are not part of HH~269, but if they are part of the two \oiii~high radial velocity features to their west that are of about the same value, then the argument for linking those features to the large scale shocks is destroyed. However, the latter association is unlikely and we designate the EC 13, 116-356, and 107-342 sequence as HH~1147.

 \subsubsection{Motions Near 126-406}
\label{sec:1150}

The region enclosed in a truncated ellipse in Figure~\ref{fig:fig41} contains evidence of several different 
flows and is designated here and in Section~\ref{sec:HH1148} as 126-406. In this section we consider only some of the features. The others are probably part of the HH~1148 system and are discussed in Section~\ref{sec:HH1148}. 

There are a series of barely resolved moving knots closed by an expanding boundary that we designate as 
123-404. The boundaries slowly widen, being 0.6\arcsec\ wide at its east end. The boundaries are clearly defined for 2.6\arcsec\ then become irregular, as shown by the dashed line extensions in Figure~\ref{fig:fig41}.The orientation of the linear boundary region is  258\arcdeg. This points at 129.9-401.6. We designate this flow as HH~1150. The probable source (129.9-401.6) has only been detected in the 2MASS survey.

There is another series of shocks near the top of the truncated ellipse that we designate as 125-401. Their orientation is 
266\arcdeg, which again points close to 129.9-401.6. Since it shares a common source with HH~1150, we include it within HH~1150.

There is a 1.0\arcsec\ wide bow shock (123-402) pointed towards  283\arcdeg. Its orientation and position  indicates that 129.9-402.6 is not its
source. The closest other candidate sources  COUP 582 at 23.3\arcsec\ and reciprocal  283\arcdeg, EC 14 at 24.8\arcsec\ and reciprocal  287\arcdeg, and EC 13 at 19.5\arcsec\ and reciprocal  289\arcdeg. In terms of the alignment the best possible source is COUP 582, which has only been 
seen as an X-ray source (COUP 582). We designate this source and shock combination as HH~1130.

The moving knot 126-403 (\vt=37 \kms, 192\arcdeg) and the moving shock 125-405 (\vt=17 \kms,  169\arcdeg) do not have identified sources and must be unrelated to either HH~1150 or HH~1130.
They are probably unrelated to the HH~1148 flow, although the other features are probably part of HH~1148 and are discussed in Section~\ref{sec:HH1148}. 

\subsubsection{Shocks Near but Probably Unrelated to Either HH~269 or HH~507}
\label{sec:others2}

A different constellation of shocks are those at 137.3-350.2, 125.5-346.5 and 094-336 seen in the F502N motions image. 
In this case all of the shocks are of similar form, none appear as bowshocks, and the orientation is
 283\arcdeg. The easternmost shock (137.3-350.2) is close to the millimeter source EC 9; but,  a projection of this axis passes only 0.4\arcsec\ south of the 6 cm radio source 140.9-351.2 \citep{fel93} and that is the most likely source. There is no obvious alignment with the stars in the vicinity of the OOS region. We designate this system as HH~1151, 

The symmetry and motion of the shocks at 090.6-337.8 and 092.5-340.1
 indicate that they have a symmetry axis of about  308\arcdeg, making them unlikely to be related to the shocks  covered in the above paragraphs. This axis passes 1.6\arcsec\ north of the star 096.5-346.7 which is superimposed on the HH~269-East complex of shocks and this cannot be their source. Otherwise, there are no nearby stars along the axis of these two shocks.
 
 \subsubsection{HH~1131}
\label{sec:HH1131}

HH~1131 is newly discovered in this study. As shown in Figure~\ref{fig:fig41}
it is composed of a well defined large shock (115-359) symmetrically oriented along the path of a series of moving knots (grouped in the figures
as 119-401 and 116-400). The symmetry of the shock and the alignment plus motion of the knots  define an axis at  288\arcdeg.

Projected backwards, the axis of HH~1131 leads to the star COUP 480. That object is an unresolved point source in X-rays \citep{get05} and 
infrared observations \citep{mlla,hil00,lad04}. There is no known optical counterpart, even on the GO~12543 F547M images.

\subsection{HH~1148}
\label{sec:HH1148}

\begin{figure}
\epsscale{1.0}
\plotone{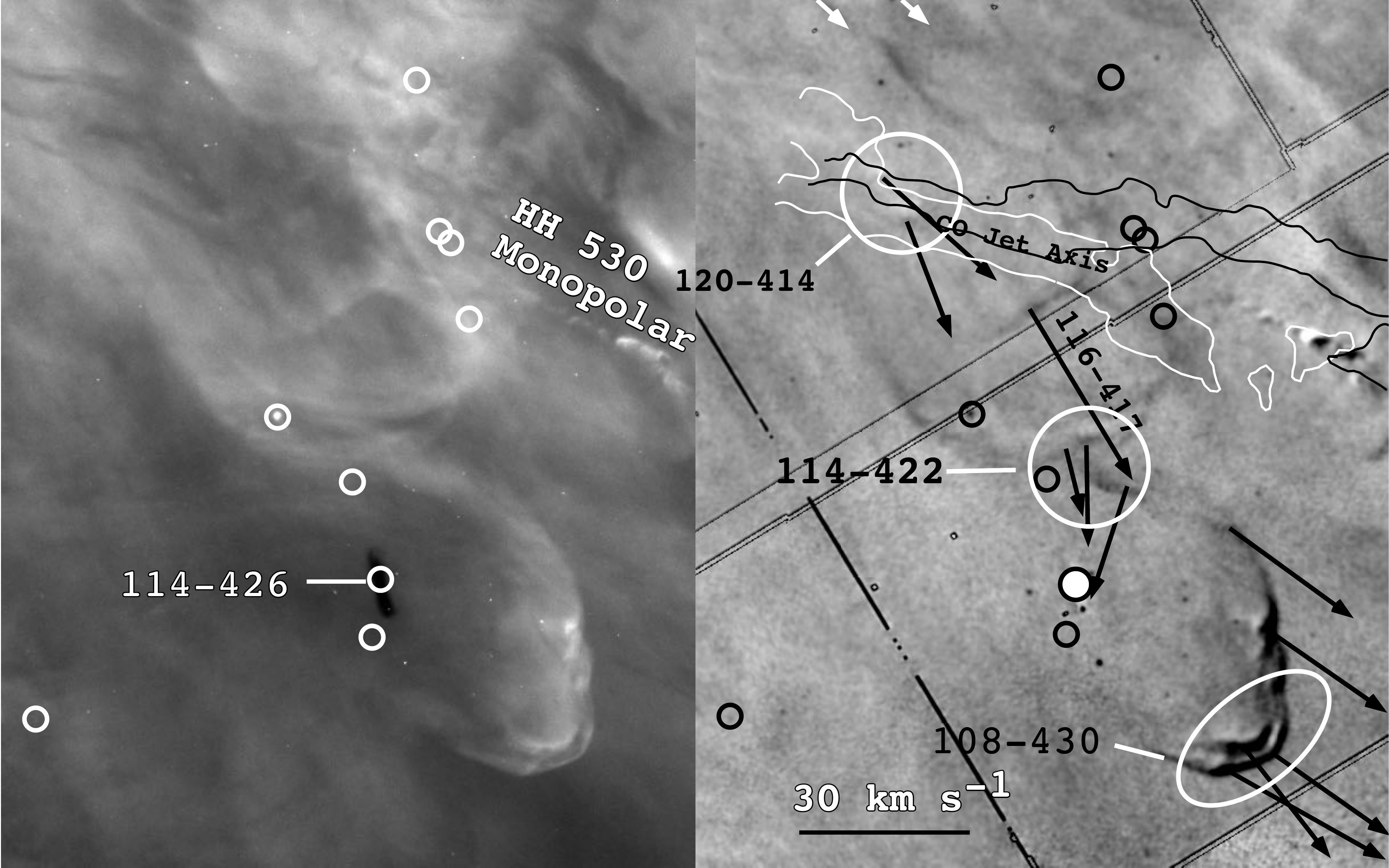}
\caption{This 24.4\arcsec$\times$30.6\arcsec\ FOV is designated as HH~1148 in Figure~\ref{fig:fig2}. It shows the F658N image on the left and the F658N motions image on the right.  The large silhouette proplyd 114-426 is shown and labeled, with only the positions of other SIMBAD sources depicted. Only two individual velocity vectors are labeled in addition to two groupings of velocity vectors.  The velocity vectors extending to the image boundary have been truncated. The velocity vectors of the components of the HH~530 Monopolar jet have not been shown. The uppermost white vectors are discussed in the text. The outer contours of the CO flows (blueshifted-white, redshifted-black) recently discovered by \citet{bal15} and discussed in Section~\ref{sec:HH530redefined} are shown in the right hand panel.
}
\label{fig:fig42}
\end{figure}

\placefigure{fig:fig42}

The nearly edge-on silhouette proplyd 114-426 and its surroundings have been a subject of interest since first announced \citep{ode96} and studied in detail \citep{mcc96,thr01,bal15}.  As illustrated in Figure~\ref{fig:fig42}, there are a host of nearby shocks 
running from ENE to the SW, with the clearest component of the 108-430 grouping having been measured multiple times\citep{bal00,ode03a,hen07}. 
At first this series of shocks appear to be like HH~626, which would indicate that they too represent interaction of a large-scale outflow interacting with the ambient low ionization gas. However, this is not likely to be correct. Our new images have
allowed determination of motions of multiple low ionization features (some well defined shocks, some linear, and some simply knots). We see in Figure~\ref{fig:fig42} that these features move along a common axis of about  229\arcdeg\ and are probably related to one another, but not to the proplyd 114-426.  The linear blueshifted CO flow found by \citet{bal15} has a spur oriented like HH~1148 that must be associated with the 120-414 shocks. The peak intensity of this spur lies immediately to the NE of the \nii\ shocks in 120-414. This argues for an association of the optical objects and a molecular flow. A counter-argument to this is the optical features "upstream" discussed in the next paragraph. However, it could be that the molecular flows just graze the MIF.

The neighboring group of shocks that we collectively refer to as 126-406 and are within the truncated ellipse in Figure~\ref{fig:fig41} has at least four overlapping flows (Figure~\ref{fig:fig41}). There is a knot at 126-403 moving at 37 \kms\ towards  192\arcdeg\ and a bowshock at 125-405 moving at 17 \kms\  towards  169\arcdeg.  These cannot be related to the HH~1148 flow.  In addition, there is the HH~1130 flow probably coming from COUP 582 and the two branches of the HH~1150 flow coming from 129.9-401.6.

However, 127-403 (23 \kms, 246\arcdeg), 126-403 (6 \kms, 239\arcdeg) and the bowshock 125-406 (19 \kms, 235\arcdeg), may be associated with HH~1148.  Their average motion is towards  240\arcdeg. Projections of the velocity vectors of 125-406 and 127-403 are shown as white arrows at the top of Figure~\ref{fig:fig42}.  The
northern well-defined component of HH~1148 (120-414) lies at  227\arcdeg\ from 126-406. 

These three shocks within 126-406 align (227\arcdeg) almost exactly with the common axis of HH~1148 (229\arcdeg)  and their tangential velocities (240\arcdeg) are directed almost towards the northern components of HH~1148 (120-414).  The major components of HH~1148, including the 126-406 group,  are shown as open circles in Figure~\ref{fig:fig2}. Their velocity vectors are shown in the upper left of the F658N motions image in Figure~\ref{fig:fig42}. It is almost certain that the three shocks of 126-406 are members of the HH~1148 flow.  

We have also looked for other members of this flow and for candidate sources. 
A projection back crosses the putative source for HH~1150 (129.9-401.6), and within the uncertainty of the flow direction, numerous candidate source stars beyond that. There are multiple molecular outflows along the projection, but none of them align with HH~1148.
The projection passes through the middle of the Figure~\ref{fig:fig11} FOV, where the number of candidate stars is a maximum. This line passes within one second of arc east of the center of the concentric arcs at 137.6-348.8 discussed
in Section~\ref{sec:OOSwest} and Section~\ref{sec:blanksAC}. It also aligns with the symmetry axis (57\arcdeg) of the large WNW pointing bow shock 141-345 (feature II in Figure ~\ref{fig:fig22}).
That bow shock has a series of otherwise invisible high negative velocity \oiii\ features leading to it at 74\arcdeg. If the high ionization shock 141-345 is part of the low ionization HH~1148 flow, this means that the source lies between the last \oiii\ radial velocity feature at 5:35:13.6 -5:23:47 and the 126-406 features. Given the large number of stellar sources along this line, we cannot make a convincing identification of the HH~1148 flow.

\subsection{Possible Outflow From DR 1186}
\label{sec:HH2014C}

In their study at 1.3 cm wavelength, \citet{zap04b} detected a microjet (about 0.5\arcsec) extending towards  324\arcdeg$\pm$4\arcdeg. 
The compact source, which is also seen at 3.6 cm \citep{zap04a} lies nearly along the middle of an irregular low ionization linear feature of about 4\arcsec\ full length oriented about  
 331\arcdeg. \citet{zap04b} argued that DR 1186 could be the source of both HH~202 and HH~528. Although we do not detect a tangential velocity along the optical filament, there is a nearby feature seen to move in the F658N motion image (Figure~\ref{fig:fig20}, but the 
direction of the motion indicates that it is probably part of the HH~269 flow. In our spectra we see a faint \Ha\ component at \vrad=-50 \kms\ and another in \nii\ at \vrad= -53 \kms, which may be an indication of an outflow from DR 1186.

\section{IONIZATION SHADOWS}
\label{sec:shadow}

In an earlier HST study, \citet{ode00a} studied linear features oriented towards the dominant ionization star in both the Orion Nebula and the Helix Nebula. It was established that these features are  ionization shadows cast by small objects (proplyds in the case of the Orion Nebula and knots in the case of the Helix Nebula). LyC radiation stops at the ionization front of the obscuring object, leaving a shadowed zone outside of it. In the case of the Orion Nebula, the shadowed region is no more than moderate ionization and \oiii\ is the primary way of detecting the gas. In the shadowed region \oiii\ is not found as that tapered column 
is subjected only to ionization by LyC photons coming from recombining ionized hydrogen. The shadowed region is only evident when there is ionized gas present, a feature used by \citet{ode09} to determine the distribution of gas in the inner parts of the
Orion Nebula.  

\subsection{A Rapidly Rotating Ionization Shadow}
\label{sec:fastshadow}
One of the radial shadows studied by \citet{ode00a} is double in our F502N motions images.  It is labeled as ''Ionization Shadow'' in Figure~\ref{fig:fig22}. This is a low ionization (\oiii\ weak-\nii\  strong) feature extending from 58\arcsec\  to 128\arcsec\  from the dominant ionizing star (\tc) and its middle lies along  256.1\arcdeg. Its width at 
66.4\arcsec\ 
from \tc\  is about  
0.87\arcsec.
It has been displaced by 0.44\arcdeg\ CW  between the first (GO~5469) and second (GO~12543) epoch observations.  The other shadows studied by \citet{ode00a} that fall within the GO~12543 FOV do not show this magnitude of rotation. There are three candidate sources for the obscuring object. They are the only objects within $\pm$7\arcdeg\ of the PA of the low ionization shadow.

The first candidate is the proplyd 158.0-326.7 (LV 6). 
The true nature of this object was revealed in the one of the first images made after the initial servicing mission of the HST, when 
\citet{ode94} 
recognized it as a proplyd. In a subsequent study \citet{bal00} described it as a dark-disk object with an optically visible central star and an ionized crescent facing \tc\ (incorrectly calling it d158-326).  The chord perpendicular to a radial line towards \tc\ is seen on the GO~12543 images to be 0.31\arcsec. Using the SIMBAD position for \tc\ (5:35:16.46 -5:23:22.9) locates the proplyd at 10.7" distance at  249\arcdeg\ from \tc\ (GO 5469 images give 10. 7\arcsec\ at  250\arcdeg). 
The second candidate is COUP 747 is a much fainter proplyd with an ionized crescent facing \tc. The chord of the crescent is 0.29\arcsec. It is 9.5\arcsec\ from \tc\ at  254\arcdeg\ (GO 5469 images give 9.6\arcsec\ at  255\arcdeg). 
The third candidate source of the ionization shadow is the proplyd 161-324 (LV 4). It has much higher surface brightness in F656N than LV 6 and is only marginally resolved, with a diameter of $\leq$0.2\arcsec.  It has a narrow low ionization tail of about 2.9\arcsec\ in the  \tc\ anti-direction. It is  6.3\arcsec\ from \tc\ at  256\arcdeg\ ((GO 5469 images give 6.1\arcsec\ at  256\arcdeg). 

The width of the ionization shadow should scale linearly with increasing distance from the obscuring source. This means
that the 0.87\arcsec\ width of the shadow (at 66.4\arcsec\ from \tc) should project to a size of 0.14\arcsec\  at LV 6, 0.13 at 158.5-325.5,  and 0.08 \arcsec\ at the distance of LV 4.

The better agreement of the inferred size of the obscuring source and the agreement of the PA of the low ionization shadow and  the orientation of the object with respect to \tc\  indicates that  the proplyd LV 4 is the object producing the shadow.

The rotation of the ionization shadow must be due to the change of orientation of the obscuring source with respect to \tc.
As noted above, this was about 0.44\arcdeg\ CW over the 16.8 yr period between the first and second epoch images. If the source is LV 4, this corresponds to a relative tangential motion of 1.2  pixels perpendicular to  256\arcdeg, corresponding to a velocity of 7 \kms.  This is a problem.

 \citet{van88} included \tc\ in their determination of proper motions of cluster members and found that it was moving about 2.3 mas/yr towards the SE, corresponding to 1.0 pixel motion in our study. If the \citet{van88} motion is correct and the proplyd is stationary, then the shadow would have moved 0.4\arcdeg\ CCW. This is of similar magnitude but in the opposite direction of the observed rotation. Photographic determination of the motion of bright stars with respect to faint stars, as was done by van Altena et al., is 
notoriously difficult and a recent, more definitive study by \citet{oli13}, using HST images, indicates that the relative motion within the Trapezium cluster stars is only a few kilometers per second, comparable to the velocity dispersion of the low mass members (\citep{jw88}). It is best to leave this discussion at this point with the conclusion that the rotation of the low ionization shadow indicates that \tc\ and LV 4 are rotating with respect to one another at a rate of 0.026\arcdeg\ per year (a period of 14000 years), which corresponds to a relative tangential velocity of 7 \kms. 
The radial velocity study of \citet{tob09} indicates a wide dispersion of velocities of stars in the declination range of the Trapezium and a deviation of 7 \kms\ is possible. If the high proper motion of \tc\ posited by \citet{tan04} is correct, we would see a pattern of motions of the radiation shadows, which we do not.

\subsection{Additional Ionization Shadows}
\label{sec:othershadows}

A well-defined non-moving ionization shadow is at  225\arcdeg. At 58.8\arcsec\ from \tc\ it is 0.8\arcsec\ wide. The PA value aligns
with the well-defined proplyd COUP 717. Using the SIMBAD position of 5:35:15.509 -5:23:37.18 the calculated PA is 224.9\arcdeg\ and its measured chord is 0.36\arcsec. This chord size projected to 58.8\arcsec\ is 1.05\arcsec, so there is adequate agreement between the observed and predicted shadow size if this proplyd is causing the ionization shadow. The 
rotational motion of the ionization shadow is less than 0.5 pixels (0.04\arcsec), corresponding to an amount of PA change of $\leq$0.019\arcdeg. This corresponds to a maximum relative velocity perpendicular to the line connecting \tc\ and COUP 717
of $\leq$ 2.5 \kms, a number compatible with the known velocity dispersion of the lower mass members of the Orion Nebula Cluster.

There is a sharp ionization shadow whose northern boundary is at  255.4\arcdeg\ from \tc.  This angle passes through the northern side of the ionization boundary of proplyd 158.5-325.5. At position 5:35:11.85 -5:23:41.0 this northern boundary
is 72.1\arcsec\ from \tc\ and the projection of the ionized chord (0.29\arcsec) of 158.5-325.5 would be 2.2\arcsec. The area 
immediately south of this point is complex and the south boundary is not visible. Even in the simpler region at 
5:35:10.75 -5:23:46.1 (at  89.7\arcsec\ from \tc), where the predicted south boundary of the projection would occur (predicted 2.7\arcsec) there is no clear evidence of a boundary. At the well defined northern boundary there is no indication of rotation of the ionization shadow.

There are a series of unresolved linear features grouped over about 1.0\arcsec\ and extending from 5:35:13.4 -5:23:42.2 to
5:35:11.1 -5:23:56.7 with  244\arcdeg. These show up best in Figure~\ref{fig:fig22}. The NE end passes less than a second of arc south of the shocks composing HH~1127, but are clearly unrelated. The projection back to \tc\ passes 0.6\arcsec\ south of the 
center of LV 6 and well outside of the ionization shadow of the proplyd.  The other proplyds in the area are even
further from the projected line to \tc. The complex structure make it impossible to tell if this is the case of a obscuring proplyd rotating rapidly around \tc.

\section{DISCUSSION}
\label{sec:discussion}

In Section~3 through Section~6 we have presented the results on a plethora of objects within the \hr. These objects have been as small as the  microjets  of less than 1\arcsec\ to shocks at up to 1000\arcsec\ from their sources. Summarizing this wealth of information is inevitably flawed, but we have ordered this section into subjects that are arguably separate from one another, usually proceeding  from large-scale to small scale, with an integrating summary at the end.

\subsection{Nature and Origin of the Major Northwest and Southeast Flows}
\label{sec:fans}

The correct identification of sources of flows depends on many factors. A tangential velocity vector indicates an origin along the reverse direction from the moving feature. Where possible, a high velocity jet produces another line useful for determining where orientation angles cross.  The smaller the distance to the intersection, the more accurately that intersection is determined and the more lines of motion available leads to a more accurate location of the source. In a system like the Major Northwest (dominated by HH~202)-Southeast Flows (dominated by HH~203 and HH~204) we have the advantage of vectors over a wide range of PA and a subsequently accurate determination of the source. In the case of the major east-west flows (various features of HH~529 moving east and the HH~269 features moving west) it is more difficult. Earlier work that combined the data from these nearly orthogonal flows led to the identification of the OOS, under the assumption of a common origin. In this paper we have shown that the two systems may be independent, yet their sources can be determined, both lying slightly outside the OOS.

\begin{figure}
\epsscale{1.0}
\plotone{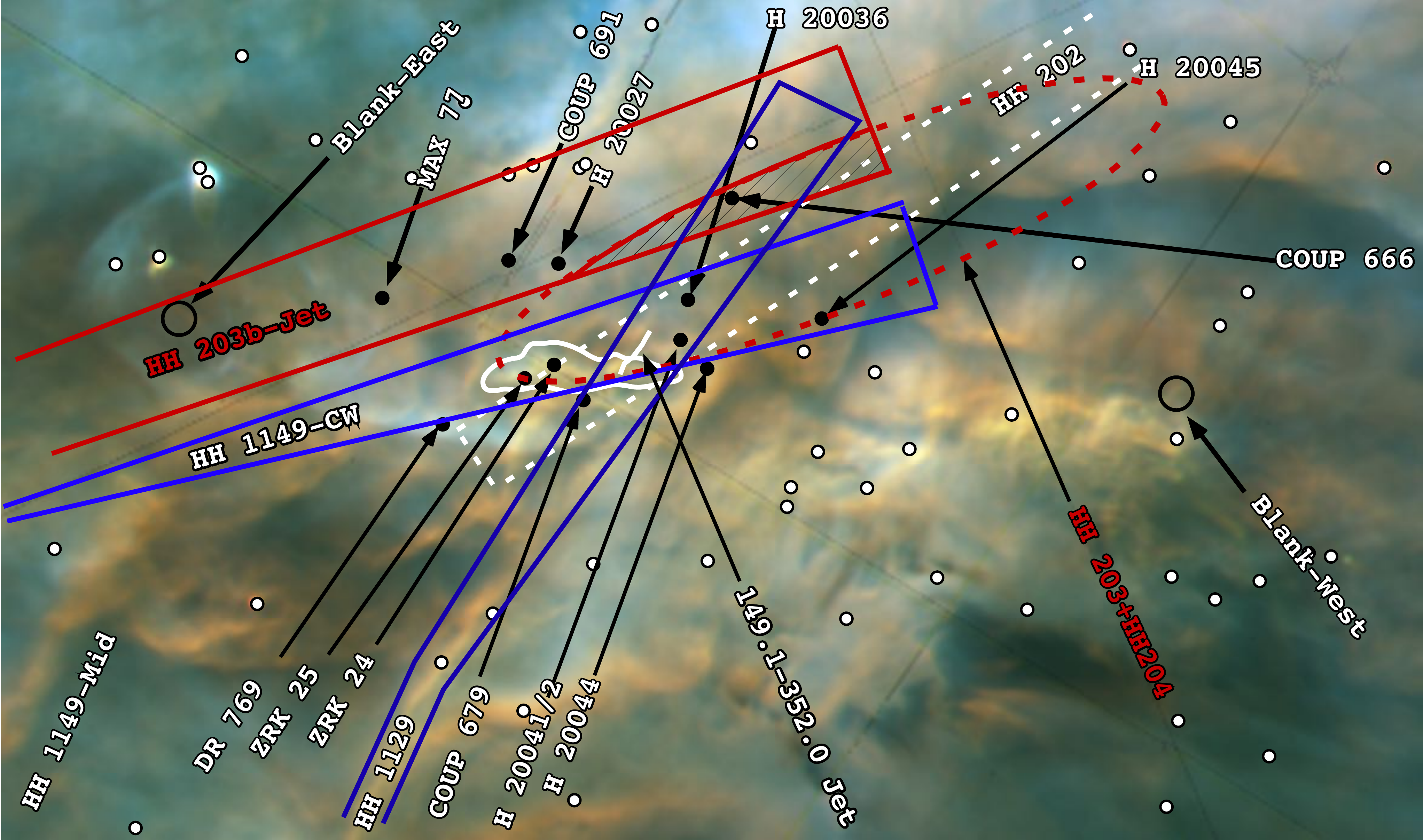}
\caption{This 47.1\arcsec $\times$27.8\arcsec\ HST image shows a field of view near the center of Figure~\ref{fig:fig5}. The color coding is Blue-F502N, Green-F656N, and Red-F658N. Stars relevant to the discussion of sources are shown as filled circles, those in black being the most relevant here. The large black open circles shown the positions of inferred positions (c. f. Section~\ref{sec:blanks} without SIMBAD sources. The colored lines are explained in the text.}
\label{fig:fig43}
\end{figure}

\citet{hen07} discussed these shocks in their earlier study. That study had to rely on less accurate motions and was done before HH~1149 was recognized. They could only argue that there was a source near the OOS.

Appreciation of what is happening demands understanding the inner features that we have been discussing thus far and the outer features beyond them. All are discussed in the following six sections. We will see that a coherent picture develops.

\subsubsection{HH~202}
\label{final202}

In Figure~\ref{fig:fig43} the dashed white band labeled HH~202 is a 2\arcsec\ band centered on the jet leading to HH~202-S. The source is most like to lie within this band near the region of highest stellar column density.  Examination of
the other shocks associated with HH~202 (at greater and smaller PA values) indicates a source in this area, but the angles are too imprecisely determined to narrow-down the source precisely. Stars H 20036, H 20041/2, and COUP 679 fall within this large area.

\subsubsection{HH~203 and HH~204}
\label{final2034}

The large red oval marked HH~203+HH~204 is the region where the lines of symmetry of these bow shocks converge. Since the range of angles is small, this area is large. Stars COUP 666, H 20036, H 20045, H 20041/2, ZRK 24, and ZRK 25 fall within this large area. 
A red trapezoid with sides divergent by $\pm$1\arcdeg\ indicate the axis of the HH~203b jet. 
Among the many stars in this large area, stars COUP 666, H 20037, and COUP 691 are closest to the area of convergence of the HH~203 and HH~204 symmetry axes.
Under the assumption that HH~203, HH~204, and the HH~203b jet share a common origin (the cross-hatched region in Figure~\ref{fig:fig43}, the best candidate source is COUP 666.  It is visible in X-rays \citep{get05,sch01}, optical \citep{dar09}, NIR \citep{jw88,her02,hil97,hil00}, and the longer wavelength infrared \citep{lad04}. The spectral type has been reported as K6-M4 \citep{hil13}.   

\subsubsection{HH~1149}
\label{sec:final1149}

The other dominant SE flow is HH~1149. Although neither boundary of its fan of linear features and bow shocks can be determined accurately, there is a well defined jet that probably forms that CCW boundary and allows identification of the source.

The shock used to define the CW boundary has a symmetry axis of 120\arcdeg.
The corresponding blue trapezoidal source box (blue lines in Figure~\ref{fig:fig43}) contains the stars H~20045, H~20036, H~20041/2, ZRK 24, and ZRK 25.
Close examination shows that this shock is complex and therefore its symmetry axis is not well defined. 
This means that the source box is probably much larger than drawn and is of limited aid in identifying a source in this crowded field.

The CCW boundary of HH 1149 varies from 168\arcdeg\ in the south and then changes to 160\arcdeg, again being depicted with blue lines in Figure~\ref{fig:fig43}.  It is notable that the center of this box passes very close to the linear jet, which points towards COUP 666. 
The jet's center is at 5:35:14.91 -5:23:52.0 (hence a designation as 149.1-352.0 jet) and crosses the East-Jet with no obvious interaction.
The jet clearly misses (Figure~\ref{fig:fig12} star H 20036 . This means that the CCW boundary of HH~1149 is caused by the sequence of objects COUP 666, the 149.1-352.0 jet, the 160\arcdeg linear feature and the 168\arcdeg\ linear feature. 

The most CW shock assigned to HH~1149 has a smaller symmetry axis PA than those to the CCW direction from it.  This progression of PA values is  what defines the group as a common flow. However, the PA of the interior shocks symmetry axes are generally difficult to determine. The bowshock 158-405 is oriented towards 144\arcdeg, while the linear feature 154-401 is moving towards 150\arcdeg. Both of these directions are similar to their PA value (141\arcdeg\ and 148\arcdeg\ respectively) from COUP 666. 
This argues that both these interior shocks originate from that star.

This leaves the most CW shock of HH~1149 in limbo, as it shares the form and low ionization of the other HH~1149 bow shocks, but its formal symmetry axis PA is too small to fit into HH~1149. Close examination shows that this shock is complex and therefore its symmetry axis may not  be well defined.

The feature HH~1149-Mid in the lower left of Figure~\ref{fig:fig43} resembles the CCW boundary of HH~1149 (designated as HH~1129). As noted before (Section~\ref{sec:HH528}), this feature projects to the HH~528 Base feature and it may be the source of the HH 528-West flow. Its origin is uncertain, although its north portion is at 166\arcdeg\ and a projection passes very close to the Blank-East region.

\subsubsection{Features Near the Bright Bar}
\label{sec:BB}

In Section~\ref{sec:SEbig} we saw that there are numerous features along the Bright Bar that are oriented towards the OOS region. These include a large low ionization pair of bow shocks inside the Bright Bar aligned with HH~203 and HH~204 (Section~\ref{sec:insideBB}) and a large number of bow shock like features that we designate as crenellations (Section~\ref{sec:crenellations}) that are found both inside and beyond the Bright Bar from the direction of HH~203 CCW to beyond the HH~528-Cap. Whether or not these are formed in the tilted surface of the Bright Bar or in the foreground Veil is unknown, however, they are likely to be associated with outflows from the OOS region. If from a single source, this would demand collimated flows in multiple directions over a confined range of PA's or if from single stars, this would demand collimated flows in similar directions from at least  twelve different stars.

\subsubsection{Outer Features}
\label{sec:outer}

 \begin{figure}
\epsscale{1.0}
\plotone{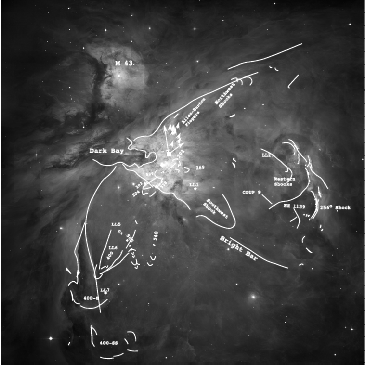}
\caption{This 30\arcmin $\times$30\arcmin\ HST image of the \hr,  M~43, and most of the Extended Orion Nebula \citep{hen07} has been annotated with white lines highlighting features outside of the \hr\ found by \citet{br01} and \citet{hen07}. The relation of the inner and outer features are discussed in Section~\ref{sec:Biggies}, with the exception of the Allen-Burton Fingers originating from the BN-KL region, which have been adequately discussed in Section~\ref{sec:fans}. }
\label{fig:fig44}
\end{figure}

\citet{hen07} argued that the distant  (633\arcsec\ from Orion-S,  146\arcdeg\ from the OOS) low ionization shock HH~400 \citep{bal01} lies on an extension 
of HH~203+HH~204+HH~528 and that they could be related. \citet{hen07} found a larger shock they called HH~400-S further out (789\arcsec\ at 148\arcdeg), and third feature composed of multiple smaller shocks designated as HH~400-SS (970\arcsec\ at 162\arcdeg).
With our new division of HH~528 into two parts and a more careful examination, we believe that these conclusions must be revised. 

The orientation of HH~203 is 123\arcdeg\ and that of HH~204 is 127\arcdeg. Although curves in flows can occur \citep{bal06}, this is quite different from HH~400's PA (146\arcdeg). It is most likely that HH~400 and HH~400-S are associated with the HH~528-East flow at 142\arcdeg\ which probably originates in COUP 632.  HH~400-SS is probably associated with the flow from HH~528-West (163\arcdeg), which is fed by HH~1149-Mid (162\arcdeg).
The latter series of features is designated as HH~1142  

\citet{hen07} pointed out a series of shocks on HST images that we determine to be at 590\arcsec\ and  321\arcdeg\ from the OOS. They  
concluded that these were associated with either what was producing HH~202 or the fingers of outflow from the BN-KL region. Their forms 
indicate an origin in the direction of the OOS. Their PA (321\arcdeg) is larger than that of the most CCW HH~202-N shock (327\arcdeg) but is close to the alignment of HH~202-S(323\arcdeg). Figure~\ref{fig:fig23}  shows that there are numerous shocks extending beyond the bright HH~202 shocks (Section~\ref{sec:HH202}) and it seems safe to conclude that the Northwest shocks are part of the HH~202 shocks. Their distance (590\arcsec) is only somewhat smaller than HH~400 (633\arcsec), but, unlike HH~400, there are shocks seen further out. This is consistent with the model of the EON \citep{ode10} where its ionization boundary is further to the SE than it is to the NW. 

\subsubsection{NW and SE Flow Summary}
\label{sec:sumNWSE}

HH~202 is a multiple member set of bow shocks with a high ionization jet, with bow shocks outside of the HH~202 features and distant shocks (Northwest Shocks) even further out. An origin has not been established because the jet does not continue into the OOS region and the varied PA bow shocks are not enough well defined to allow backwards triangulation within the band of possible sources of the jet.  The star COUP 666 lies slightly outside the path of the HH~202 jet, although it should be noted that in Figure~\ref{fig:fig23} the jet curves towards this star. 

HH~203 and HH~204 are more complex than previously appreciated. HH~203 is actually two superimposed bow shocks, components HH~203a and HH203b. The latter is aligned with a high ionization jet. HH~204a is a well defined bow shock with a faint irregular shock further out designated as HH~204b (at a slightly larger PA). 
The region of convergence of the symmetry axes of HH~203a and HH~204a overlaps with the extension of the jet pointing towards COUP 666.  

HH~1149 is a series of partial and complete bow shocks. 
The CCW boundary is ill defined but probably is due to the outflow 149.1-352.0 jet that originates in COUP 666. Although the well defined interior bow shocks also seem to arise from COUP 666, the most CW boundary complex bow shock is oriented at too small a PA. The HH~1149-Mid linear feature aligns with the HH~528 Base feature and may give rise to the HH~528-West flow. Its origin may be in the Blank-East region.

The above information means that COUP 666  appears to have SE outflows towards HH~203b (123\arcdeg), HH203a (124\arcdeg), HH~204a (127\arcdeg), the linear jet at 160\arcdeg, the CCW boundary of HH~1149,
and crenellation features at 127\arcdeg --135\arcdeg\ and 140\arcdeg --160\arcdeg.
 In addition, COUP 666 is likely to be producing the multiple bow shocks within HH~1149.  
 This means that COUP 666 has multiple outflows over 123\arcdeg --160\arcdeg and if the SW-Group of features are included, to 184\arcdeg. 
 This means that COUP 666 has produced a lobe of collimated outflows to the SE.
If COUP 666 is also the source of the HH~202 shocks, then this means that it is producing shocks to the NW at 296\arcdeg\ to 327\arcdeg. If the SW-Group of features is not included with the COUP 666 features, the range of angles of the NW flow (296\arcdeg\ to 327\arcdeg) is very similar to the SE flow (123\arcdeg\ to 160\arcdeg), with the center of the ranges in nearly reciprocal directions (142\arcdeg\ and 312\arcdeg).
 
COUP~632 is producing the outflow HH~528-East (142\arcdeg), which eventually connect with HH~400 (147\arcdeg) and HH~400-S (148\arcdeg).

These considerations lead us to the important conclusion that COUP 666 is producing multi-fingered lobes of bipolar flow, with the lobes pointed NW and SE. 

All of the measured components of the two sides of the bipolar flow are blue shifted. This is probably because the outflows are bipolar over a wide range of directions in space, although concentrated to the NW and SE. We see this motion in the plane of the sky as a pair of fans of shocks over multiple directions. The red shifted components of the outflow will be directed to behind the MIF and because of the jump in extinction that is expected upon passing through the PDR, the blue shifted components are not seen optically. 

We see a similar pattern in the Allen-Burton fingers arising from the BN-KL explosive event. There is a rich literature of studies of the Allen-Burton fingers and the BN-KL source \citep{hu96,schu99,kai00,lee00,van01,gus03,kris03,ode03a,doi04,col08,ode08a,fuk09,fur09,pla09,zap09,bal11,zap11,zap12}. The Allen-Burton fingers are simply the NW portion of a multi-element bipolar flow, with a less well characterized optically invisible similar flow\citep{mcc97} to the SE. 
These fingers are the result of a single event estimated to have occurred 500 years ago (by the tangential motions of the radio sources involved \citep{rod05}) or no more than 1000 years (by the tangential velocities of the optical shocks \citep{doi02}). 

Unlike the Allen-Burton fingers, it is not likely that the Northwest and Southeast features discussed here arose at the same time and from a single event. We see only one driving jet for each of the two best defined shock systems (HH~202 and HH~203). It appears more likely that we are seeing the result of collimated bipolar flows occurring at different times and directions.

\begin{figure}
\epsscale{1.0}
\plotone{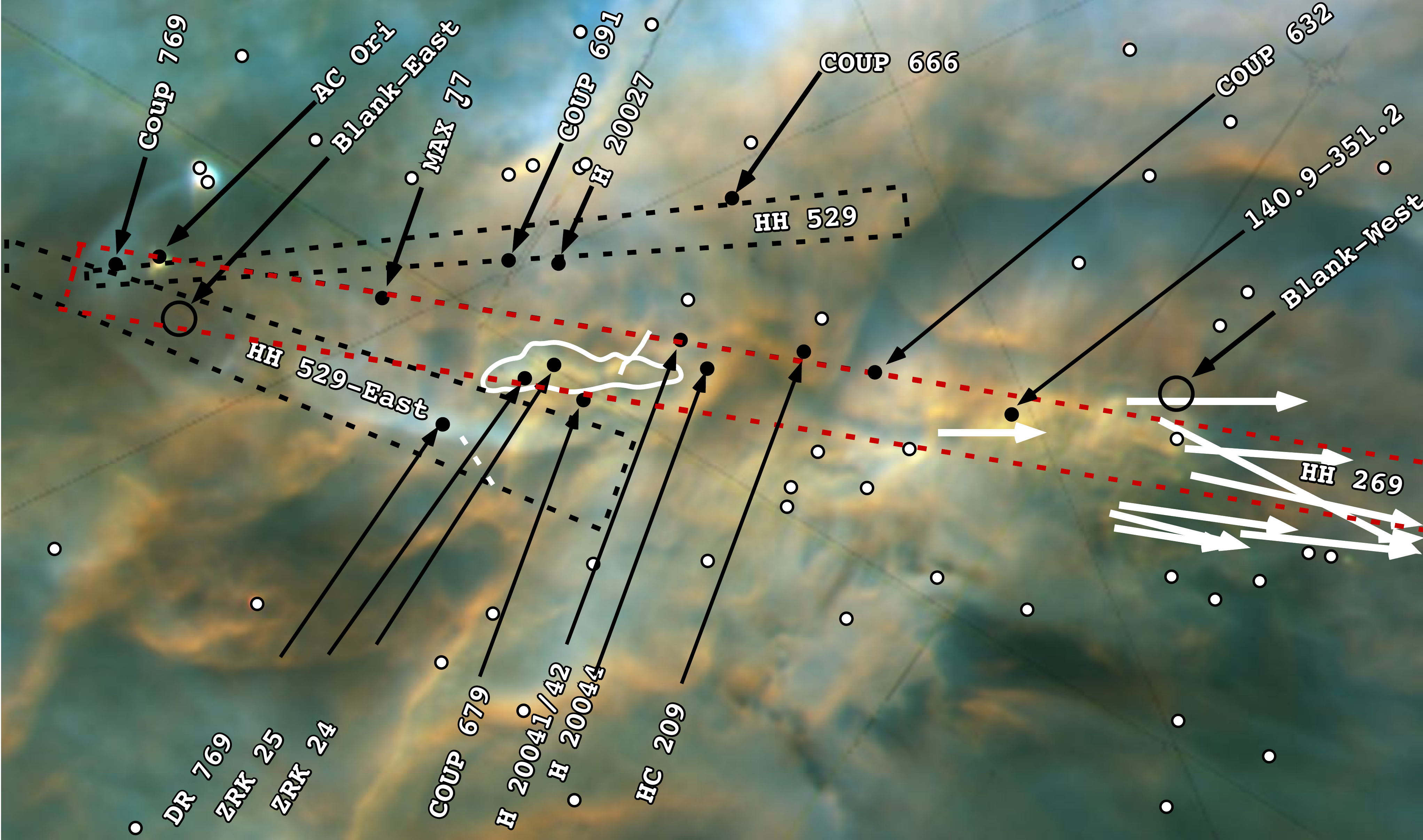}
\caption{Like Figure~\ref{fig:fig43} except now the colored lines discussed in the text are those relevant to understanding the east-west flows.}
\label{fig:fig45}
\end{figure}

\subsection{Nature and Origin of the HH~269 and the HH~529 East-West Shocks}
\label{sec:269529}

\subsubsection{HH~269}

The westward moving HH~269 feature has been shown in this paper (Section~\ref{sec:HH269}) to begin with the feature HH~269-East and to extend beyond the new HH~269-Westmost features to join with \htwo\ knots even further west. These features are of low ionization, although \oiii\ is sometimes seen, and a series of small shocks to the east of HH~269-East are likely to be caused by the same invisible flow at 276\arcdeg\ (Section~\ref{sec:HH269}). Their motion is shown by the white arrows in Figure~\ref{fig:fig45}. The westernmost velocity vector is important since it indicates that the source lies to the east of 140.9-351.2. When radial velocities are available, their are all blue shifted . In Figure~\ref{fig:fig45} we show a band $\pm$1\arcsec\ from the path of the well-defined HH~269 axis as parallel dashed red lines. The source of this flow is most likely to occur within the central region, which contains stars
H 20044, HC 209, ZRK 24, ZRK 25, MAX 77, AC Ori, and COUP 769.  Detection in multiple wavelength ranges is a good indication of a very young star. As shown in Table~\ref{tab:SIMBAD}
only two are seen outside of a single wavelength range. AC Ori is seen in all wavelength bands and COUP 769 is seen in the X-ray \citep{get05} and infrared \citep{lad04}, these are the best candidate sources. 

\subsubsection{Outer Shocks Possibly Associated with HH~269}

\citet{hen07} have shown that there are a large number of shocks near the edge of the optical EON that lie west of HH~269. They indicate that an association with HH~269 is unlikely unless a strong change of PA occurs. The shocks are found from about 256\arcdeg\ to 283\arcdeg\ from the OOS.  This includes the PA of 276\arcdeg\ for HH~269, but there are no outer shocks at that angle. 

\subsubsection{Origin of the HH~529 shocks}
\label{sec:origin529}

In this study we have shown that in addition to the HH~529-I, HH~529-II, and HH~529-III high ionization shocks (axis 109\arcdeg) there is an apparently independent shock we have designated as HH~529-East (axis 84\arcdeg). In Figure~\ref{fig:fig45} we show trapezoidal boxes that begin with the center of the defining shock and diverge by $\pm$1\arcdeg. These are not extended beyond the concentration of known stars. 

The box for HH~529 (white dashed lines) contains no stars, although stars 
COUP 666, H 20027, 
COUP 691, AC Ori, and COUP 769 fall on the edge of the box. 
COUP 666 and AC Ori are seen at all wavelengths, H 20027 and COUP 691 only in the infrared, and COUP 769 in X-rays and the infrared. 
The star COUP 632 identified earlier by \citet{riv13} on the basis of extrapolations from a large distance falls well outside the box for candidate stars.

The three stars seen in more than one wavelength window are the best prospects for the source.
COUP 666 has already been invoked as the most likely source of the large-scale SE flows. If this is the source of the HH~529 shocks, then this would be simply an extension of the PA range of its outflow.  AC Ori has a jet with orientation ( 47\arcdeg) \citep{ode08a}, which does not align with HH~529-East's symmetry axis of 109\arcdeg, making it a less
likely source. COUP 769  is well located within the source, but the nearby series of moving knots headed at 82\arcdeg\ argues against an association.

These considerations leave COUP 666 and AC Ori as the prime candidates. The HH~529-I shock is fully developed and AC Ori is centrally located within the shock and slightly offset. These considerations, in addition to the non-alignment with the  known AC Ori jet means that COUP 666 is the most likely source. This star is not deeply imbedded (we see it as an optical star) and that is attractive because the HH~529 shocks are high ionization and are formed within the \oiii\ region of the nebula \citep{bla06}.  Placing the source (COUP 666) in front of the MIF explains the bow shock's ionization.

\subsubsection{Origin of the HH~529-East shock}

The HH~529-East high ionization shock's potential source box (again with $\pm1$\arcdeg\ divergence and as black dashed lines) contains only the inferred source Blank-East (159-355) that has no known SIMBAD stars and the millimeter source \citep{dar09} 
DR 769. The star COUP 769 falls on the edge of the box and as noted previously 
a line of 82\arcdeg\ from the center of the defining bow shock through several west moving intervening high ionization knots passes indistinguishably close to COUP 769 (seen in X-rays and the infrared) and the Blank-East center. 

This means that HH~529-East can be explained either by collimated flows from either COUP~769 or a hidden source in Blank-East. Since no source is seen in Blank-East, any source there must be highly embedded and it would be difficult to explain the high ionization of HH~529-East. This leaves COUP~769 as the most likely source for HH~529-East.  

\subsubsection{Summary of the Likely Sources of the East-West Systems}
\label{sec:EastWestSummary}

HH~269 has as its best candidate sources AC Ori and COUP 769. The lack of alignment of the reciprocal of the AC Ori jet (227\arcdeg) and the HH~259 axis (276\arcdeg) means that by default COUP 769 is the more likely source. 

HH~529 is mostly likely driven by a collimated outflow from COUP 666. This would extend the PA range of SE shocks driven by this star. The different nature of these shocks (high ionization) indicates that this region of the outflow quickly moves into gas ionized by \tc.

HH~529-East is most likely driven by COUP 769. The knots seen nearby at 82\arcdeg\ and the HH~529-East shock (84\arcdeg) are the strongest arguments. 

The above conclusion means that HH~529 is simply one more flow from COUP 666, whereas COUP 769 is now revealed as the most likely source of HH~269 (276\arcdeg, reciprocal 96\arcdeg) and a counter flow (HH~529-East, 84\arcdeg).  If both of these flows originate from the same source as a nearly bipolar flow, it must mean that the the opposite flows are not exactly aligned. The alternative is that even for COUP 769 we are seeing a multi-branched outflow. Unfortunately, there are no radial velocities for HH~529-East. If it is blue shifted, like the HH~269 features, then this strengthens the argument for a multi-flow coming from COUP 769.

\subsection{Nature and Origin of HH~1132 and HH~1141}
\label{sec:EastAndSecond}

\begin{figure}
\epsscale{1.0}
\plotone{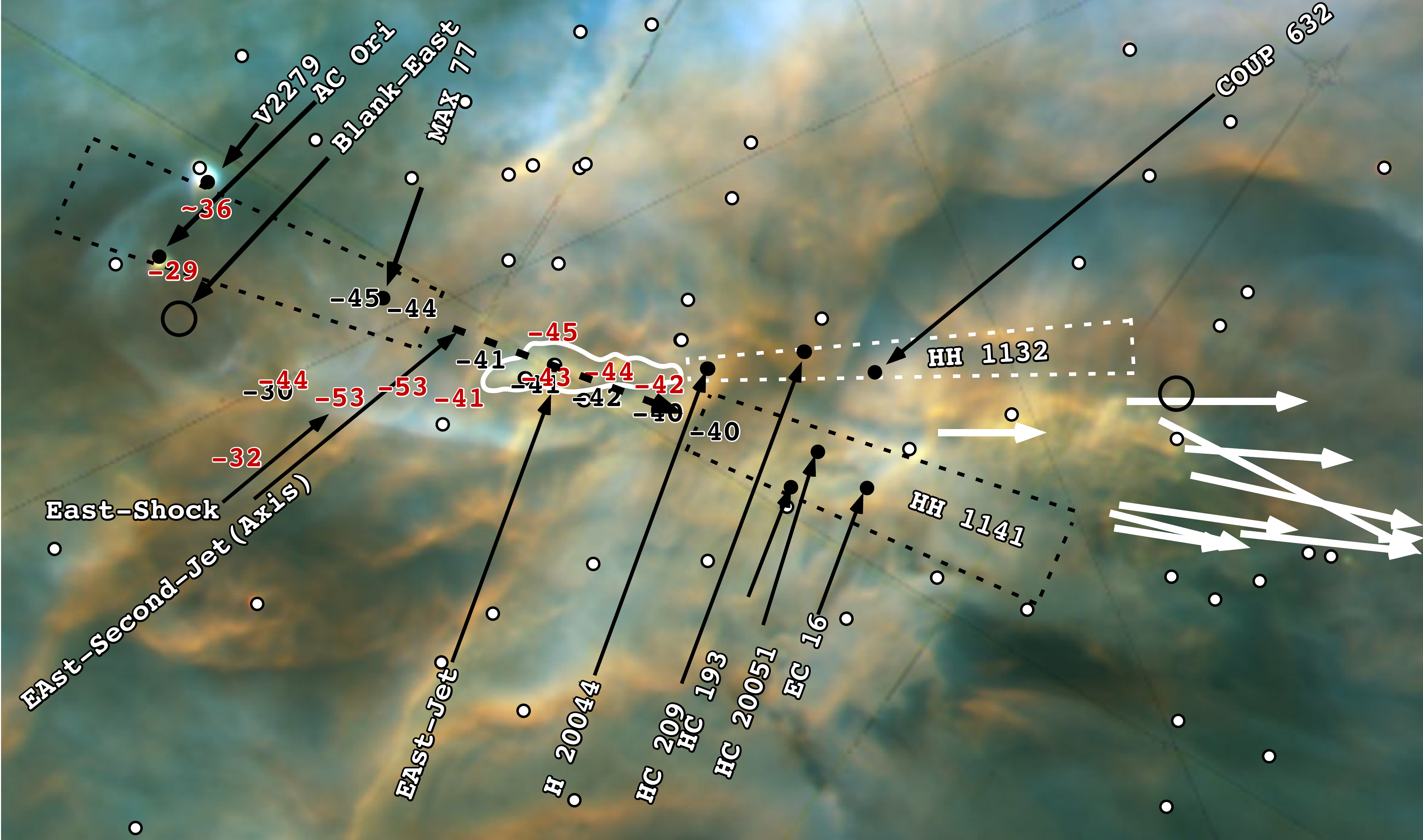}
\caption{Like Figure~\ref{fig:fig43} except now the black and white features discussed in the text are those relevant to understanding the East-Jet and East-Second-Jet flows. The black numbers indicate radial velocities in \nii\ and the red numbers indicate radial velocities in \oiii.}
\label{fig:fig46}
\end{figure}

In Figure~\ref{fig:fig46} we see both the East-Jet and the axis of the East-Second-Jet (Section~\ref{sec:OOSeast} as extracted from Figure~\ref{fig:fig14}. The axis (107\arcdeg --108\arcdeg) of the east moving East-Jet projects into the white trapezoidal box with $\pm$2\arcdeg\ divergence. The East-Second-Jet  (axis 87\arcdeg) is less well defined by radial velocities, so we have projected its trapezoids (divergence$\pm$3\arcdeg)
in both directions. 

\subsubsection{The East-Jet (HH~1132)}
\label{sec:EastJet}

The alignment of the East-Jet argues that the source of this jet is either HC 209, H 20044 or the more distant COUP 632. 
There are no other nearby compact sources lying within the small uncertainty of the angle of the jet.
 H 20044 lies at the end of the East-Jet form and its radial velocity features, while objects HC 209 and COUP 632 become candidates as one moves further to the west. HC 209 and H 20044 are relatively faint and seen only in near-IR surveys \citep{rod09}. Although H 20044 is listed in the early X-ray catalog of Orion sources \citep{fei02}, it does not appear in the later COUP \citep{get05} catalog.
  
 In contrast with the other candidate stars, COUP 632 is a luminous source seen in the full energy range of surveys \citep{rob05,get05,mlla,smi04,lad00,lad04}. 
COUP 632's characteristics have been summarized well in appendix A9 of \citet{riv13}. \citet{riv13} establish that it suffers high extinction in the infrared, even more in the X-ray, and represents emission from a star of 0.5--1.5\Msun, most likely being near the larger value. These characteristics make it the most attractive candidate source for the East-Jet. 

\subsubsection{Features Related to the East-Jet}
\label{sec:EastJetShock}

As seen in Figure~\ref{fig:fig46} there is a high ionization partial bow shock whose apex lies 5.8\arcsec\ along an extension of the East-Jet axis. This is labeled East-Shock in Figure~\ref{fig:fig14}.
Its tangential motion is difficult to determine but one sees in Figure~\ref{fig:fig14} that lies along the projection of the East-Jet.
It has numerous high velocity \oiii\ features that have about the same velocity as the East-Jet. The high ionization is consistent with the transition of ionization as the driving jet emerges. There are no obvious related features further along the line of projection.

Therefore, we have a sequence of features arising from COUP 632. As the collimated flow breaks through an ionization front (either the MIF or the ionization front of the Orion-S cloud, as discussed in Section~\ref{sec:OOSeast}) it forms the high tangential and radial velocity features and finally produces the East Shock. This sequence can be designated as HH~1132.

\subsubsection{The East-Second-Jet (HH 1141)}
\label{sec:SecondJet}
HH~1141 is defined only by \nii\ radial velocities. Since there are no associated tangential velocities we do not know if it is moving east or west and the trapezoids probably containing its source star are shown projected in both directions from the feature.

If it is moving west, there are three candidate sources MAX 77 (infrared only) , AC Ori (X-rays, visual, infrared, radio continuum), and V2279 (X-rays, visual, infrared). It does not align with the microjet (47\arcdeg) from AC Ori and V2279 actually falls slightly outside the possible source trapezoid. MAX 77 is the only good candidate, especially as the \nii\ high velocity features begin at that point. 

If it is moving east there are again three candidate sources HC~193(infrared), HC 20051 (infrared), and EC 16 (infrared). All are relatively weak sources seen only in the infrared, with HC 20051 and EC 16 being more favorably located. 

This means that the only good candidate source is MAX 77 and the East-Second-Jet features are expected to be moving to the west. We can define the new HH object HH~2014ESJ as the source, MAX 77 and the seven \nii\ radial velocity samples to its west. The axis of this flow is drawn as a heavy black vector in Figure~\ref{fig:fig46}.

\subsection{An outer flow (HH 1139)}
\label{sec:outershock}
In Figure~\ref{fig:fig44} we show most of the boundary of the Extended Orion Nebula. One of the best defined shocks is labeled as ``256\arcdeg\ Shock'' in this figure. It aligns well with the bright (V=12.57) star COUP 9. The spectral type of COUP 9 has been reported as spectral type K1, K0IV, K3III-IVe, and K1 \citep{hil13}. It is seen in X-rays, optical, and near IR wavelengths. If this is the source producing the 256\arcdeg\ shock. We designate the COUP 9 + 256\arcdeg\ shock as HH 1139.

\subsection{Relation of Huygens Region Outflows to 21-cm Absorption in the Veil}
\label{sec:21cm}

In a recent study of high velocity HI absorption lines in the Veil (angular resolution about 6\arcsec), \citet{pvdw} detected several regions of blue shifted material that are probably related
to the large-scale outflows detected at the boundaries of the \hr. 

Their broad (about 60\arcdeg) velocity system L is centered on the HH~203 and HH~204 shocks and extends SW to the end of the HH~528 shocks. The fan of shocks identified as part of HH~1149 overlaps in PA with feature L, but HH~1149's most CW PA coincides with HH~203's jet and its CCW PA limit goes beyond feature L. The radial velocity of the L feature is 15 \kms, quite different than that of HH~203 (-46$\pm$15 \kms) \citep{doi04} and HH~204 (-18$\pm$18 \kms) \citep{doi04}. These shocks are probably interacting with the foreground Veil, but the neutral material is in a mass-loaded region ahead of the shocks.

Their system F was found around the NW boundary of the series of shocks that include HH 202 and the width of the feature coincides with the NW flows identified in Section~\ref{sec:HH202} and Figure~\ref{fig:fig23}. System F has the partial shell structure expected from a series of shocks striking and accelerating the foreground Veil.  Feature F's HI absorption radial velocity (about \vrad =-1 \kms, contrasts with the optical shock feature velocity of -39$\pm$2 \kms \citep{doi04} again indicating the the HI absorption comes from a decelerated region ahead of the shock.

 Samples taken on van der Werf's features C (lying to the SW within the \hr), D, and G show no optical features in \nii\ or \oiii\ that resemble the HI absorption features. This indicates that these HI features are not related to the optical features we see nearby. 
 
\subsection{Relation of the Southwest Shock to the Orion-S Region}
\label{sec:Biggies}

In Figure~\ref{fig:fig44} we see that the ``Southwest Shock'' is at 246\arcdeg\ from the molecular outflow from EC 14. This direction is only slightly less than the direction of the outflow, which begins with   245\arcdeg\ and changes to  238\arcdeg\ after deflection. Perhaps the deflection began later than the jet driving the ``Southwest Shock". If this is the case, then it is noteworthy that the well defined shock 108-430  (c. f. Figure~\ref{fig:fig23})
lies at  247\arcdeg\ from EC 14 and this shock may be a result of the same outflow as the ``Southwest Shock''. \citet{bal00} assigned this shock to the group called HH~530 close to the silhouette proplyd 114-426, but concluded that it was likely to be driven by a source in Orion-S. We are able to reach the same decision, but with more accuracy as the outflow from EC 14 has been only recently been defined in detail  \citep{zap10}. 

\subsection{Flows Without Candidate Sources}
\label{sec:blanks}

There are three cases in which there seem to be flows without an apparent source. In the first two the angle on the sky back to the sources is small.  In the third a point of origin is not obvious. The first is in the West region and appears in Figure~\ref{fig:fig11} and Figure~\ref{fig:fig19}, the second is near AC Ori (Figure~\ref{fig:fig19}). It is remarkable that the features are most visible in \oiii, which must lie further from the MIF than the \nii\ layer, making it harder to conceal the source. The third feature is HH~1154 (Figure~\ref{fig:fig38}), which is quite different from the first two.

\subsubsection{Flows Originating Within the Dark Arc}
\label{sec:blanksDarkArc}
We note that in the West F502N motion-image (Figure~\ref{fig:fig11}) there are three groups of nearly circular moving arcs, the first lying NNE of compact source EC 9, the second lying ENE of that same object, and the third to the NNW. Their direction of motions are shown by the thin dashed white arrows in Figure~\ref{fig:fig11}. Their point of intersection (5:35:13.76 -5:23:48.8, with an uncertainty of about 1\arcsec) clearly lies north of source West-A and there is no candidate for a common source. We will call this position Blank-West. Although there are large arcs in WFPC2 and WFC3 images due to fringing in the filter-detector combination, these occur at a much larger scale and cannot produce the small-scale fringes that we see. The better-defined north series is moving about 16 mas/yr, corresponding to 34 \kms. If they arise from a source near the intersection, this time was about 300 years ago. This structure is unique. It is unlikely to be a light-echo from a long period star because it shows up only in the \oiii\ emission. It may be that it is the result of varying intensity large-scale pre-stellar wind from an object whose obscuring disk is aligned with the direction to the observer.

\subsubsection{Flows Originating From Near AC Ori}
\label{sec:blanksAC}

In Section~\ref{sec:HH998} we presented the arguments that the earlier conclusion of \citet{ode08a} linking the HH 998 shock feature to a putative counter jet from AC Ori was possibly incorrect and that COUP 769 had to be considered as a candidate source. Neither one is a good explanation and we discuss here other features that share a common origin with the HH~998 shocks. 

There is another well defined shock in the vicinity that is identified as 156-352 in Figure~\ref{fig:fig19}. Lines drawn perpendicular to the motion of the 156-352 shock and along the symmetry axis of the HH~998 shock intersect at 5:35:15.91 -5:23:54.5 with an uncertainty of less than 1\arcsec. We will designate this as Blank-East.
There are no sources in the SIMBAD catalog at this position. 

Other, more distant shocks may also originate from this source-free area. The well-defined shocks at 155-354  lie at 
5.9\arcsec\ along  277\arcdeg, 148-346  14.9\arcsec\  at 265\arcdeg, and 153-359 9.4\arcsec\ 24.2\arcdeg\ (essentially on an extension of HH~998). All of these show symmetry and motions along a line originating in the source-free area.  Their radial velocities are uncertain since they lie along an extension of the East-Jet and the local numbers near 155-354 are similar in value to those in the East-Jet.

In Section~\ref{sec:HH998} we discussed the fact that there are a series of faint moving knots 8.3\arcsec\ east of the obscured star COUP 769 and are shown as a dashed vector in Figure~\ref{fig:fig19}. Although their axis passes close to COUP 769, it also passes into the 
empty source region identified above and may also arise from there. We call these HH~1152.

In addition to the above multi-angle flows, we have shown in Section~\ref{sec:final1149} that the linear features HH~1149-Mid and HH~528-West may also arise from Blank-East. If so, it is one of the most elaborate regions of outflow in the \hr.

If the HH~998 shock arises from the undetected source, it is about 145 years
old and if the 156-352 shock arises from the same region, it is about 450 years old. 
The other shocks would be much older (155-354, 2700 yrs; 148-346, 4600 yrs; 153-359, 2200 yrs). These numbers indicate that collimated outflow in different westerly directions are occurring essentially at the present and began at least about 4600 years ago. This source or sources have undergone collimated ejections in two quite different directions about 300 years apart. These properties are not unlike other sources found in the \hr, but the big difference is that the source or sources have not been detected on any studies. 

\subsubsection{HH~1154}
\label{sec:HH1154disc}

The curious object HH~1154 has the highest \Vomc\  (200 \kms) of all the \hr\ objects. It is only 5.2 \arcsec\ from the 
proplyd 139.2-320.3 (COUP 593), which has been detected in X-rays through infrared radiation. But, the most straightforward explanation of the change of shape and position of HH~1154 indicates that it is moving exactly towards the proplyd. There are no candidate sources along the axis of motion. If we have interpreted the two epoch images correctly, the source of HH~1154 has not been detected.

\subsection{Types of Flows}
\label{sec:individuals}

Even though this study has analyzed in detail only a fraction of the \hr, it does include most of the core of the ONC. 
This makes it possible to consider this a representative sample of outflows from young stars in a rich cluster. The results 
can be generalized to similar clusters, commonly thought to be the primary source of field stars. In this section we categorize the outflows of different types and where possible identify their sources. We have divided them into five categories: Monopolar, Bipolar, Multi-Direction, Unassigned, Hot Star in Table \ref{tab:FLOWS}. In each case we give the section(s) where the flow is discussed, the figure(s) where it is presented, and the candidate source. In some cases a flow appears in more than one category. Two previously known flows (HH~512 and HH~888) that have been studied in detail elsewhere are included in Table~\ref{tab:FLOWS} for completeness. 

There are 27 monopolar flows, 21 have been discovered in this study and two of the previously known flows are reassignments of portions of previously known flows. There are 9 bipolar flows.  One is a combination of a well known flow plus a new feature, two are newly recognized portions of a known flow, and three are newly discovered flows. There are nine multidirection flows, four of them are newly discovered and the others are known flows or reclassified known flows. Three flows are unassigned, one of which was known before. 

A  Monopolar flow may have an aligned counterpart that is aimed in a direction where there is insufficient material to produce a optically visible shock. Where we have good radial velocities for a Monopolar feature, with the exception of HH~518, they are blueshifted.  This argues that the redshifted flow penetrates behind the obscuring PDR of the MIF, or that the flows are moving into the dense material of the foreground Veil, with the redshifted flow passing into the lower density regions within the cavity lying between the Veil and the MIF.  All theories of star formation call for a polar ejection in opposite directions, so the Monopolar flows are probably the result of these selection effects.

Some of the Monopolar and Bipolar flows may actually be parts of Multi-Direction flows. Those objects listed in that category in Table~\ref{tab:FLOWS} have identifiable flows in more than one nearby direction in the plane of the sky and where there are radial velocities, they are both blueshifted.  The detection of paired blueshifted or redshifted flows in opposite components of the Bipolar flows would move them into the Multi-Direction category. 
It is worth noting that the SE redshifted component of the HH~726 flow from LV~2 shows a range of angles of its outer shocks from 117\arcdeg\ to 122\arcdeg\ even though the flow is very linear at 120\arcdeg\ in its inner region.

In summary we can say that observational selection effects may account for the Monopolar objects not having the Bipolar features expected theoretically. Adding these two categories (Monopolar and Bipolar) yields 32 systems of flows, where there are nine Multi-Direction flows. More complete data may move some of the former flows into the latter. 

This gives the surprising result that Multi-Direction flows represent a significant (about one-fith) of all known flow systems in this study and correction for the observational selection effects can only increase this fraction. This means that in addition to the well-known intermittency of collimated outflows, there must be either multiple unresolved stars forming with different orientations in space, but at essentially the same time, or effects are present that cause the collimated jets to have rapidly varying directions. In private correspondence John Bally points out that some of these multi-shock features could be due to wide-angle outflows (like in HH~626) undergoing instabilities or having had to punch through obstructing media.

\subsection{The Shocks Near the Trapezium in Context.}
\label{sec:NonHH}

In section \ref{sec:theta1C} we saw that there are multiple shocks to the east, west, and north of the Trapezium.
Most of these can be assigned to designated HH flows, although the sources are sometimes ambiguous. However, there are two sets of multiple shocks lying almost parallel  to one another, the Northern Shocks and the Western Shocks.  The individual shocks in both sets are oriented away from \tc.  Either \tc\ has had multiple collimated 
outflows at various PA values but all about the same time, or the more likely explanation, these shocks are at instabilities in zones of interaction of a large-scale wind from \tc\ with the ambient nebula gas.

\tc\ is known to be a triple system with component masses of 36 \Msun, 8 \Msun, and 1 \Msun~\citep{vit02,kra09,leh10} and the strong stellar characterstics of very hot stars \citep{how89}. This means that there is ample opportunity for instability produced phenomena in several directions. 
 
These two sets of shocks are very different from structures seen further away from \tc\ that arise from the large-angle stellar wind from that star. They are much closer than the "\oiii\ Shell`` discussed in \citet{ode09}, where it is argued that this marks a boundary of material shocked by the high 
speed stellar wind from \tc. This is in agreement with arguments presented in \citet{mas95}. None of these shock structures coincide with the larger but inner-nebula C-shaped high ionization zone identified by \citet{ode09}. this C-shaped feature
is open in the 
direction (SW) of the Orion-S cloud, and has an irregular radius of about  34\arcsec, and was designated there as the \oiii\ Shell \citep{ode09}.   It may be that this feature is actually circular as projected on the plane of the sky, rather than open, because Orion-S obscures it in the SW. It is clearly an ionization phenomenon since where \oiii\ is strong, \nii\ is weak, but it is not an ionization boundary because there is no enhancement of \nii\ on the outside.  \citet{ode09} argue that inside the \oiii\ Shell is a region of lower gas density that results from the high speed stellar wind from \tc. Sample B (Section~\ref{sec:HH1149} is our only spectroscopic sample of high S/N ratio for the \oiii\ Shell. Its \vrad(\oiii)=18.6 \kms\ while our other samples average to 15.5$\pm$2.6 \kms\ and \citet{gar08} found a \hr\ average of 16.3$\pm$2.8 \kms. Our 
Sample-B gave \vrad(\nii)=22.5 while our other samples average to 20.0$\pm$1.9 \kms\ and \citet{gar08} found a \hr\ average of 20.5$\pm$2.9 \kms. The small differences of velocity (one sigma high in both \oiii\ and \nii) of the \oiii\ shell relative to the ambient gas agrees with the models presented in \citet{ode09}. The \tc\ collimated outflow shocks are generated within this inner zone of high ionization. They are much more visible in \oiii, but some do have \nii\ present on the 
concave side of the shocks.

Even further there is a high ionization feature similar to the \oiii\ shell in terms of its ionization properties. \citet{ode09} divide it into two portions a curving bright \oiii\ feature that they call the Big Arc East (to the SE from \tc)  and another longer and linear feature (the Big Arc South) beginning at the CW end of the Big Arc East and running nearly east-west. The portion of the latter directly south of \tc\ is at 69\arcsec from \tc. Unlike the \oiii\ Shell, there is a region of \nii\ enhanced emission associated with both the Big Arc and Big Arc East features and the \nii emission peaks further from \tc, which is consistent with the expectations of photoionization. 

In summary we can say that while the inner systems of shocks must be related to a recent period of high velocity wind  from \tc, the much larger \oiii\ Shell was formed by an earlier period of a similar wind. \tc\ is well known to have strong variations in its spectrum and derived radial velocities. This makes it more plausible to interpret these
two features as caused by two different periods of intense wind. The Big Arc South and Big Arc East features are likely to
be caused by an even earlier period of strong stellar winds from \tc. 

As noted in Section~\ref{sec:theta1C}, there are also shock features aligned with $\theta^1$~Ori~D. 156-256 at 34.5\arcsec\ and 309\arcdeg\ (plus a more symmetric shock 3\arcsec\ outside it).
$\theta^1$~Ori~D is quite different from \tc\ in that at  spectral type O9.5V it is much cooler and the star lies close to the MIF. It is close to the Nye-Allen infrared source and shows unusually strong He I 388.9 nm absorption \citep{ode93}. \tc\ is a much hotter spectral type O6V star and lies about one-fourth parsec in front of the MIF. $\theta^1$~Ori~D is known to be a binary \citep{gre13}. If a collimated flow from $\theta^1$~Ori~D is real, it is the first known from a star in this temperature range.

\placetable{tab:FLOWS}
\begin{deluxetable}{lccl}
\tabletypesize{\tiny}
\tablecaption{Summary of Data on Flows\label{tab:FLOWS}}
\tablewidth{0pt}
\tablehead{
\colhead{Designation in Text} &
\colhead{Section} & 
\colhead{Figure} &
\colhead{Source}}
\startdata
\multicolumn{4}{c}{Monopolar Flows} \\ \hline 
$\mathbf{HH ~507}$ &\ref{sec:HH507} & \ref{fig:fig41} & Unknown, many candidates\\
$\mathbf{HH ~512}$ & Bally et al. 2000, O'Dell $\&$~Henney 2008a&O'Dell $\&$~Henney 2008a, figure 14& COUP 728\\ 
$\mathbf{HH ~528-East}$& \ref{sec:HH528}, \ref{sec:final1149}, \ref{sec:outer} & \ref{fig:fig23}&COUP 632?\\
$\mathbf{HH ~528-West}$& \ref{sec:HH528}, \ref{sec:final1149}, \ref{sec:outer} & \ref{fig:fig23}&Blank-East?\\
$\mathbf{HH ~529}$&\ref{sec:HH529}, \ref{sec:sources}, \ref{sec:origin529}&\ref{fig:fig16}, \ref{fig:fig17}, \ref{fig:fig19},\ref{fig:fig45}&Probably COUP 666\\
$\mathbf{HH~530}$&\ref{sec:HH530redefined}& \ref{fig:fig34}&EC 14\\
$\mathbf{HH~ 625}$&\ref{sec:HH625},\ref{sec:HH625source}&\ref{fig:fig25},\ref{fig:fig26}& COUP 555\\
$\mathbf{HH~1127}$&\ref{sec:1127}& \ref{fig:fig36}, \ref{fig:fig37}&MAX 46 or the more distant COUP 602\\ 
$\mathbf{HH~ 1128}$&\ref{sec:MotionsDarkArc}&\ref{fig:fig36}& EC 9\\ 
$\mathbf{HH~1129}$&\ref{sec:HH1149}, \ref{sec:final1149}, \ref{sec:sumNWSE}&\ref{fig:fig19}, \ref{fig:fig20}, \ref{fig:fig43}& COUP 666\\ 
$\mathbf{HH~1130}$&\ref{sec:1150}& \ref{fig:fig41}& COUP 582\\ 
$\mathbf{HH~1131}$&\ref{sec:HH1131}&\ref{fig:fig41}&COUP 480\\ 
$\mathbf{HH~1132}$&\ref{sec:OOSeast},\ref{sec:EastJet}&\ref{fig:fig12}, \ref{fig:fig13}, \ref{fig:fig14}, \ref{fig:fig15}, \ref{fig:fig16},\ref{fig:fig46}&COUP 632\\ 
$\mathbf{HH~1133}$&\ref{sec:HH1133}& \ref{fig:fig39}& HC 236\\ 
$\mathbf{HH~1134}$&\ref{sec:NorthOfTc}& \ref{fig:fig40}& $\theta^1$~Ori~E\\  
$\mathbf{HH~1135}$& \ref{sec:HH1135}& \ref{fig:fig39}& EC 17 or H 20018\\ 
$\mathbf{HH~1136}$& \ref{sec:HH1136} & \ref{fig:fig39}& COUP 717\\ 
$\mathbf{HH~1137}$&\ref{sec:HH1137} and \ref{sec:SumOfThree}& \ref{fig:fig39}&LV~4\\ 
$\mathbf{HH~1138}$& \ref{sec:NorthOfTc}& \ref{fig:fig40}&HC 341\\ 
$\mathbf{HH~1139}$&\ref{sec:outershock}&\ref{fig:fig44}&COUP 9\\ 
$\mathbf{HH~1140}$&\ref{sec:nearHH507}&\ref{fig:fig41}&COUP 478\\ 
$\mathbf{HH~1141}$ &\ref{sec:OOSeast}, \ref{sec:SecondJet}& \ref{fig:fig14},\ref{fig:fig46}& MAX 77\\ 
$\mathbf{HH~1142}$ &\ref{sec:HH998}, \ref{sec:blanksAC}, \ref{sec:HH1149}, \ref{sec:final1149}, \ref{sec:HH528}, \ref{sec:outer}& \ref{fig:fig19}, \ref{fig:fig20}, \ref{fig:fig23}, \ref{fig:fig44}& Blank-East\\ 
$\mathbf{HH~1143}$& \ref{sec:NorthOfTc}& \ref{fig:fig40}&COUP 900\\ 
$\mathbf{HH~1144}$& \ref{sec:NorthOfTc}& \ref{fig:fig40}&HC 271\\ 
$\mathbf{HH~1145}$&\ref{sec:HH1145} and \ref{sec:SumOfThree}& \ref{fig:fig39}&COUP 747\\ 
$\mathbf{HH~1146}$&\ref{sec:HH1146}and \ref{sec:SumOfThree}& \ref{fig:fig39}&HC 292\\ 

\multicolumn{4}{c}{Bipolar Flows} \\ \hline 
$\mathbf{HH~269 + HH~529-East}$&\ref{sec:HH269}, \ref{sec:motions}, \ref{fig:fig8}, \ref{sec:HH529}, \ref{sec:origin529}, \ref{sec:269529}&\ref{fig:fig5}, \ref{fig:fig6}, \ref{fig:fig7}, \ref{fig:fig16},  \ref{fig:fig17}, \ref{fig:fig18}, \ref{fig:fig19}, \ref{fig:fig20}, \ref{fig:fig45}& Most likely source, COUP 769\\
$\mathbf{HH~540}$&\ref{sec:SWSF}& \ref{fig:fig24}&Most likely source Beehive Proplyd 181-826\\
$\mathbf{HH~510}$ &\ref{sec:HH510}&\ref{fig:fig29}, \ref{fig:fig30}, \ref{fig:fig31}, \ref{fig:fig32},\ref{fig:fig33}& edge-on dark proplyd 109.4-326.7\\
$\mathbf{HH~1157}$&\ref{sec:HH1157}& \ref{fig:fig34}& V2202\\
$\mathbf{HH~726}$& \ref{sec:eastoftc}& \ref{fig:fig40}&LV~2\\ 
$\mathbf{HH~888, LL 1}$&\ref{sec:individuals}& \ref{fig:fig44}& LL Ori--COUP 245\\
$\mathbf{HH~1147}$&\ref{sec:13574082objects}& \ref{fig:fig41}&Likely source EC 13\\ 
$\mathbf{HH~1148}$& \ref{sec:HH1148}&\ref{fig:fig2}, \ref{fig:fig42}&Undefined, too many candidate sources\\ 

\multicolumn{4}{c}{Multi-Direction Flows} \\ \hline 
$\mathbf{HH~202 + HH ~203 ~and ~HH ~204}$& \ref{sec:HH202}, \ref{sec:HH2034}, \ref{sec:SEbig}, \ref{sec:crenellations}, \ref{sec:Biggies}& \ref{fig:fig22}, \ref{fig:fig23}, \ref{fig:fig24}, \ref{fig:fig43}, \ref{fig:fig44}&Most likely source COUP 666\\
$\mathbf{HH~518a~ \&~ HH ~518b}$& \ref{sec:HH518}& \ref{fig:fig22}, \ref{fig:fig39},  \ref{fig:fig40}& V2279\\
$\mathbf{HH~626}$& \ref{sec:HH626}&\ref{fig:fig35}& V1328\\
$\mathbf{HH~998}$& \ref{sec:HH998}, \ref{sec:blanksAC}& \ref{fig:fig14}, \ref{fig:fig19}&Probably Blank-East\\
$\mathbf{HH~1149}$& \ref{sec:HH1149}, \ref{sec:final1149}& \ref{fig:fig20}, \ref{fig:fig43}&Unidentified\\ 
$\mathbf{HH~1150}$& \ref{sec:1150}& \ref{fig:fig41}&Probably 129.9-401.6\\ 
$\mathbf{HH~1151}$& \ref{sec:others2}& \ref{fig:fig41}&140.9-351.2\\ 
$\mathbf{HH~1152}$&\ref{sec:HH998}, \ref{sec:blanksAC}& \ref{fig:fig19}, \ref{fig:fig46}&Unknown\\ 

\multicolumn{4}{c}{Unassigned Flows} \\ \hline 
$\mathbf{HH~1153}$&\ref{sec:GroupShocks}& \ref{fig:fig11}, \ref{fig:fig19}&Unknown\\
$\mathbf{Near~Dark ~Arc}$&\ref{sec:MotionsDarkArc}& \ref{fig:fig36}, \ref{fig:fig37}&Unknown\\
$\mathbf{HH~1154}$& \ref{sec:HH1154},& \ref{fig:fig23}, \ref{fig:fig38} &Unknown*\\
\enddata
\tablecomments{~*Source is COUP 593 if the shock is moving westward.}
\end{deluxetable}

\newpage
\subsection{Sources of the Outflows.}
\label{sec:sources}

A major goal of this program was to determine patterns in the outflows from young stars in the central part of the ONC.
The intent of this was to see if one can narrow down the properties of the stars with outflows in order to inform the process by which stars and protoplanetary disks are formed.

\subsubsection{Sources of Outflows in this Study.}
\label{sec:SourcesOurs}

It is worth examining the properties of the sources of the HH flows in order to establish patterns.
Of the 31 flows where we have made an identification there were two that had an ambiguity of the source and one (HH~626) that was not a source of collimated outflows, leaving 28 with single candidate sources. 
Of these eleven were stars visible in the optical and 18 of the total were COUP sources. This means that 
about one-third of the sources were not subject to heavy foreground extinction and two-thirds were young enough to be X-ray sources. Nine of the ten visible sources were also COUP sources.

We would expect to find molecular outflows where there is high extinction, otherwise the molecules would be photo dissociated into atomic flows. If one sees an optical HH object arising from a molecular outflow, one may see features where the outflow passes through an ionization front. There are three known optical HH objects with molecular outflows driving them [EC 14 with the Southwest Shock, EC 13 (137-408) with HH~1147, COUP 555 with HH 625].  HH 625 clearly appears to be passing through an ionization front. The HH~1132 outflow arising from COUP 632  also has the characteristics of collimated outflow passing through an ionization front, but there is no known associated molecular outflow. 

In our study we identify COUP 666 with the large bipolar flow HH~202, HH~203~and~HH ~204, the multiple high ionization shocks composing HH~529, and  the linear feature HH~1129. The star is bright optically and is seen in X-rays and the infrared. None of the associated outflows show break-through features associated with passing through an ionization front, which indicates that their source is not imbedded. It is very important to study the optical spectrum of this star as it is the optically brightest candidate for the source of outflows. 

\subsubsection{Comparison of These Results with those in an Earlier Study.}
\label{sec:SourcesRivilla}

\citet{riv13} recently reported a study intended to compare outflows in the Orion-S region with X-ray stars in the COUP list \citep{get05}. Their sample came from within the North FOV area of our GO 12543 FOV, which means that we have high quality optical data of the area they investigated. Their major emphasis was to establish if the sources of molecular outflows determined from radio observations were also high-extinction COUP stars. They established that six of the seven sources were COUP stars, the only exception being EC 13 (135.7-408.2), which is seen only as a millimeter source. This establishes that
COUP stars with high-extinction can produce molecular outflows. 

They then looked for at whether the sources of optical HH objects were also COUP sources. For this purpose they used the positions of HH objects from a ground-based a Fabry-Perot image published by \citet{smi04}. That study 
identified only HH 202, HH 203, HH 204, HH 269, HH 528, HH 529, HH 530, and HH 625. As we have shown in this study, these HH objects are much more complex that indicated in the ground-based image they used and the best available spatial resolution and derivation of motions is necessary to get accurate results. They concluded that HH 529 was driven by COUP 632 and argued more generally that the other outflows were driven by non-specified high-extinction sources in Orion-S.  As we have determined in the present study, the three previously known bow shocks in HH 529 are probably driven by COUP 666 rather than COUP 632. The sources we posit for the other HH objects they consider do lie within the Orion-S region, but, the devil is in the detail. 

In their appendices they give an in depth discussion of three well studied and luminous sources in the Orion-S region (COUP 554, COUP 555, and COUP 632) all of which are sources of molecular outflows. In this study we identify COUP 555 as the probable source for HH 625. We identify COUP 632 as the source for HH~1142. We find no optical HH objects with COUP 554 as a source. 

The statement in Rivilla et al. (2013) "Although some of the HH objects might be produced by stars of the ionized nebula ONC seen at optical wavelengths, it is expected that most of them are powered by stars still embedded in the dense molecular core in ONC1-S." merits examination in the light of our new results.
If this were rigorously true, one would not expect to detect sources of outflows in the visual, but in the preceding section we see that one-third of the identified sources were detected there. 

Evidently the process of collimated stellar outflow can occur both in a high-extinction region like Orion-S and also in a region of low density ionized gas. The presence of X-ray emission also occurs in optically visible stars and this cannot be used as proof that the Orion-S sources are younger.

\subsection{Summary of the Conclusions}
\label{sec:conclusions}

We have found that our new data have allowed identification of the source of most of the optical outflows. In the case of the seven molecular outflows, three of these can be related to optical outflows, one of which (HH~625) clearly has the characteristics of breaking through an ionization front. We have also found that a significant fraction of the outflows not related to molecular outflows have optically visible sources. This argues that the period of collimated outflows survives even though the star lies in a ionized region. There are two regions from which outflows seem to originate that do not have a candidate source that is currently identified. Multiple shocks diverging from a single sources are not uncommon, indicating that some stars have outflows directed in many directions, at different but similar times.  This explains why many apparently aligned outflows are always blue shifted. 

\acknowledgements

The late Robert H. Rubin, our colleague of many years, was the original Principal Investigator of HST program GO-12543 which produced the primary images used in this study. 
He was a career-long student of the Orion Nebula and the physical processes occurring in ionized nebulae. He made many contributions in his numerous research papers and his presence at professional meetings was always a source of pleasure and stimulation.

In this study we have made extensive use of the SIMBAD database, operated at CDS, Strasbourg, France and its mirror site at Harvard University.

We are grateful to the referee, Professor John Bally, for his helpful comments on this paper.
 GJF acknowledges support by NSF (0908877; 1108928; and 1109061), NASA (10-ATP10-0053, 10-ADAP10-0073, and NNX12AH73G), JPL (RSA No 1430426), and STScI (HST-AR-12125.01, GO-12560, and HST-GO-12309).  MP received partial support from CONACyT grant 129553. WJH acknowledges financial support from DGAPA--UNAM through project PAPIIT IN102012. CRO's participation was supported in part by HST program GO 12543.

{\it Facilities:} \facility{HST {(WFC3)}}

\appendix
\section{ERRORS IN POSITION COMPILATIONS}
\label{AppendixA}

During the examination of the stars within Figure~\ref{fig:fig39} it was noted that a star at 5:35:15.76 -5:23:38.4 is not listed in any of the sources summarized in SIMBAD. The closest star is COUP 734 (5:35:15:73 -5:23:37.9), which is 0.67\arcsec\ to the NW. Our new star (257.6--338.4) is present on both the first and second epoch images within 0.04\arcsec\ and the discrepancy cannot be explained as its being of high tangential velocity.

The source DR 1186 was reported as seen in ground-based images by \citet{dar09}.
It was not detected in previous visual images and does not appear in GO 12543 F547M images. Similarly, it does not appear in the HST Legacy Program catalog of visual and NIR images compiled by \citet{rob13}. It should be noted that in their
image of the inner Orion Nebula (Figure~11 of \citet{rob13}) the stars and the nebula are in registration, but the coordinates scales shown are in error.

In the examination of our new images we noted two stars in the \citet{dar09} list that are probably mistaken positions for nearby
stars. Their star 305 (5:35:10.71 -5:23:46.1) is probably \citet{hil00} 224 (5:35:10.73 -5:23:44.6). Their star 1814 (5:35:11.77 -5:23:53.4) is probably the star designated in \citet{hil00}  as 205 (5:35:11.70 -5:23:51.3) and in \citet{rob13} as 3405 (5:35:11.73 -5:23:51.7).

\section{NEW TANGENTIAL VELOCITIES IN THE HUYGENS REGION}
\label{AppendixB}
\placetable{tab:TanVel}
\begin{deluxetable}{lcccc}
\tabletypesize{\small}
\tablecaption{Tangential velocities* \label{tab:TanVel}}
\tablewidth{0pt}
\tablehead{
\colhead{Position} &
\colhead{PA(\arcdeg)~\oiii} &
\colhead{\vt \oiii*} & 
\colhead{PA~(\arcdeg)~\nii} &
\colhead{\vt~ \nii* }\\
\colhead{Designation} &
\colhead{---} &
\colhead{---} &
\colhead{---} &
\colhead{---}}
\startdata
053-509&317&98&---&---\\
053-509&318&115&---&---\\
053-510&341&83&---&---\\
057-516&319&50&---&---\\
065-343&---&---&284&27\\
076-343&---&---&265&93\\
077-342&296&67&---&---\\
077-343&---&---&284&64\\
077-343&---&---&284&69\\
077-344&---&---&266&88\\
077-348&284&59&---&---\\
077-348&---&---&256&51\\
078-345&---&---&266&75\\
079-343&284&30&---&---\\
085-531&---&---&320&25\\
090-339&284&21&---&---\\
090-340&274&52&302&39\\
094-336&296&61&---&---\\
096-329&274&59&---&---\\
097-345&---&---&279&32\\
096-346&---&---&270&48\\
096-347&---&---&254&47\\
098-330&250&65&---&---\\
098-412&---&---&212&15\\
101-329&252&68&---&---\\
102-330&---&---&233&15\\
102-416&279&69&---&---\\
103-313&279&67&---&---\\
103-418&---&---&198&18\\
104-417&---&---&274&28\\
105-325&257&24&---&---\\
106-322&284&47&---&---\\
106-325&---&---&265&26\\
107-314&276&38&---&---\\
107-322&309&28&---&---\\
107-342&272&49&---&---\\
107-430&---&---&232&25\\
108-307&278&96&---&---\\
108-418&111&65&---&---\\
108-430&---&---&246&34\\
108-430&---&---&246&34\\
109-313&---&---&296&28\\
109-321&---&---&299&40\\
109-323&302&31&---&---\\
109-426&---&---&259&30\\
109-430&---&---&232&25\\
109-430&---&---&232&25\\
109-431&---&---&255&50\\
110-321&---&---&298&26\\
110-345&---&---&264&26\\
110-346&---&---&270&42\\
110-347&---&---&242&37\\
110-347&---&---&270&33\\
110-351&---&---&286&29\\
110-416&---&---&262&95\\
110-423&---&---&248&27\\
110-423&---&---&333&15\\
110-423&---&---&248&27\\
112-320&---&---&333&24\\
112-324&---&---&284&32\\
112-349&---&---&298&41\\
111-323&---&---&302&40\\
111-347&250&43&---&---\\
112-351&---&---&262&16\\
112-352&---&---&252&19\\
112-423&---&---&237&27\\
113-324&---&---&69&50\\
113-324&---&---&282&50\\
113-324&37&97&---&---\\
113-346&---&---&284&40\\
113-400&277&15&---&---\\
113-422&---&---&176&21\\
113-357&260&50&---&---\\
113-422&---&---&176&21\\
114-421&---&---&195&18\\
114-421&---&---&208&12\\
114-422&---&---&208&12\\
115-352&---&---&293&26\\
115-359&---&---&276&29\\
116-327&---&---&242&17\\
116-332&279&30&---&---\\
116-344&298&40&---&---\\
116-346&269&48&---&---\\
116-348&279&54&---&---\\
116-352&---&---&282&38\\
116-356&289&20&---&---\\
116-359&277&29&269&28\\
116-359&272&30&272&30\\
116-417&---&---&225&35\\
117-342&---&---&284&23\\
117-400&---&---&279&29\\
117-401&---&---&297&18\\
118-320&288&27&---&---\\
118-346&278&46&279&53\\
118-348&261&81&---&---\\
119-400&297&18&---&---\\
119-415&---&---&215&22\\
120-414&---&---&242&27\\
121-325&---&---&275&16\\
121-353&---&---&257&11\\
121-408&211&14&---&---\\
122-319&338&21&---&---\\
123-402&259&12&---&---\\
123-408&208&19&---&---\\
123-552&---&---&240&15\\
124-405&---&---&201&17\\
125-316&318&22&---&---\\
125-405&---&---&169&17\\
125-406&230&18&239&19\\
125-548&---&---&238&14\\
126-346&---&---&270&40\\
126-403&---&---&192&37\\
126-403&---&---&239&6\\
127-403&---&---&246&23\\
130-433&232&15&---&---\\
130-544&---&---&290&11\\
132-342&256&38&258&26\\
133-342&245&70&255&40\\
133-344&308&43&---&---\\
133-352&---&---&282&20\\
134-353&244&26&---&---\\
135-332&---&---&265&21\\
135-347&319&39&---&---\\
135-352&---&---75&15\\
136-317&---&---&296&159\\
136-428&213&14&---&---\\
136-508&---&---&6&13\\
136-353&---&---&278&15\\
137-351&---&---&280&14\\
137-352&---&---&271&49\\
137-408&---&---&155&25\\
137-514&---&---&150&37\\
138-254&---&---&275&10\\
138-350&---&---&257&33\\
138-519&---&---&191&41\\
139-350 & 251&15&273&15\\
139-353&---&---&276&15\\
139-516&---&---&190&30\\
141-421&342&18&342&20\\
141-425&234&30&---&---\\
142-356&349&9&---&---\\
142-357&24&6&---&---\\
142-358 & 21 & 8 &---&---\\
143-325&275&24&---&--\\
143-351&---&---&284&8\\
143-357&43&12&---&---\\
143-418&9 & 19 &---&---\\
143-428&197&36&227&18\\
148-354&---&---&239&15\\
149-352&---&---&104&101\\
149-354&---&---&129&95\\
149-356&241&38&---&---\\
151-351&17&34&---&---\\LV
151-353 &---&107&85\\
151-356&---&---&187&9\\
151-359&---&---&244&19\\
152-353 & 89&87&---&---\\
152-354&---&---&106&96\\
152-358&---&---&201&20\\
153-359 & 255&44&---&---\\
154-305&320&58&---&---\\
154-305&325&62&---&---\\
154-328&---&---&246&16\\
154-344 & ---&---&225&10\\
154-401 ---&---& 160&25\\
155-327 & 254&39&305&13\\
157-327&246&38&---&---\\
155-354&257&13&---&---\\
158-324 & 290 & 38 & ---&---\\
158-326 & 276&38&---&---\\
158-333&39&5&---&---\\
166-308&23&78&---&---\\
178-328&132&161&---&---\\
179-326&106&64&---&---\\
182-336&159&37&---&---\\
183-327&106&74&---&---\\
184-329&90&68&---&---\\
\enddata
\tablecomments{~*Velocities are in \kms.}
\end{deluxetable}

\newpage
\section{RADIAL VELOCITIES OUTSIDE THOSE OF THE MIF}
\label{AppendixC}
\placetable{tab:FLOWS}

\begin{deluxetable}{lcc}
\tabletypesize{\small}
\tablecaption{Radial velocities outside those of the MIF \label{tab:RadVel}}
\tablewidth{0pt}
\tablehead{
\colhead{Position} &
\colhead{\vrad (\oiii)*} & 
\colhead{\vrad (\nii)* }\\
\colhead{Designation} &
\colhead{---} &
\colhead{---}}
\startdata
106-310 & -24& ---\\
106-337 &-45& ---\\
106-416 & -52&---\\
108-315  &-18&---\\
108-343&  -40&---\\
108-351&  +2&---\\
108-413&  -49&---\\
109-314 & -20&---\\
109-323 &---& -4\\
109-345  &-39&---\\
109-351 &---& +2\\
110-322&---&  -5\\
110-345 & -40&---\\
112-323&---&  -4\\
112-346&  -41&---\\
113-324&---&   -12\\
113-348 &-40&---\\
114-348 & -40&---\\
114-426 & +5&---\\
115-323 & -32&---\\
117-309 & -20&---\\
117-319  &-31&---\\
117-354 & -39&---\\
  118-345 &---& 0\\
119-309&-54&    ---\\       
121-310& -55&---\\
 121-327&  -25&---\\
 121-334&  -34&---\\
 121-336 & -23&---\\
122-314  & -20&---\\
122-315 & -58&---\\
122-437&---&   +25\\
123-313 & -21&---\\
 123-315&  -19&---\\
 123-358 & -58&---\\
 123-323&  -17&---\\
 123-337 & -30&---\\
123-428  & -19&---\\
 125-317& -56&---\\
 125-319 & -16&---\\
 125-428 & -17&---\\
126-319 & -53&---\\
126-320 & -13&---\\
126-338 & -28&---\\
126-342 & -30&---\\
 126-346 & -45&---\\
 126-348 & -29 &---\\
 126-419 & -33&---\\
126-429 & -19  &  ---\\                       
127-318 & -11&---\\
127-321&  -46&---\\
 127-338 & -27&---\\
127-341 & -33&---\\
 127-342&  -38&---\\
 127-344 & -45&---\\
127-347 & -46&---\\
 127-348&  -22&---\\
129-320 & -14&---\\
 129-325 & -43&---\\
129-339 & -29&---\\
129-341 & -34&---\\
129-342 & -39&---\\
129-344&  -41&---\\
 129-347 & -43&---\\
 130-327 & -43&---\\
 130-333 & -37&---\\
130-339 & -31&---\\
130-340 & -33&---\\
130-341 & -38&---\\
130-344 & -44&---\\
131-329 & -39&---\\
131-342 & -36&---\\
  133-330&  -36&---\\
  133-341&  -35&---\\
                                     134-232 & -34&---\\
                                     134-339 & -34&---\\
                                     134-429 & -35&---\\
                                     135-316&  -41&---\\
                                     135-333 & -30&---\\
                                     135-401 & -47&---\\
135-432&---&  -37\\
 137-315&---&   -100\\
                                       137-330&  -30&---\\
                                       137-335&  -20&---\\
                                       137-337&  -32&---\\
                                       137-348 & -32&---\\
                                        138-317 & -38&---\\
                                        138-333 & -30&---\\
                                        138-340 & -32&  ---\\                                     
                              139-318 & -37&---\\
                              139-321 & -23&---\\
                              139-332 & -25&---\\
                              139-336 & -28&---\\
                              139-340 & -31&---\\
                              139-346 & -21&---\\
                              139-433 & -32&---\\
                               141-318 & -34&---\\
                               141-319 & -45&---\\
                               141-324 & -21&---\\
                               141-332 & -22&---\\
                               141-336 & -26&---\\
                               141-342  &-32&---\\
                               141-347 & -34&---\\
                                142-328 & -37&---\\
                                142-329 & -21&---\\
                                142-330 & -23&---\\
                                142-427 & -22&---\\
                                 143-327 & -35&---\\
                                 143-339 & -40&---\\
                                 143-425&  -18&---\\
                                 143-427 & -24&---\\
                                  145-330 & -44&---\\
                                  145-338  &-43&---\\
                                  145-344 & -31&---\\                           
                                   146-331 & -41&---\\                                 
                                    147-324&  -8&---\\
                                    147-333  &-32&---\\
147-354 &---&  -42\\
                                    147-428 & -20&---\\
                                    149-323 & -19&---\\
                                    149-324 & -25&---\\
                                    149-326 & -42&---\\
149-353&---&   -40\\
149-355 &---&  -42\\
                                     150-324 & -25&---\\
                                     150-333 & -27&---\\
                                     150-352 & -44&---\\
 150-354&---&  -42\\
                                      151-323 &  -30&---\\
                                      151-326  & -7&---\\
                                      151-336   &-15&---\\
151-354&---&  -41\\
                                      151-352&  -45&---\\
                                      151-353 & -48&   ---\\                             
                                       153-323 & -30&---\\
                                       153-326&  -4&---\\
                                       153-336 &  -16&---\\
153-354&---&   -42\\
                                       153-355 & -47&---\\
                       154-322 & -25&---\\
                        154-339 & -13&---\\
                        154-355 & -53 &   -45  \\                        
                        154-404 & -8   &  ---\\                                                                 
                         155-323  & -25&---\\
                         155-341 & -15&---\\
                         155-355 & -53 & -45\\
                         155-426 & -14&---\\
                          157-337&  +93&---\\
                          157-339 & -16&---\\
                          157-356 & -44 & -38 \\                    
                            158-324 & +82&---\\
                            158-332 & -30&---\\
                            158-337 & -27&---\\
                            158-339 & +67&---\\
                            158-359 &  -32 &---\\
                             159-351 & -36&---\\
                             159-353& -29 &  -25\\
                             159-423 & +24&---\\
                              161-343 &+122&---\\
                              161-353 & -33&---\\
                              161-354 & -24 & -27\\
                  162-328 &  -11&---\\
                  162-343 &  +103&---\\
                  162-352 & -16&---\\
                  162-353 & -39 &    ---\\                            
                   163-326 & -39&---\\
                   163-338&  +70&---\\
                   163-343 & +95&---\\
                   163-355 & -8&---\\
                   163-359 & -6&---\\
                   163-403 & -4&---\\
                   163-455 & -67 &  ---\\                                                     
                    165-358 & -22&---\\
                    165-405 & -4&---\\
                     166-400 & -28&---\\
 166-358&---& -33\\
                       167-327 & +25&---\\
 167-358 &---&  -27\\
                       167-359 &  -30&---\\
                        169-339 & -9&---\\
                        169-359 & -29 & -22\\
                          170-328 & +115&---\\
                          170-330 & +108&---\\
                          170-332  & +110&---\\
                          170-337&  +151&---\\
                          170-339 & -4&---\\
                          170-3256 & -18&---\\
                          170-357 & -31&---\\
                          171-336 & -37&---\\
                          171-339 & -54&---\\
                           173-339  & +52&---\\
                           173-339  & -57&---\\
                             174-336 & -40&---\\
                             174-345  & -51&---\\
                              176-332 & +5 &---\\                        
\enddata
\tablecomments{~*Velocities are in \kms, Heliocentric. Subtract -18.1 \kms\ for L.S.R.}
\end{deluxetable}

\section{USING THE STRENGTH OF LINES TO ANALYZE HH~1149}
\label{AppendixD}

It is in the ratio of signals of different emission line components of the high resolution spectra that the greatest deviations occur. Therefore, we derived the ratio
of different components for the various samples. The signal units (S) as we have measured them are in arbitrary units (numerous normalizations have been made since extracting the data from the Atlas). The results for this signal study are also shown in Table~\ref{tab:HH1149}. Unfortunately, in the \oiii\ NEAR sample the association of the line components is unclear.

If the standard blister model applies, the ionization ratio of  \oiii\ to \nii\ is expected to decrease with the distance from \tc.
The MIF signal ratio is S(\oiii)/S(\nii)~=~0.29 for SHOCKS and 0.44 for NEAR if the 9.7 \kms\ and 20.9 \kms\ components are added together (if only the 20.9 \kms\ component is counted the signal ratio is 0.08).
These are the not the type of changes expected in moving outward from \tc, indicating that the two nearby samples are quite different regions. The average of five of the large samples (without C and D)
is 0.60$\pm$0.11 and for the average of C and D is 0.06$\pm$0.02.  This indicates that the signals from C and D come from lower ionization zones
 than the other large samples. This is consistent with their both being near where jets are producing shocks in the foreground Veil.
The SHOCKS sample  is significantly lower than the large samples while the NEAR  is closer to the SHOCKS sample than the Large Samples. The NEAR sample 20.9 \kms\ component ratio is much closer to the values of samples C and D  and the 9.7 \kms\ component is closer to the large samples without C and D.

For the HH~1149 samples the \nii\ S(BLUE-\nii)/S(MIF-\nii) ratios are similar  for SHOCKS (0.02) and NEAR (0.06)
This ratio is 0.07$\pm$0.05 for the average of all the large samples (except for 
sample E, where the value is a deviant 0.15 (sample E falls on a foreground high extinction feature, which may be the cause). We can conclude that the NEAR sample is coming from a region similar to the greater part of the \hr\ and that the SHOCKS sample deviates in that the BLUE component is anomalously weak.

The differences are greater when examining the BLUE  \oiii\ values using the S(BLUE-\oiii)/S(MIF-\oiii) ratios. 
 In the SHOCKS sample this ratio is 0.06. Only four of the large samples could be averaged, giving 0.17$\pm$0.09. 
For the two large samples (A and G) lying on the Orion-S molecular cloud the ratio was 0.21 and 0.29. Sample-C's value was much greater than unity and probably indicates that the LOW component is actually the result of being on the edge of the HH 204 shock.  Sample-D had no identifiable LOW \oiii\ component. Sample-B, which lies in the high ionization zone of \citet{ode09}, had a value of only 0.02. The SHOCKS sample value of 0.06  is well less than the average of the four samples, but much higher than in Sample-B.
These numbers indicate that the \oiii\ BLUE component in the SHOCKS sample including HH~1149 is 
intermediate in value from most of the large samples and the high ionization sample B.

As noted above the NEAR \oiii\ line profile is divided into three components. The component at 9.7 \kms\ has a relative signal of 1.00, the 20.9 \kms\  is weaker at 0.09, and the backscattered 33.4 \kms\ has a signal ratio of 0.12. If the 9.7 \kms\ component is the blue component then S(BLUE-\oiii)/S(MIF-\oiii) is highly anomalous at 11.2. If the 9.7 \kms\ component is the MIF component, then the ratio is 0.0.

\subsubsubsection{Testing the Methodology Using \vrad}
\label{sec:HH1149tests}

Several tests of the association of the SHOCKS and NEAR velocity components with similar components can be made. The first are the \vrad\ values. The \vrad\ values for the strong MIF emission in the individual large samples are probably accurate to better than 2 \kms. Variations between the samples can be larger than this because of local variations.  Except when the BLUE component is strong compared with the MIF component, the \vrad\ values of the weak individual BLUE components are probably accurate to about 4 \kms.

 All of the \vrad 's for the MIF in SHOCKS and NEAR are similar to the averages of the large samples (Table~ \ref{tab:HH1149}), except
 for the NEAR \oiii\ sample. 
 In the case of the weak SHOCKS-BLUE \nii\ component, its value of -4.7 \kms\ is only slightly different from that for the large samples (1.3 $\pm$2.5 \kms). The NEAR-BLUE \nii\ value 2.9 \kms\ is closer to the large samples.  
The SHOCKS-BLUE \oiii\ velocity of 1.1 lies within the scatter of the large 4 samples. At 9.7 \kms\ for the  bluer NEAR \oiii\ component, this strongest feature is more positive than the other BLUE components (2.6$\pm$2.2 \kms). 
 
 \subsubsubsection{Testing the Methodology Using \vrad\ Differences}
 
 Another test is the difference in velocity of the \oiii\ and the \nii\ MIF components. This difference is due to material flowing through the PDR into the MIF and becoming more highly ionized on the \tc\ side of the PDR. 
The average \vrad(\oiii) - \vrad(\nii) for the seven large samples is -4.5$\pm$1.6 \kms. This can be compared with -4.9 \kms\  and -10.3/-0.1 \kms\ in the SHOCKS and NEAR samples respectively. The SHOCKS \vrad\ difference is similar to the seven large samples, but the NEAR sample differs either high or low according to the assignment of its components, again pointing to a special nature of the nebula in this region.

\subsubsubsection{Testing the Methodology Using Derived PDR \vrad 's}

A final test is that the average of the MIF and the Red scattered-light component should equal the velocity of the PDR (\vrad(PDR)), within the approximation of fitting the Red component to a single Gaussian line. For our seven big samples \vrad(PDR) was 27.2$\pm
$2.9 \kms, in good agreement with the OMC velocity (25.8 \kms).  \vrad(PDR) was 29.8 \kms\ and 28.0 \kms\ for SHOCKS in \nii\ and \oiii\ respectively.  \vrad(PDR) was 29.1 \kms\ for NEAR in \nii, indicating that the SHOCKS sample the MIF and RED components are correctly identified in \nii. In the case of the NEAR \oiii\ sample the shorter (9.7 \kms) and longer (20.9 \kms) components give \vrad(PDR) values of 21.6 \kms\ and 25.1 \kms\ respectively, which argues that the longer and weaker component is due to the MIF.
 The signal ratio S(Red-\nii)/S(MIF-\nii) is 0.06 and 0.07 for SHOCKS and NEAR, while it is 0.13$\pm$0.04 for the average of the Big Samples, so that both regions have weaker \nii\ Red components than the surrounding region. 
S(Red-\oiii)/S(MIF-\oiii) is 0.10 and 0.12/1.38 for the SHOCKS and NEAR samples, whereas it is 0.10$\pm$0.04 for the Big Samples. This means that the SHOCKS sample is indistinguishably the same as the Big Samples in both ions,  and in the SHOCKS sample the agreement is only in \nii. By the criterium of the RED to MIF flux ratio, the NEAR 9.7 \kms\ component is more likely the MIF feature, in contradiction with the \vrad(PDR) results. 

All the above indicates that we cannot clearly identify the sources of the two shorter velocity components of the NEAR \oiii\ features. 

 In a study now in progress we find multiple other regions with strong, even dominant, \oiii\  BLUE velocity components having velocities that associate them with the BLUE velocity components that are seen throughout the \hr. These strong \oiii\ BLUE velocity areas must be samples of the material seen throughout the nebula, but are under different conditions of illumination by \tc. They can also be the result of forward scattering from previously undetected material. The multiple shocks that constitute HH~1149 appear to be encroaching on one of these peculiar areas.

\clearpage

\end{document}